\documentclass[12pt]{article}
\usepackage{url} 

\newcommand{\blind}{0}

\usepackage{xr}
\makeatletter

\newcommand*{\addFileDependency}[1]{
\typeout{(#1)}
\@addtofilelist{#1}
\IfFileExists{#1}{}{\typeout{No file #1.}}
}\makeatother

\usepackage[utf8]{inputenc}
\usepackage{afterpage}
\usepackage{enumerate}
\usepackage[OT1]{fontenc}
\usepackage{bm}
\usepackage{multirow}
\usepackage{diagbox}
\usepackage{amsmath,amssymb,mathrsfs}
\usepackage{natbib}
\usepackage{bbm, dsfont}
\usepackage{makecell}
\usepackage{upgreek}
\usepackage{mathtools}
\usepackage[usenames]{color}
\usepackage{dsfont,makecell,chngcntr}
\usepackage{smile}
\usepackage{multirow,stfloats,booktabs}
\usepackage{setspace}
\usepackage[utf8]{inputenc}
\usepackage[english]{babel}
\usepackage{graphicx} 
\usepackage{xcolor}
\definecolor{darkred}{RGB}{150,50,50}
\definecolor{brown}{RGB}{250,100,100}
\definecolor{green}{RGB}{000,150,100}
\definecolor{navy}{RGB}{000,000,150}

\usepackage{subcaption}
\usepackage{makecell}
\usepackage[colorlinks,
linkcolor=red,
anchorcolor=blue,
citecolor=blue
]{hyperref}

\def\tr{\mathop{\text{tr}}\kern.2ex}

\def\bX{{\Xb}}

\def\bc{{\boldsymbol{c}}}
\def\bPhi{{\boldsymbol{\Phi}}}
\def\U{{\Ub}}
\def\bTheta{{\boldsymbol{\Theta}}}
\def\bLambda{{\boldsymbol{\Lambda}}}
\def\Vb{\mathbf{V}}

\def\P{{\mathbb P}}
\def\E{{\mathbb E}}

\def\supp{\mathop{\text{supp}}}

\long\def\comment#1{}

\def\tr{\mathop{\text{Tr}}}

\newcommand{\bel}{\begin{eqnarray}\label}
\newcommand{\eel}{\end{eqnarray}}
\newcommand{\bes}{\begin{eqnarray*}}
	\newcommand{\ees}{\end{eqnarray*}}

\let\hat\widehat
\let\tilde\widetilde
\def\mid{\,|\,}

\def\P{{\mathbb P}}

\def\supp{\mathop{\text{supp}\kern.2ex}}

\def\tr{{\rm{Tr}}}

\def\supp{\mathop{\text{supp}}}

\def\tr{\mathrm{Tr}}

\def\PMIbb{\mathbb{PMI}}
\def\PMIbbhat{\widehat{\PMIbb}}

\usepackage{mathrsfs}

\usepackage{fullpage}

\usepackage{tabularx,array}
\newcolumntype{M}{>{\centering\arraybackslash}m{0.17\textwidth}}

\newcommand{\zx}[1]{#1}
\usepackage{hyperref}
\usepackage[    protrusion=true,
expansion=true,
final,
babel
]{microtype}
\def\##1\#{\begin{align}#1\end{align}}
\def\$#1\${\begin{align*}#1\end{align*}}
\usepackage{enumitem}

\theoremstyle{plain}

\theoremstyle{mytheoremstyle}

\date{}

\begin{document}


\if0\blind
  \title{\bf Inference of Dependency Knowledge Graph for Electronic Health Records}
{
  \author{
  Zhiwei Xu$^{1\sharp}$, Ziming Gan$^{2\sharp}$\footnote{Xu and Gan contributed equally.}, Doudou Zhou$^{3}$, Shuting Shen$^{3}$, \\
  Junwei Lu$^{4*}$, Tianxi Cai$^{4,5}$\footnote{Corresponding authors Lu and Cai contributed equally.}\bigskip \\
\small 
$^1${Department of Statistics, University of Michigan, Ann Arbor, MI, USA} \\
\small 
$^2${Department of Statistics, University of Chicago, Chicago, Il, USA} \\
\small 
$^3${Department of Statistics and Data Science, National University of Singapore}\\
\small
$^4${Department of Biostatistics, Harvard T.H. Chan School of Public Health, Boston, MA, USA}\\
\small 
$^5${Department of Biomedical Informatics, Harvard Medical School, Boston, MA, USA}}
  \maketitle
} \fi

\if1\blind
{
  \bigskip
  \bigskip
  \bigskip
  \begin{center}
    {\LARGE\bf Inference of Dependency Knowledge Graph for Electronic Health Records}
\end{center}
  \medskip
} \fi

\bigskip
\begin{abstract}
The effective analysis of high-dimensional Electronic Health Record (EHR) data, with substantial potential for healthcare research, presents notable methodological challenges. Employing predictive modeling guided by a knowledge graph (KG), which enables efficient feature selection, can enhance both statistical efficiency and interpretability. While various methods have emerged for constructing KGs, existing techniques often lack statistical certainty concerning the presence of links between entities, especially in scenarios where the utilization of patient-level EHR data is limited due to privacy concerns. In this paper, we propose the first inferential framework for deriving a sparse KG with statistical guarantee based on the dynamic log-linear topic model proposed by \cite{arora2016latent}. Within this model, the KG embeddings are estimated by performing singular value decomposition on the empirical pointwise mutual information matrix, offering a scalable solution. We then establish entrywise asymptotic normality for the KG low-rank estimator, enabling the recovery of sparse graph edges with controlled type I error. Our work uniquely addresses the under-explored domain of statistical inference about non-linear statistics under the low-rank temporal dependent models, a critical gap in existing research. We validate our approach through extensive simulation studies and then apply the method to real-world EHR data in constructing clinical KGs and generating clinical feature embeddings.
\end{abstract}

\noindent%
{\it Keywords:}  low-rank models, non-linear structure, knowledge graph embedding, hypothesis testing
\vfill

\newpage

\section{Introduction}

Electronic Health Record (EHR) data is increasingly recognized for its potential to revolutionize healthcare research \citep{Ahuja2022,wen2023multimodal}. Its high-dimensional nature, however, poses significant methodological challenges. Constructing clinical knowledge graphs (KGs) from EHR data has become a prevalent approach, offering an effective way to understand the complex interrelations among diverse EHR features. This understanding is crucial for enhancing the efficiency of predictive modeling tasks in various medical applications, such as drug analysis and disease diagnosis \citep{sang2018sematyp,abdelaziz2017large}. The utility of these graphs extends to improving clinical decision-making \citep{bauer2013network,finlayson2014building,nelson2022embedding} and facilitating the integration and sharing of EHR data \citep{hong2021clinical,abu2022healthcare}. 

In recent years, a range of methods for constructing KGs has emerged, notably in the form of KG embeddings. These methods, including translation-based models \citep[e.g.]{bordes2013translating,wang2014knowledge}, tensor factorization-based models \citep[e.g.]{nickel2011three,yang2014embedding}, and neural network-based models \citep[e.g.]{socher2013reasoning,bordes2014semantic}, effectively map entities and relations into a low-dimensional vector space, capturing the semantics and structure of KGs. Following the advent of word embedding algorithms in natural language processing \citep{mikolov2013distributed}, these KG embeddings utilize distributed representation technology and address data sparsity and computational inefficiency challenges \citep{dai2020survey}.
However, a critical limitation of existing techniques is the lack of statistical certainty in the presence of links between entities. This uncertainty quantification is particularly crucial in healthcare applications where accurate and reliable data interpretations are vital for patient care and treatment decisions. Most KG algorithms in EHR data analysis \citep{rotmensch2017learning,chen2019robustly,zhang2020hkgb,shang2021ehr,harnoune2021bert,roy2021incorporating,jiang2022gated}, however, can not provide this level of certainty, especially when the use of patient-level data is constrained by privacy concerns. 

To fill this gap, we propose the first inferential framework for deriving a sparse KG with uncertainty quantification. This framework is based on the dynamic log-linear topic model proposed by \cite{arora2016latent}. Under this model, the KG embeddings of the EHR entities can be efficiently approximated via a low-rank representation of the population pointwise mutual information (PMI) matrix. As such, these embeddings can be estimated by performing a singular value decomposition (SVD) of the empirical PMI matrix. We derive the asymptotic properties of this estimator to address estimation uncertainty and provide statistical guidance for edge selection in KGs. 
To estimate the variance for the estimator, we propose two methods. The first leverages patient-level co-occurrence data for precise variance estimation. While this method offers high accuracy, it raises concerns about data privacy and requires substantial computational resources, particularly with large datasets. We hence propose the second method for situations where patient-level data is not available. Here, we approximate the variance of the estimator under a global null hypothesis of no dependency between EHR entities. This method, prioritizing patient privacy, involves aggregating and summarizing patient data before analysis. It offers improved computational efficiency and enhanced privacy protection. This alternative has shown considerable promise in EHR data applications, balancing the need for privacy with computational efficiency.
Leveraging the asymptotic normality of our results, we conduct hypothesis testing on the existence of the edges and incorporate the Benjamini-Hochberg (BH) procedure under dependence \citep{benjamini2001control} to regulate the False Discovery Rate (FDR).

\subsection{Our Contributions}

The dynamic log-linear topic
model, foundational to our work, has been adapted in several studies \citep{arora2018linear,xxx2021,zhou2021multi}. However, these adaptations have primarily focused on estimation, leaving a gap in uncertainty quantification. Undertaking inferential analysis on KGs presents three major challenges. For data generation, the model's hierarchical structure entangles several time series and brings the data with a complex dependency structure. For estimation, two layers of non-linearity brought about by the logarithmic and division operators shown later in \eqref{eq: pmi def} further complicate the analysis. For data privacy, in many situations, we do not have access to patient-level data thus causing the estimation of covariance structure a problem.

Our work contributes to the field of statistical inference in low-rank models, a domain fraught with challenges. Performing statistical inference in such models is particularly difficult due to the intricate nature of low-rank structures and the complexity of their estimation processes. Recent advances have been made in linear low-rank models, including matrix completion \citep{foucart2017biasing,chen2019inference},
principal component analysis \citep{xia2021normal,xia2022inference} and low-rank matrix
regression \citep{carpentier2019uncertainty,chernozhukov2023inference}. \zx{These work assume that the population matrix itself is low-rank. In contrast, our setting is fundamentally different - the population PMI matrix is not exactly low-rank but emerges from a generative model. We assume that the embedding dimension is low compared to the vocabulary size, which is widely used in both theoretical analysis and in reality where the embedding dimension is only several hundreds and the vocabulary size is more than tens of thousands. We prove that under our generative model, the population PMI matrix can be well approximated by a low-rank matrix, aligning with empirical observations in previous studies \citep{NIPS2014_feab05aa, arora2016latent}. We develop a novel inference framework that leverages a second-order approximation of the PMI matrix and employs
advanced perturbation techniques, such as Davis-Kahan bounds, to establish theoretical guarantees. Unlike existing approaches, our method extends inferential techniques to non-linear PMI estimation under temporal dependent generative models, where the ground truth matrix is derived from the generative process rather than assumed to be low-rank.}

Our research makes several pivotal contributions.  First, we enhance the understanding of the \zx{near} low-rank nature of the population PMI by deriving a shaper approximation rate using KG embeddings. \zx{While prior works \citep{arora2016latent, xxx2021} have explored the approximate low-rank structure of the population PMI matrix, existing analyses primarily focus on two aspects: the empirical PMI estimator without a low-rank decomposition and the estimation consistency, without extending to statistical inference.  The key challenge preventing inference lies in the variance-bias decomposition. Previous approaches relied on a first-order Taylor expansion to approximate the PMI matrix with an inner product of the embeddings. However, the bias term from the first-order expansion is larger than the standard deviation of the estimator, making inference infeasible. A key contribution of our work is the discovery that a second-order Taylor expansion still results in a low-rank structure but significantly reduces the bias, making inference possible.} \zx{Moreover,} instead of assuming certain prior of the embeddings in the existing works \citep{arora2018linear,xxx2021}, the embeddings in our model are deterministic vectors and thus make hypotheses on the cosine similarities between embeddings well-defined.  The sharper rate is also crucial for valid inference.

Second, we provide a method for uncertainty quantification of the estimator by demonstrating its asymptotic normality, addressing a significant gap where no previous valid inference method for non-linear low-rank models \zx{with complex data dependency}. The complexity of this analysis stems from the high non-linearity of the generative model and the dependencies inherent in longitudinal EHR datasets. \zx{The entries of the PMI matrix are non-linear transforms of the co-occurrence counts involving the softmax and log transformations. In the Supplementary  S.1 and S.2.1, we develop a series of novel concentration inequalities for the nonlinear transformation of samples from a geometrically ergodic Markov chain. This inequality plays a crucial role in establishing the asymptotic normality of our PMI estimator. As the empirical PMI matrix is obtained from the occurrence data generated by a latent Markovian process, we cannot apply the standard proof technique to show the asymptotic normality. In Supplementary S.2.2, we propose a new theoretical analysis to decouple the temporal-dependence of the discourse vectors and show the asymptotic normality as long as quantifying the scale of their asymptotic variances. }

Lastly, our study introduces an inference procedure for the KG using only summary-level EHR data. \zx{This approach has significant practical value in healthcare settings where sharing patient-level data is restricted due to privacy regulations. By enabling rigorous statistical inference while preserving privacy, our method provides both a theoretically sound and practically implementable solution for analyzing multi-source EHR data.}

In summary, our work goes beyond existing literature by (i) developing a second-order approximation that enables statistical inference on PMI matrices, (ii) addressing variance estimation and asymptotic normality in the presence of non-linearity and time dependencies, and (iii) proposing a privacy-preserving inference framework for knowledge graph construction. These contributions make our findings both theoretically novel and practically relevant.

\subsection{Paper Organization}

The rest of the paper is organized as follows. Section \ref{sec:method} introduces the generative model of EHR data and proposes the method for KG inference. Section \ref{sec:thm} provides the theoretical properties of the empirical PMI matrix and the entry-wise asymptotic normality of the low-rank estimator. Section \ref{sec:sim} shows the simulation results and Section \ref{sec:app} applies the proposed method to construct a KG of four domains of EHR entities based on the empirical PMI matrix along with other summary statistics derived from EHR data of over $12$ million patients in the Veteran Affairs (VA) health system. The conclusion is given in Section \ref{sec:con}. 
In addition, the proof of the theorems is developed in the supplementary, and the implementation code is available at \url{https://github.com/junwei-lu/WordVec_Inference}.

\section{Method}
\label{sec:method}

\subsection{Notation}
 We use $\boldsymbol{1}_p$ to represent a $p$-dimensional vector with all entries equal to $1$ and $\Ib_p$ to represent the $p \times p$ identity matrix. Let $\mathcal{O}^{p \times p} = \{\Rb\in \mathbb{R}^{p \times p}: \Rb^{\top}\Rb = \Ib_p \}$ denote the set of all $p \times p$ orthonormal matrix. For a vector $\ba \in \mathbb{R}^{p}$, we use $\|\ba\| = \sqrt{\sum_{i=1}^p a_i^2}$ to represent its $l_2$ norm. For a matrix $\Ab \in \mathbb{R}^{n \times m}$, we denote by $\Ab_{i,\cdot} = (A_{i1},\cdots,A_{im})$ (respectively $\Ab_{\cdot ,i}$) its $i$th row (respectively column), $\Ab_{i,i'}$ its $(i,i')$th element, $\|\Ab\|_{\max} = \max_{i,j}|\Ab_{ij}|$ its matrix max norm, $\|\Ab\|_{2, \infty} = \max_{i} \|\Ab_{i,\cdot}\|$ its $2$-to-$\infty$ norm, and $\|\Ab\|$ its matrix operator norm. The Hadamard product of matrices $\Ab$ and $\Bb$ is denoted by $\Ab \circ \Bb$. If a symmetric matrix $\Ab \in \mathbb{R}^{n \times n}$ has eigen-decomposition $\Ub \bLambda \Ub^{\top}$, where $\bLambda = \text{diag}(\lambda_{1},\cdots,\lambda_{n})$, and $|\lambda_{1}|\geq |\lambda_{2}| \geq \cdots \geq |\lambda_n| $ (sort by magnitude), then we use $\lambda_{i}(\Ab) = \lambda_{i}$ to represent the $i$th largest absolute eigenvalue of $\Ab$. For two sequences $\{x_n\}$ and $\{y_n\}$, we say $x_n = O(y_n)$ or $x_n \lesssim y_n$ if there exists some constant $C>0$ such that $x_n \leq C y_n$ for all $n$. We say $x_n = o(y_n)$ if $\lim_{n \rightarrow \infty} x_n / y_n = 0$, and $x_n = \Omega(y_n)$ if $y_n = O(x_n)$. Denote $[d]$ as the set $\{1,2,\cdots,d\}$. We let $\Phi(\cdot)$ be the cumulative distribution function of the standard normal distribution.

\subsection{Generation of the Longitudinal EHR Feature Occurrences}\label{sec:dat_generation}

In this section, we introduce the generative model for the observed longitudinal EHR data. 
Suppose that there are a total of $n$ patients and $d$ unique EHR entities. We use the set $[d] = \{1,2,\ldots,d\}$ to represent the vocabulary of all possible EHR entities (e.g., diagnosis codes, medications, procedures) observed across the patient cohort. We assume that the entities in the EHR are generated from the dynamic log-linear topic model \citep{arora2016latent,xxx2021}. Specifically, for the $i \in [n]$th patient, we observe the longitudinal EHR data as the sequence of EHR entities $\{w_{i,1}, \ldots, w_{i,t}, \ldots,  w_{i,T_i}\}$, where $T_i$ is the last observation time, and $w_{i,t}$ denotes  a single EHR entity (or code) observed at time $t$. An entity \(w_{i,t} \in [d]\)  could represent a diagnosis code, such as an ICD code, a medication code, or any other discrete piece of information recorded in the patient's record at time $t$. For example, if \(w_{i,t}\) is the diagnosis code of Hypertension, it might represent that the patient was diagnosed with ``Hypertension'' at time $t$. Similarly, if \(w_{i,t}\) corresponds to a procedure code for ``Blood pressure measuremen'',  it indicates that the patient underwent a blood pressure measurement at time $t$. These discrete entities provide granular information about the patient’s medical history.

We assume that $w_{i,t}$ follows the following multinomial distribution  
\begin{equation}\label{eq: code prob}
    \PP (w_{i,t} = w|\bc_{i,t}) = \frac{\exp (\langle \Vb_w , \bc_{i,t} \rangle)}{\sum_{k=1}^d\exp (\langle \Vb_{k} , \bc_{i,t} \rangle)}, \text{ for all } w \in [d], 
\end{equation}
where $\Vb_w$ is the KG embedding of the entity $w$, and $\bc_{i,t}$ is the latent discourse vector following  a slow autoregressive (AR) process: 
\begin{equation}\label{eq:hidden_mp}
\boldsymbol{r}_{i,t} \overset{\rm i.i.d.}{\sim} N(0, \Ib_p/p), 
\bc_{i,1} = \br_{i,1}, 
\bc_{i,t+1} =  \sqrt{\alpha} \bc_{i,t} + \sqrt{1-\alpha} \boldsymbol{r}_{i,t+1}, \quad t\geq 1,
\end{equation}
where $p$ is the dimension of the embedding, $\alpha = 1 - (\log d)/p^2$ is a measure of the walking rate aligning with that in \cite{xxx2021},
 $\boldsymbol{r}_{i,t}$  is independent of $\{\bc_{i,k}\}_{k=1}^{t-1}$; and the discourse vectors are independent across patients. This discourse process is stationary in such a setting. 
 We consider that the sequence $\{\bc_{i,t}\}_{t=1}^{T_i}$ follows an AR model different from the random walk of the sphere as in \cite{arora2016latent} to incorporate the heterogeneity of the temporal variance of the discourse. Although this assumption facilitates theoretical analysis, our method is empirically robust to variations in the distribution of $\bc_{i,k}$, as demonstrated in Figure \ref{fig:robust_of_discourse}.

\begin{remark}[Suitability of the Model for EHR Data]
Although the model in \eqref{eq: code prob} was originally developed for text generation \citep{arora2016latent}, it is particularly well-suited for modeling EHR data. Like text data, EHR data consist of sequences of discrete events or entities, such as diagnoses, procedures, or medications, collected over time. The dynamic log-linear topic model effectively captures the temporal dependencies and latent structures inherent in these sequences, making it a natural choice for analyzing longitudinal EHR data.

A key strength of this model is the inclusion of a latent discourse vector $\bc_{i,t}$, which evolves over time to reflect changes in a patient’s medical status. This feature is particularly important in healthcare, where disease progression and treatment responses unfold over time. By modeling $\bc_{i,t}$ as an autoregressive process, the framework accommodates heterogeneous temporal variances in patient records, allowing for more accurate modeling of real-world medical trajectories. 

Finally, the integration of knowledge graph embeddings $\Vb_w$ enhances the interpretability of the model. These embeddings enable the inference of relationships between medical entities, facilitating the discovery of hidden connections among diagnoses, treatments, and outcomes. This makes the model especially useful for extracting clinically meaningful insights from large-scale EHR datasets. 
\end{remark}

Denote the KG embedding matrix by $\Vb = (\Vb_1,\ldots,\Vb_d)^{\top}$. We see from \eqref{eq: code prob} that the probability distribution over the entities is invariant to a constant shift of the rows of $\Vb$, i.e., for any constant vector $\bmu \in \mathbb R^p$, replacing $\Vb$ with $\Vb - \boldsymbol{1}_d\bmu^{\top}$ does not change the probability of occurrence.  In order to have identifiable embeddings, we assume without loss of generality that $\Vb$ is centered, i.e., $\Vb^{\top} \bp = \mathbf{0}$, where $\bp = (p_1,\cdots,p_d)^{\top} \in \mathbb{R}^{d}$ with $p_w = \E_{\bc_{i,t}\sim N(0,\Ib_p/p)}[\P(w_{i,t}=w|\bc_{i,t})]$ being the marginal occurrence probability of feature $w$ over the stationary discourse process. Thus, the identifiability condition $\Vb^{\top} \bp =0$ implies that the feature embeddings are centered at zero under the marginal occurrence probability~$\bp$.

Under this generative model, the KG can be inferred from the embeddings. This is based on the principle that clinically interconnected entities often appear together. For instance, if the latent discourse vector $\bc_{i,t}$ pertains to Alzheimer’s Disease, there is a high probability of observing both the diagnosis code for ``dementias'' and a prescription for ``memantine.'' This suggests that the knowledge embeddings for these entities are spatially proximate to each other and also near $\bc_{i,t}$ in the embedding space. More precisely, the determination of whether an edge exists between any two entities $w$ and $w'$ hinges on hypothesis testing as follows:
\begin{equation}
    {\rm{H}}_0:  \Vb_w^{\top} \Vb_{w'} = 0 \text{\quad versus\quad} {\rm{H}}_1:  \Vb_w^{\top} \Vb_{w'} \neq 0.
    \label{h0}
\end{equation}

The utilization of KG embeddings for predicting edges has gained popularity in constructing KGs. For instance, \cite{hong2021clinical} employed cosine similarity, defined as $\frac{\Vb_w^{\top} \Vb_{w'}}{\|\Vb_w\| \|\Vb_{w'} \|}$, to ascertain the edge weight between $w$ and $w'$. \cite{nickel2011three} introduced the relational learning approach RESCAL, using the score $\Vb_w^{\top} \Rb \Vb_{w'}$ for weight prediction, where $\Rb \in \RR^{p \times p}$ is a relation matrix to be learned. \cite{yang2014embedding} later simplified this by constraining $\Rb$ to be a diagonal matrix. However, these methods lack a mechanism to quantify uncertainty in the existence of an edge between $w$ and $w'$. They primarily focus on calculating and selecting high-scoring entity pairs, without a clear threshold to differentiate linked from unlinked entities, nor a way to control the false discovery rate in link discovery.

Our approach, encapsulated in the hypothesis testing procedure \eqref{h0}, addresses this gap. By applying this hypothesis testing to all entity pairs $(w,w')$ where $w \neq w'$ within the set $[d] \times [d]$, we can construct a sparsely connected KG. This testing is equivalent to assessing the cosine similarity 
$${\rm{H}}_0: \frac{\Vb_w^{\top} \Vb_{w'}}{\|\Vb_w\| \cdot \|\Vb_{w'} \|} = 0 \text{\quad versus\quad} {\rm{H}}_1: \frac{\Vb_w^{\top} \Vb_{w'}}{\|\Vb_w\| \cdot \|\Vb_{w'} \|} \neq 0\,,$$
thus bridging the gap with methods that use cosine similarities \citep[e.g.]{beam2019clinical,hong2021clinical,zhou2022multiview}. It also aligns with semantic matching models \citep[e.g.]{nickel2011three,yang2014embedding} by treating the relation matrix $\Rb$ as an identity matrix.

\subsection{Inferential Procedure through Low-Rank Approximation}
\label{sec:est}

The challenge for testing \eqref{h0} lies in the fact that the embedding matrix $\Vb$ is not directly observable and must be inferred from the data $\{w_{i,1}, \ldots, w_{i,t}, \ldots,  w_{i,T_i}\}_{i=1}^n$. Utilizing likelihood for this purpose is impractical due to the extensive number of parameters and the complex non-linearity of the probability mass function. This issue is further exacerbated when patient-level data are off-limits due to privacy concerns. Fortunately, it has been demonstrated that these embeddings can be effectively recovered from the PMI matrix, offering a viable solution to this problem. Specifically, the PMI between two entities $w$ and $w'$ is defined as \begin{equation}\label{eq: pmi def}
    \PMIbb(w,w') = \log \frac{p_{w,w'} }{ p_{w} p_{w'}},
\end{equation}
where $p_{w,w'} = \sum_{u=1}^q \E_{(\bc_{i,t},\bc_{i,t+u})\sim F_u}[\P(w_{i,t} = w, w_{i, t+u} = w'|\bc_{i,t}, \bc_{i,t+u})]/q$
 denotes the co-occurrence probability of entities $w$ and $w'$ within a pre-defined window of size $q$, and 
 $F_u$ is the stationary distribution of $(\bc_{i,t}, \bc_{i,t+u})$. With the generation model \eqref{eq: code prob} in place, we can show the relationship between the KG embeddings 
and the PMI matrix as follows: 
 \begin{equation}\label{eq:tildeV}
 \PMIbb \approx \alpha_p \Vb \Vb^{\top}, \text{ where } \PMIbb  = [ \PMIbb(w,w')]_{w,w'\in[d]} \in \RR^{d \times d} \text{ and } \alpha_p = \frac{\sqrt{\alpha}(1 - \alpha^{q/2})}{q p (1 - \sqrt{\alpha})} . 
 \end{equation}
 Here $\alpha$ is the scaling factor of the AR process \eqref{eq:hidden_mp} and $\sqrt{\alpha_p}$ is the scaling factor of the embeddings. A more formal argument is presented in Theorem  \ref{thm0:PMI_lowrank}. 
The relationship between the PMI matrix and the KG embedding implies that we can derive the KG of the associations among different EHR entities from the magnitude of the PMI matrix entries.

To estimate the PMI matrix, we first define the co-occurrence between the entities  $w$ and $w'$ within the window of size $q$ in all patients' health records as 
\begin{equation*}
\begin{aligned}
   & \CC_{w,w'}   = \sum_{i=1}^n \CC^{(i)}_{w,w'},  \text{ where } \\
  &\CC^{(i)}_{w,w'}  = |\{(t,s) \in [T_i]\times[T_i]: 0 < |t-s|\leq q, \min\{t,s\} \le T_i - q \text{ and } w_{i,t} = w,w_{i,s} = w' \}|    .
\end{aligned}
\end{equation*}
We then denote $\CC^{(i)} = [ \CC^{(i)}_{w,w'}]_{w,w' \in [d]}$ as the $i$th patient's co-occurrence matrix,  $\CC = [ \CC_{w,w'}]_{w,w' \in [d]}$ as the total co-occurrence matrix, $\CC^{(i)}_{w}=\sum_{w^{\prime}} \CC^{(i)}_{w, w^{\prime}}$ the individual marginal occurrence of $w$, and  $\CC_{w} = \sum_{w'=1}^d \CC_{w,w'}$ the marginal occurrence of $w$, and finally, $\overline{\CC}^{(i)} = \sum_w \CC^{(i)}_{w}$ and $\overline{\CC} = \sum_w \CC_{w}$.

An empirical estimation for the occurrence probabilities and co-occurrence probabilities is
\begin{equation}\label{eq:est_of_pw_pww'}
        \hat p_{w} = {\CC_{w}}/{
    \overline{\CC}} \text{ and }
        \hat p_{w,w'} = {\CC_{w,w'}}/{\overline{\CC}}, \quad \text{for } 1 \leq w,w' \leq d.
\end{equation}
We plug in these estimations in the definition of PMI matrix (\ref{eq: pmi def}), and obtain an empirical estimator $\widehat\PMIbb = [\widehat\PMIbb(w,w')]_{w,w'\in [d]}$ for the population PMI matrix $\PMIbb$ based on the co-occurrence matrix $\CC$, i.e.,
\begin{equation}\label{eqcal:def_emprical_pmi}
\widehat\PMIbb(w,w') = \log\frac{ \hat p_{w,w'} }{  \hat p_{w}  \hat p_{w'} } = \log\frac{\overline{\CC} \cdot \CC_{w,w'}}{\CC_{w} \CC_{w'}}.
\end{equation}
\begin{remark}
    Considering that some co-occurrence entries may be equal to zero, we can set a threshold to $\widehat\PMIbb(w,w')$ to guarantee that it is well defined. In particular, we could use $\log (\max \{ \overline{\CC} \cdot \CC_{w,w'}/(\CC_{w} \CC_{w'}), \eta \} )$ to avoid the singularity case when the co-occurrence is zero. Here $\eta > 0$ is a threshold to be chosen. For the rest of the paper, we only consider the estimator in \eqref{eqcal:def_emprical_pmi} for simplicity by showing the co-occurrence is non-zero with high probability. 
\end{remark}
While we can directly apply $\hat\PMIbb$ to do inference on $\Vb \Vb^{\top}$, this method lacks efficiency as it overlooks the low-rank structure of $\Vb \Vb^{\top}$. 
To utilize the low-rank structure, we propose to use the rank-$p$ approximation of $\widehat\PMIbb$ as the ultimate estimator of $\alpha_p \Vb \Vb^{\top}$. Namely, denote the eigendecomposition of $\PMIbbhat$ as
\begin{equation*}
    \begin{split}
\hat{\PMIbb} &=\left[\begin{array}{ll}
\hat{\Ub} & \hat{\Ub}_{\perp}
\end{array}\right]\left[\begin{array}{cc}
\hat\bLambda & \mathbf{0} \\
\mathbf{0} & \hat\bLambda_{\perp}
\end{array}\right]\left[\begin{array}{l}
\hat{\Ub}^{\top} \\
\hat{\Ub}^{\top}_{\perp}
\end{array}\right],
    \end{split}
\end{equation*}
where $(\hat{\Ub}, \hat{\Ub}_{\perp}) \in \mathbb{R}^{d \times d}$ are the eigenvectors with the eigenvalues sorted in descending order by their magnitude and $\hat{\Ub} \in  \mathbb{R}^{d \times p}$. Then we let 
\begin{equation}
\tilde{\PMIbb} = \hat{\Ub}\hat\bLambda\hat{\Ub}^{\top}
\label{def:titlde}
\end{equation}
be the low-rank estimator of $\alpha_p\Vb \Vb^{\top}$. In the following section, we demonstrate that, under certain assumptions, $\tilde\PMIbb$ exhibits entry-wise asymptotic normality and the column space spanned by $\hat{\Ub}$ closely aligns with the column space spanned by $\Vb$. We will show that the low-rank estimator $\tilde\PMIbb$ possesses lower variance in comparison to the empirical estimator $\PMIbbhat$ and minimal bias relative to $\alpha_p\Vb \Vb^{\top}$. This finding suggests that $\tilde\PMIbb$ is a more efficient estimator, and we will employ it as the foundation for subsequent inferential procedures.

\subsubsection{Variance Estimation with Patient-level Data}\label{sec:2.2.1}
We now discuss how to estimate the covariance of the estimators using the patient-level co-occurrence matrices $\{\mathbb{C}_i\}_{i=1}^n$.
Denote the error matrix by $\Wb = \widehat{\PMIbb} - \alpha_p\Vb \Vb^{\top}$. 
We can express the leading term of the error  as $ \Wb_{w,w'} \approx  \frac{1}{n}\sum_{i=1}^n  \hat \Wb^{(i)}_{w,w'}$, where
\begin{equation*}
 \hat \Wb^{(i)}_{w,w'} = \frac{\CC^{(i)}_{w,w'}}{\CC_{w,w'}/n} - \frac{\CC^{(i)}_{w}}{\CC_{w}/n} - \frac{\CC^{(i)}_{w'}}{\CC_{w'}/n}+1,\quad 1\leq i\leq n.
\end{equation*}
The detail of this approximation is given in Supplementary \ref{sec:proof_of_sec221}.
Therefore, we can estimate the covariance between $\hat{\PMIbb}_{w,\cdot}$ and $\hat{\PMIbb}_{w',\cdot}$, i.e., $\Cov(\hat{\PMIbb}_{w,\cdot}, \hat{\PMIbb}_{w',\cdot})$ by
\begin{equation}\label{eq:hatPMI-cov}
\hat{\bSigma}_{w,w'} = \frac{1}{n(n-1)}\sum_{i=1}^n \hat\Wb_{w,\cdot}^{(i) \top} \hat\Wb_{w',\cdot}^{(i)} \,. 
\end{equation}
Then the variance of the low-rank estimator $\tilde{\PMIbb}$ can   be estimated by
\begin{equation}\label{eq:lowrank_entry}
    \widehat{\rm{Var}}(\tilde{\PMIbb}(w,w')) = \hat\Pb_{w,\cdot}\widehat\bSigma_{w',w'}\hat\Pb_{w,\cdot}^{\top} + \hat\Pb_{w',\cdot}\widehat\bSigma_{w,w}\hat\Pb_{w',\cdot}^{\top} + 2 \hat\Pb_{w,\cdot}\widehat\bSigma_{w',w} \hat\Pb_{w',\cdot}^{\top}
\end{equation}
where $\hat\Pb = \hat{\Ub}\hat{\Ub}^{\top}$. We refer to Theorem \ref{thm2} for more details.
However, due to data privacy, we sometimes have no access to each patient's co-occurrence information, which prevents us from estimating the covariance via \eqref{eq:hatPMI-cov}. In the following section, we will propose a summary statistic-based covariance estimator.

\subsubsection{Variance Estimation without Patient-level Data}\label{sec:2.2.2}
If we cannot access the patient-level data, we can still test the hypothesis on the significance of $\Vb_w^{\top} \Vb_{w'}$  by testing the global null hypothesis that the occurrence of all EHR entities are
independent, i.e.,
\begin{equation*}
    H_0:w_{i,t}  \stackrel{i.i.d}{\sim} {\rm Multinomial}(1, \{p_{w}\}_{w=1}^d) \text{ for }   { 1 \leq t \leq T_i }, 1 \leq i \leq n.
\end{equation*}
Under the global null distribution, we can estimate the covariance structure of $\Wb$ without the need for patient-level data using the following formulas (see details in Supplementary \ref{sec:b.5}):
\begin{equation}\label{eq:var_null_dist}
    \begin{split}
        &\hat{\text{Cov}}\big(\Wb_{w \cdot} \big) = \frac{1}{nT_0 \hat p_{w}}\Big(\boldsymbol{1}\boldsymbol{1}^{\top} (\hat p_{w} - \frac{1}{2}) - \frac{1}{2}\boldsymbol{1}\be_w^{\top} - \frac{1}{2} \be_w \boldsymbol{1}^{\top} + \text{diag} (\hat \bp)^{-1}\frac{1-\hat p_{w}}{2} + \frac{1}{2 \hat p_{w}}\be_w \be_w^{\top}\Big),\\
        &\hat{\text{Cov}}\big(\Wb_{w \cdot} , \Wb_{w' \cdot} \big) = \frac{1}{nT_0}\Big(\boldsymbol{1}\boldsymbol{1}^{\top} - \frac{1}{2 \hat p_{w}}\boldsymbol{1}\be_w^{\top} - \frac{1}{2 \hat p_{w'}}\be_{w'}\boldsymbol{1}^{\top} - \frac{1}{2} \text{diag} (\hat \bp)^{-1} + \frac{1}{2 \hat p_{w} \hat p_{w'}}\be_{w'}\be_w^{\top} \Big),
    \end{split}
\end{equation}
where $\hat \bp = (\hat p_1, \cdots, \hat p_d)^{\top}$, $T_0 = T q - q^2$ and $T = \sum_{i=1}^n T_i / n$. To estimate the covariance matrices of $\Wb$, we only need to know parameters $n, T, q,$ and the estimate of $\{p_{w}\}_{w=1}^d$, which can be obtained from the summary-level co-occurrence matrix $\CC$ by \eqref{eq:est_of_pw_pww'}. 
With $\hat{\text{Cov}}(\Wb_{w,\cdot},\Wb_{w',\cdot})$ obtained, we can further estimate the variance of $\tilde{\PMIbb} (w,w')$ by \eqref{eq:lowrank_entry} with $\widehat\bSigma_{w,w'}$ replaced by $\breve \bSigma_{w,w'} = \hat{\text{Cov}}\big(\Wb_{w \cdot}, \Wb_{w' \cdot} \big)$ as 
\begin{equation}\label{eq:lowrank_entry_without}
    \widetilde{\rm{Var}}(\tilde{\PMIbb}(w,w')) = \hat\Pb_{w,\cdot} \breve \bSigma_{w',w'} \hat\Pb_{w,\cdot}^{\top} + \hat \Pb_{w',\cdot} \breve \bSigma_{w,w}\hat\Pb_{w',\cdot}^{\top} + 2 \hat\Pb_{w,\cdot} \breve \bSigma_{w',w} \hat\Pb_{w',\cdot}^{\top} ,
\end{equation}
and conduct the hypothesis testing based on the asymptotic normality of $\tilde{\PMIbb}$. The KNowledge graph Inference and Testing (KNIT) procedure is then summarized in Algorithm \ref{em4}, where the rank of KG embeddings $p$ is estimated by Algorithm~\ref{em4 rank}.

\begin{algorithm}[htb!]
\caption{
KNowledge graph Inference and Testing (KNIT)}\label{em4}
\begin{algorithmic}
\STATE \textbf{Input:} the summary-level co-occurrence matrix   $\CC \in \RR^{d\times d}$, the window size $q$, the number of patients $n$, the rank of KG embeddings $p$, the average  observation time $T = \sum_{i=1}^n T_i/n$, and the nominal FDR level $\alpha$. 
\STATE \textbf{Estimation:} Calculate $\widehat\PMIbb$ and $\tilde{\PMIbb}$ based on $\CC$ using \eqref{eqcal:def_emprical_pmi} and \eqref{def:titlde}.
\STATE \textbf{Variance estimation:} Obtain $ \widetilde{\rm{Var}}(\tilde{\PMIbb}(w,w'))$ by \eqref{eq:lowrank_entry_without} for $1 \leq w < w' \leq d$. 
\STATE \textbf{Inference:} Obtain the $p$-value $\alpha_{w,w'} = \Phi\big( \tilde{\PMIbb}(w,w')/ \sqrt{ \widetilde{\rm{Var}}(\tilde{\PMIbb}(w,w')) }  \big)$ for $1 \leq w < w' \leq d$. 
Order $\alpha_{w,w'}$'s as $\alpha_{(1)} \leq \alpha_{(1)} \leq \ldots \leq \alpha_{(J)}$ with $J = d(d-1)/2$ and set $\alpha_{(0)} = 0$. Let $j_{\max} = \max\{0 \leq j \leq J: \alpha_{(j)} \leq \alpha j /J/(\log J+1)\}$. 
\STATE \textbf{Output:}  $\tilde{\PMIbb}$ and the edge set $\{(w,w'): \alpha_{w,w'} \leq \alpha_{j_{\max}} \}$. 
\end{algorithmic}
\end{algorithm}

\begin{algorithm}[htb!]
\caption{Estimating the rank of knowledge graph embeddings
}\label{em4 rank}
\begin{algorithmic}
    \STATE \textbf{Input:} Eigenvalues of $\widehat\PMIbb$ ($\hat\lambda_1, \ldots, \hat\lambda_d$), thresholding parameter $\eta_0 =  {8d^3 p \log^2 d}/{\sqrt{n T}} $.
    \STATE \textbf{Estimation:} Compute  $r = \max\{k \in [d]: \hat\lambda_k \ge \eta_0\}$.
    \STATE \textbf{Output: } The estimated rank of knowledge graph embeddings, $r$.
\end{algorithmic}    
\end{algorithm}
The following proposition guarantees that Algorithm~\ref{em4 rank} recovers the true rank of KG embeddings with high probability.

\begin{proposition}\label{prop: est rank}
    Under the condition that ${d^3 p \log^2 d}/{\sqrt{n T}} \ll \kappa^2 \xi^2$, with probability at least $1 - \exp(-\Omega(\log^2 d))$, we have that $r = p$.
\end{proposition}

The proof follows directly from the proof of Proposition~S.23 in the Supplementary Material.

\section{Theoretical Properties}
\label{sec:thm}
In this section, we establish the inferential results by characterizing the asymptotic distribution of $\tilde{\PMIbb}$. Given our data generative model, only the left singular space and the singular values of $\Vb$ are identifiable, since for any orthonormal matrix $\Ob \in \mathbb{R}^{p \times p}$, $\{\bc_t\}$ and $\{\Ob\bc_t\}$ have the same distribution, and replacing $\Vb$ with $\Vb\Ob$ does not change the occurrence probabilities in (\ref{eq: code prob}). Hence without loss of generality, we assume $\Vb$ has the singular value decomposition $\Vb = \Ub \bLambda^{1/2}$.  We denote $\bLambda^{1/2} = \text{diag}(\kappa_1/\sqrt{\alpha_p},\cdots,\kappa_p/\sqrt{\alpha_p})$ {with $\kappa_1/\sqrt{\alpha_p} \ge \ldots \ge \kappa_p/\sqrt{\alpha_p} > 0$ being the corresponding singular values}. Denote $\kappa := \max_{1\leq j\leq p}|\kappa_j|, \xi := \min_{j}|\kappa_j| / \kappa$.
We require the following assumptions on these scaling parameters.

\begin{assumption}[Scaling Conditions]\label{assump:ndTp}
There exist constants $0<c<C$ such that $ p \ge c\log^2 d$ and $\log p \le C\log(\log d)$. Additionally, we assume that $T_i = \Omega(p^2\log (d))$ for $i \in [n]$.
\end{assumption}
\begin{remark}
    Assumption~\ref{assump:ndTp} states the necessary scaling conditions to guarantee that the estimator converges to the truth at a fast enough rate. The lower bound for $p$ is to control the potential information loss caused by low-rank approximation, and the upper bound for $p$ is to control the variance of the estimator $\tilde\PMIbb$ and maintain the ``near low-rank'' structure of the population PMI matrix. The intricate Markov dependency structure in \eqref{eq:hidden_mp} necessitates the lower bound assumption on the sequence length $T_i, 1\le i \le n$, ensuring the mixing property of probability density functions of the \(p\)-dimensional Markov chain \(\{\bc_{i,t}\}_{t=1}^{T_i}\).
\end{remark}

\begin{assumption}\label{assump1}
Assume that there exists constants $\mu_1 \geq 1, \mu_2>0$ such that $ \|\Ub\|_{2,\infty} \leq \mu_1 \sqrt{p/d}$ and $\min_{i \in [d]}\|\Ub_{i,\cdot}\| \geq \mu_2p^2/\xi d$.
\end{assumption}
\begin{remark}
     The upper bound for $\|\Ub\|_{2,\infty}$ imposes the incoherence condition \citep{candes2010power} on the eigen-space of $\Vb$.  Concurrently, the lower bound for $\min_i \|\Ub_{i,\cdot}\|$ is weak since it is of order $p^2/\xi d$, which decays much faster than the upper bound combined with Assumption \ref{assump:eigen_bound}. These two inequalities together guarantee that the eigenvectors are roughly uniformly distributed in the eigen-space.
\end{remark}

\begin{assumption}\label{assump:eigen_bound}

 Assume that $\xi = \Omega(1/\log d)$, $\kappa^2 = \Omega(d^6\log^4 d(nT)^{-1/2}\xi^{-6})$, and $ \kappa^2 = O( d p^{-1/2} (nT)^{-1/4})$.
\end{assumption}
\begin{remark}
    The upper bound for $\kappa$ arises from the variance-bias tradeoff. We need the bias, i.e., the error between $\PMIbb$ and its low-rank approximation to be comparable to the variance. 
 The lower bound assumption for $\kappa$ is necessary for reasons similar to the lower bound for $\min_i \|\Ub_{i,\cdot}\|$ in Assumption \ref{assump1}: an eigen-gap is required to prevent the collapse of the low-rank estimator and to control the distance between two singular spaces according to the Davis-Kahan theorem \citep{davis1970}. An upper bound for the ratio of singular values $\xi$ ensures that the eigen-gap is relatively large compared to the maximum distance within non-zero singular values. 
\end{remark}

 Under the assumptions above, we can measure the distance between the population PMI matrix and the inner product of the scaled KG embeddings, and show the asymptotic normality of the empirical estimator and the low-rank estimator. The following theorem demonstrates the ``near low-rank'' structure of the population PMI matrix. Its proof is given in Supplementary \ref{proof of 3.4}.

\begin{theorem}\label{thm0:PMI_lowrank}
Suppose Assumptions \ref{assump1} and \ref{assump:eigen_bound} hold. Then we have
$$
 \|\PMIbb - \alpha_p \Vb\Vb^{\top} \|_{\max} \leq C \frac{\kappa^4 p^{2}}{d^{2}}
$$
for some constant $C>0$. Besides, we have 
$
\lambda_{1}(\PMIbb) \leq 2\kappa^2
, 
\lambda_{p}(\PMIbb) \geq  {\kappa^2 \xi^2}/{2}
,$ and $
|\lambda_{p+k}(\PMIbb)| \leq C  {\kappa^4 p^{2}}/{d}
$
for $1 \leq k \leq d-p$. 
\end{theorem}

\begin{remark}
This theorem provides the rate of the bias, which is sharper than the bound $\|\PMIbb -  {\Vb}{\Vb}^{\top}/p \|_{\max} = O(1/p)$ given in \cite{arora2016latent} and the bound $\|\PMIbb -  {\Vb}{\Vb}^{\top}/p \|_{\max} = O(\sqrt{\log (d) / p})$ given in \cite{xxx2021}, both of which are not sufficient for the inference of PMI estimator. In comparison, our result reveals a better low-rank approximation when $p = o(d^{2/3})$ as implied by Assumption \ref{assump:ndTp}. Moreover, our low-rank approximation differs from \citet{arora2016latent, xxx2021}'s as they assumed a mean-zero Gaussian or spherical Gaussian prior on $\Vb$, while our study does not impose any prior on $\Vb$, yielding more generalized results.
\end{remark}

With the bias bound above, we have the following lemma on the asymptotic normality of the error term.

\begin{lemma}\label{thm1}
Suppose Assumptions~\ref{assump:ndTp}-\ref{assump:eigen_bound} hold \zx{ and \(nT \ge T_i d^2 \log^2d\) for all \(i \in [n]\)}, then we have the entry-wise asymptotic normality of the residual matrix $\Wb$, i.e.,  for any entry $(w,w') \in [d] \times [d]$, we have
\begin{equation}\label{eq:entrywise_asym_normality}
    \frac{\sqrt{n}}{\sigma_{w,w'}} \Wb_{w,w'} \stackrel{d}{\longrightarrow} N(0, 1), 
\end{equation}
where \zx{\[
\sigma_{w,w'}^2 = \frac{1}{n}\sum_{i=1}^n (\frac{T_i-q}{T-q})^2\text{Var}\big(\CC_{w,w'}^{(i)}/\E[\CC_{w,w'}^{(i)}] - \CC_{w}^{(i)}/\E[\CC_{w}^{(i)}] - \CC_{w'}^{(i)}/\E[\CC_{w'}^{(i)}]\big)\] 
is of order $d^2/T$}. 
In the special case where \(T_i = T\) for all \(i \in [n]\), the asymptotic normality in \eqref{eq:entrywise_asym_normality} holds under Assumptions~\ref{assump:ndTp}-\ref{assump:eigen_bound} without requiring a lower bound on the sample size \(n\). In other words, $n$ does not necessarily need to approach infinity for the result to hold.
\end{lemma}

Lemma \ref{thm1} gives the asymptotic normality of the residual matrix $\Wb$. The proof of this lemma can be found in  Supplementary \ref{sec:asym_normality_empirical_PMI}. We denote the covariance matrix of the error term in \eqref{eq:entrywise_asym_normality} as $\bSigma_{E}$ where $ \bSigma_{E}(w,w') = \sigma_{w,w'}^2$ for each $w,w' \in [d]$. The following theorem gives the asymptotic distribution of the low-rank estimator.

\begin{theorem}\label{thm2}
Let $\tilde\bSigma = (\Pb^{\star}\circ\Pb^{\star})\bSigma_{E} +  \bSigma_{E}(\Pb^{\star}\circ \Pb^{\star})$. Under Assumptions~\ref{assump:ndTp}-\ref{assump:eigen_bound} \zx{ and \(nT \ge T_i d^2 \log^2d\) for all \(i \in [n]\)}, for the low-rank estimator $\tilde{\PMIbb}$, we have for any entry $(w,w') \in [d] \times [d],$ 
\begin{equation*}
    \frac{\sqrt{n}}{\tilde\bSigma^{1/2}_{w,w'}} ( \tilde{\PMIbb}_{w,w'} - 
\alpha_p \Vb_w^{\top}\Vb_{w'}) \stackrel{d}{\longrightarrow} N(0, 1)\text{ as } n, T, d \rightarrow \infty,
\end{equation*}
where $\|\tilde\bSigma\|_{\max} = O(\frac{dp^2}{T\xi^4})$.
\end{theorem}

\begin{remark}
Although both $n$ and $T$ need to go to infinity, the growth requirement for $T$  is significantly weaker compared to $n$. The key constraint arises from the bias term \(\mathbb{PMI} - \alpha_p\mathbf{V}\mathbf{V}^{\top}\), which depends only on $d$. Additionally, since the mixing properties of the Markov chain  \(\{p_{i,w}(t)\}_{t=1}^{T_i}\) depend on $d$,  the lower bound  $T_i = \Omega(p^2 \log(d))$ ensures that $T_i$ must increase with $d$. However, this increase occurs at a much slower rate compared to $n$, which scales as $\Omega(d^2 \log^2(d))$ when $T_i \approx T$. This distinction highlights that while $T$ grows with $d$, its growth is far outpaced by that of $n$, aligning with practical data scenarios.

In particular, the EHR dataset analyzed in Section~\ref{sec:app} consists of $120,518$ patients, with sequence length quantiles of $370$, $1090$, and $2642$ at the $25$th, $50$th, and $75$th percentiles, respectively, and an average sequence length \(T = 2197.7\). The vocabulary size \(d\) is $9,586$. These statistics confirm that the sample size is much larger than the vocabulary size and both the average and typical sequence lengths, supporting the practical validity of our assumptions.
\end{remark}

The proof of  Theorem \ref{thm2} can be found in Supplementary \ref{B3}.
In fact, the proof of  Theorem~\ref{thm2} provides a Bahadur decomposition of $ \tilde{\PMIbb} - \alpha_p \Vb_w^{\top}\Vb_{w'}$
    to establish the entrywise asymptotic normality for the low-rank estimator $\tilde{\PMIbb}$, which can be generalized for multiple testing. Compared with the variance of the empirical estimator $\sigma_{w,w'}^2 = d^2/T$ shown in Lemma \ref{thm1}, the variance of the low-rank estimator $\tilde\bSigma_{w,w'} = O(dp^2/(T\xi^4))$ is much smaller. This suggests the superior performance of the low-rank estimator over the empirical estimator.

\section{Simulation}
\label{sec:sim}

In this section, we conduct synthetic simulations to demonstrate the performance of our algorithm. In Section \ref{sec:41}, we compare the numerical performance of the low-rank estimator with the empirical estimator. In Section \ref{sec:42}, we generate data under the global null distribution and calculate type I error for edge selection. It shows that the inference with and without patient-level data both have empirical type I errors close to the nominal significance level. In Section \ref{sec:43}, we show the asymptotic normality of the standardized estimators.

\subsection{Numerical Performance of PMI Estimators}\label{sec:41}

In this section, we compare $\tilde\PMIbb$ with $\widehat\PMIbb$ in terms of estimation errors. We set the number of entities $d = 100$, embedding dimension $p = \lfloor \log^2 d \rfloor + 1$, $T = 1000$, and the sample size $n \in \{ 200,400,600,800,1000,2000\}$. The construction of the embeddings $\Vb$ involves three steps:
\begin{enumerate}
    \item[1.] Generate an orthonormal basis of a $d \times p$ matrix, with each element independently drawn from a standard normal distribution. Denote it as $\Vb^{\text{init}}$.
    \item[2.] Estimate the marginal occurrence probabilities $\bp$ using the Monte-Carlo (MC) method with $10^7$ i.i.d MC samples. Denote it as $\bp_{\text{mc}}$.
    \item[3.] Let $\Vb = (\Vb_1,\cdots,\Vb_d)^{\top}$, where each $\Vb_w = \Vb^{\text{init}}_w - \Vb^{\text{init}\top} \bp_{\text{mc}}$ for $w \in [d]$.
\end{enumerate}
Here the last step is to guarantee that $\Vb^{\top}\bp = 0$. Then we generate $100$ i.i.d replicates. For each replicate, the sequence $\{w_{i,t}\}_{1\leq i \leq n, 1\leq t \leq T}$ is generated in accordance with \eqref{eq: code prob}, and $\|\hat{\PMIbb} - \alpha_p \Vb \Vb^{\top}\|_{\max}$,  $\|\tilde{\PMIbb} - \alpha_p \Vb \Vb^{\top}\|_{\max}$, and $\|\PMIbb - \alpha_p\Vb \Vb^{\top}\|_{\max}$  are calculated.

\begin{figure}[htbp]%
    \centering
\begin{tabular}{ccc}
\includegraphics[width=5cm]{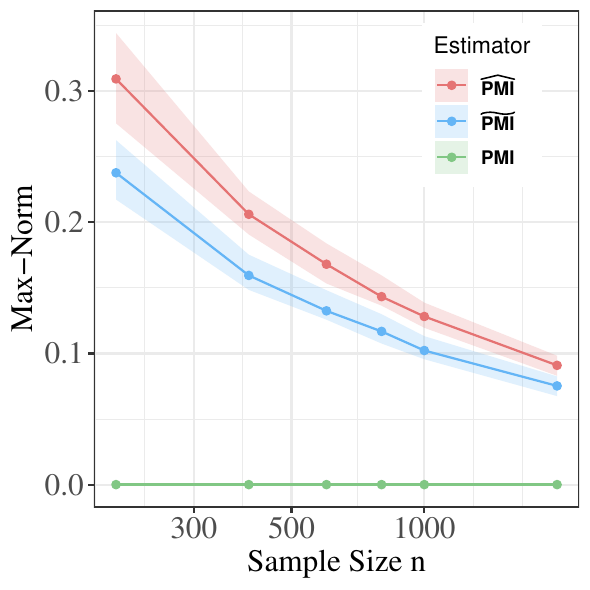}&
\includegraphics[width=5cm]{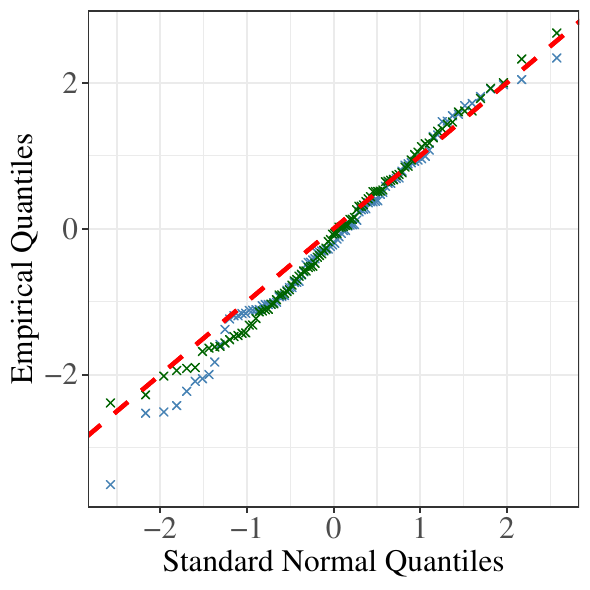} & \includegraphics[width=5cm]{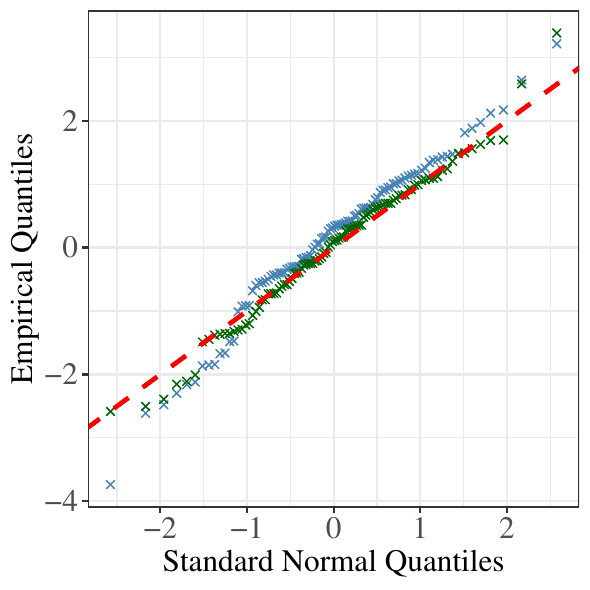}
\end{tabular}
    \caption{Left plot is the median and the $20\%$-$80\%$ quantile range of max norm distances between EHR entity embeddings and PMI estimators across 100 independent runs. Middle and right plots are Q-Q plots of one standardized entry for the empirical (green) estimator and the low-rank (blue) estimator. The Middle: $(d,n,T,\kappa) = (100,1000,800,1),(w,w')=(1,1)$. Right: $(d,n,T,\kappa) = (100,1000,1200,1),(w,w')=(1,1)$. }%
    \label{fig:qqplot}%
\end{figure}

As evidenced by the first plot in Figure \ref{fig:qqplot}, the low-rank estimator has less variability and outperforms the empirical estimator, and the bias $\|\PMIbb - \alpha_p\Vb \Vb^{\top}\|_{\max}$ is negligible compared to  $\|\hat{\PMIbb} - \alpha_p \Vb \Vb^{\top}\|_{\max}$ and   $\|\tilde{\PMIbb} - \alpha_p \Vb \Vb^{\top}\|_{\max}$, which are consistent with our theoretical results. 

\begin{figure}[htpb!]%
    \centering
\includegraphics[width=14cm]{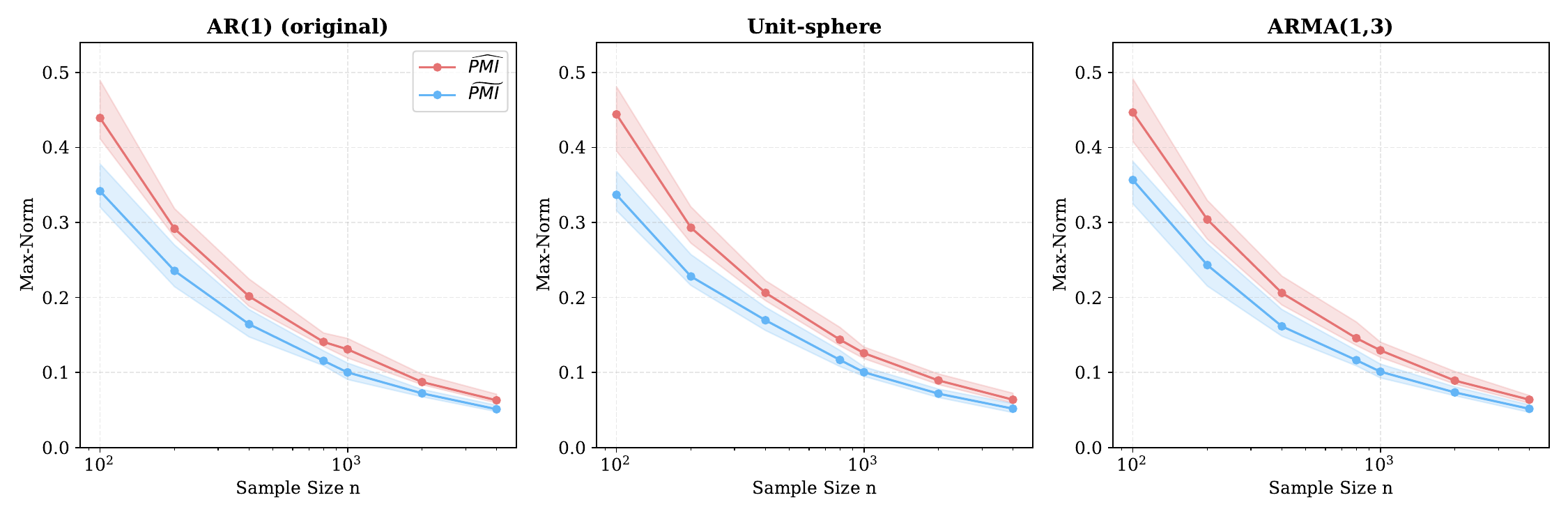} \\
\includegraphics[width=14cm]{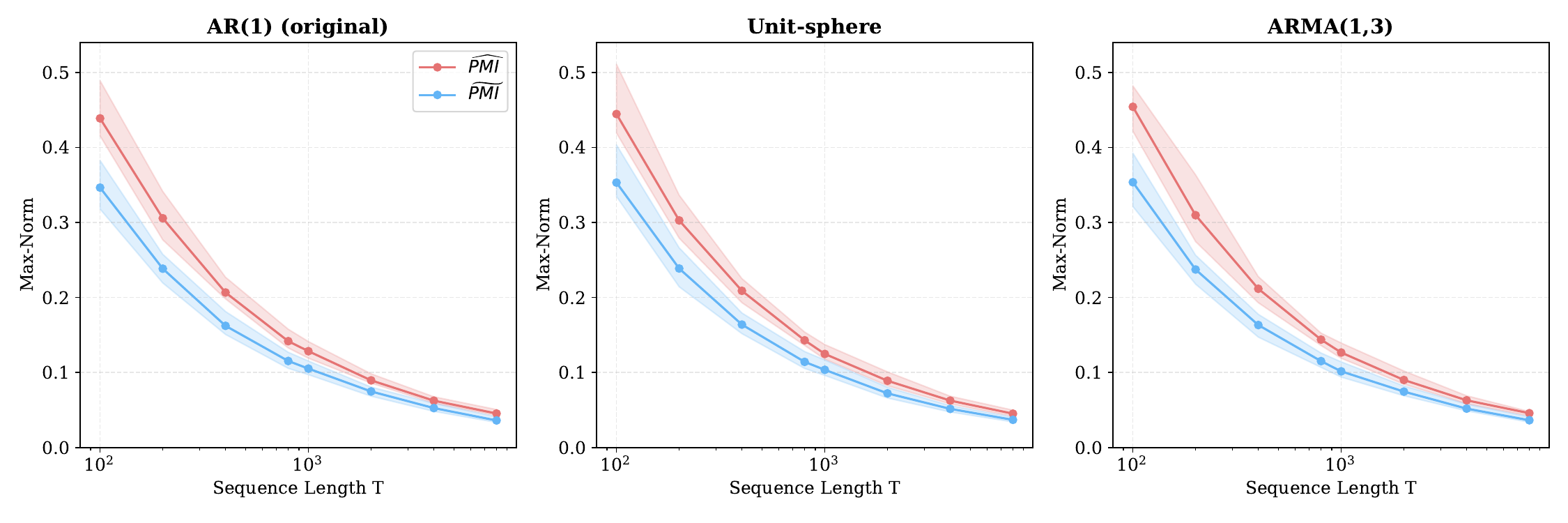} 
    \caption{The max-norm error between PMI estimators and EHR entity embeddings. In the first row, we fix \(T=1000\) and change \(n\). In the second row, we fix \(n=1000\) and change \(T\). Each column represents a distribution of the discourse vectors used in the generative model \eqref{eq: code prob}.}
    \label{fig:robust_of_discourse}%
\end{figure}

To evaluate the robustness of our method under model mis-specification, we conducted additional experiments where the original AR(1) process for the discourse vectors was replaced with alternative processes. Specifically, we considered a random walk on a unit sphere, following the approach in \cite{xxx2021}, and an ARMA(1,3) process: \(\bc_{t+1}=\sqrt{\alpha}\bc_t + \sqrt{1-\alpha}\br_{t+1} + 0.2\br_t + 0.1\br_{t-1}+0.05\br_{t-2}.\) 
The results, presented in Figure~\ref{fig:robust_of_discourse}, indicate that these changes have minimal impact on PMI estimation, demonstrating the robustness of our method.

\begin{figure}[htpb!]%
    \centering
\includegraphics[width=14cm]{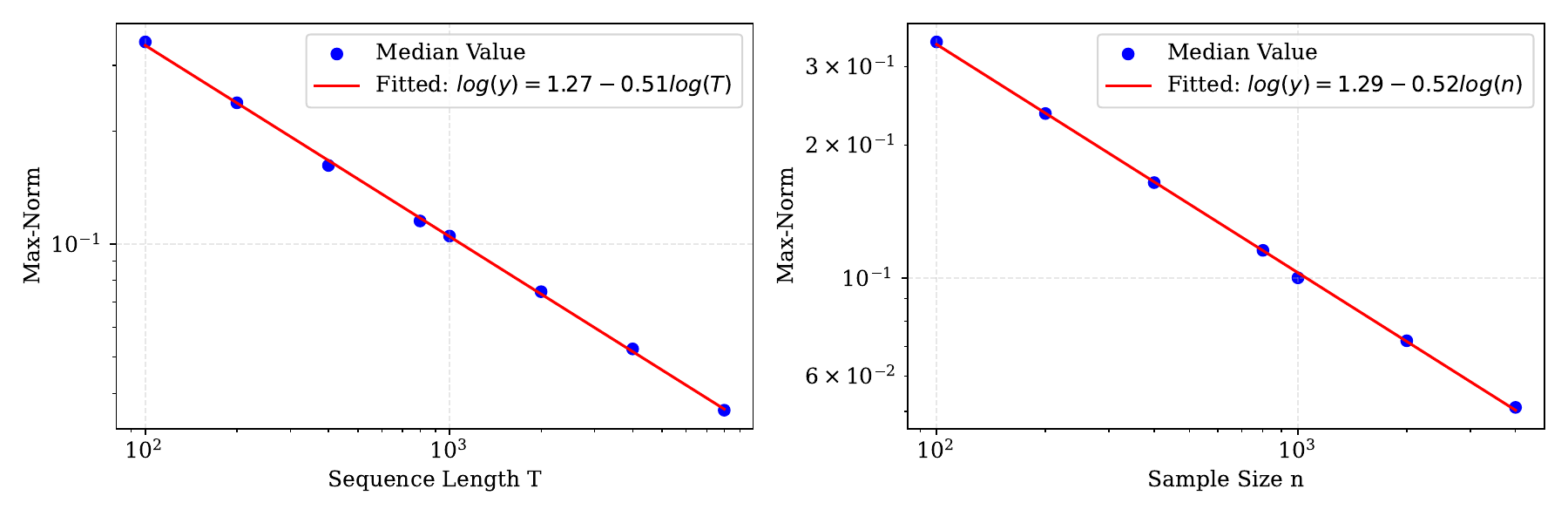}
    \caption{The data points in this figure represent the median value of the max-norm difference between the low-rank PMI estimator and the population PMI matrix, as shown in the first column of Figure \ref{fig:robust_of_discourse}. }
    \label{fig:decay_rate_lowrank}%
\end{figure}

Further, we analyzed the decay rate of the max-norm error $\|\tilde{\PMIbb} - \alpha_p \Vb \Vb^{\top}\|_{\max}$ for AR(1) by taking the logarithm of this quantity along with the sample size $n$ and  sequence length \(T\). A linear model was fitted to the median values shown in  Figure \ref{fig:decay_rate_lowrank}.  The results confirm that $\|\tilde{\PMIbb} - \alpha_p \Vb \Vb^{\top}\|_{\max}$ is approximately proportional to \((nT)^{-1/2} \rightarrow 0 \), which is consistent with the theoretical results in Theorem~\ref{thm2}.

\subsection{Numerical Results on Hypothesis Testing}\label{sec:42}

In this section, we simulate data following the global null distribution given in Section \ref{sec:2.2.2}. For simplicity, we assume that each feature occurs with equal probability, such that $p_w = 1/d$ for $1\leq w \leq d$. The sets $\{w_{i,t}\}_{1\leq i \leq n, 1\leq t \leq T}$ are independently and identically drawn from a $\text{Multinomial}(1,(1/d,\cdots, 1/d))$ distribution, considering \(d=100\) and \(n \in \{400, 800, 1600\}.\) 
Under these conditions, it is not reasonable to define the embeddings, and thus our analysis is confined to the empirical estimator. We apply the methods described in Sections \ref{sec:2.2.1} and \ref{sec:2.2.2} to estimate the variance of each entry in $\PMIbbhat$ using the Monte-Carlo approach.
This involves generating $n$ i.i.d longitudinal code sequences $100$ times, and subsequently calculating the family-wise error rate (FWER) for each method. When patient-level data becomes accessible, we estimate the variance of each entry using \eqref{eq:hatPMI-cov}. In contrast, when patient-level data is unavailable, we estimate the variance of each entry via \eqref{eq:var_null_dist}.

\begin{figure}[htpb!]
    \centering
\includegraphics[width=17cm]{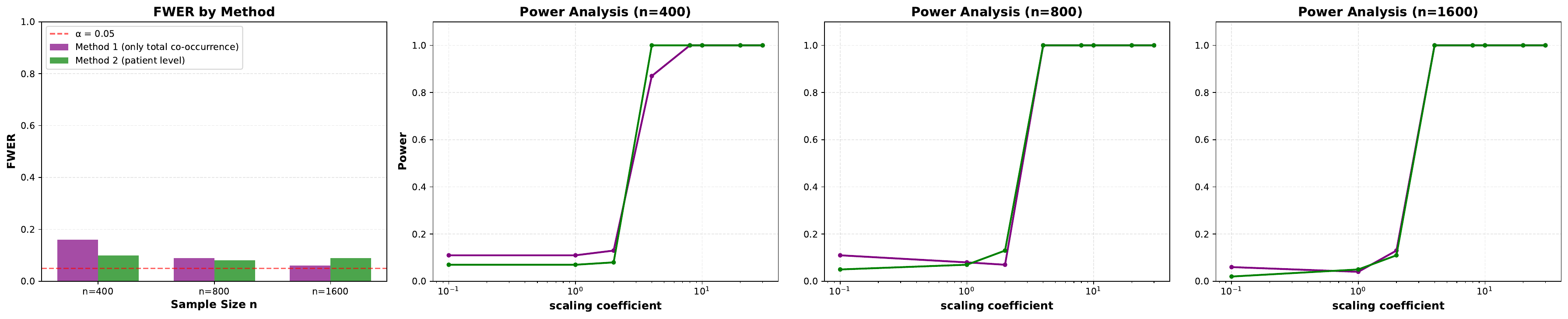}
    \caption{The left plot compares the FWER of our proposed method with and without patient-level data under the global null hypothesis; the remaining three plots analyzes the relation between power and signal strength of the proposed method under the alternative hypothesis.}%
    \label{fig:newtable1_and_power_analysis}%
\end{figure}

The testing is conducted on all unique edges \((w,w'), 1 \le w \neq w' \le d\), with the nominal significance level set as $\alpha = 0.05$ for each edge. We compare methods with and without using patient-level data for variance estimation and use Bonferroni correction to control the FWER. The FWER results are presented in Figure \ref{fig:newtable1_and_power_analysis} (Left). We can see that the FWER goes to \(\alpha\) as the sample size increases.

\subsection{Numerical Results on the Asymptotic Normality}\label{sec:43}

In this section, we generate data from \eqref{eq: code prob}. The embeddings are constructed using $\Vb = d^{-\kappa}\Ub$, where $\kappa \in \{1,2\}$ and $\Ub$ is formed from the orthonormal bases of a $d \times p$ matrix with each element independently drawn from a standard normal distribution. Notably, in this context, we do not utilize $\Vb\Vb^{\top}$, and there is no distinction between using $\Vb$ and $\Vb - \boldsymbol{1}\mu^{\top}$ for data generation. Consequently, the $2$nd and $3$rd steps outlined in Section \ref{sec:41} are unnecessary for embedding construction.

\begin{figure}[h]%
    \centering
\includegraphics[width=14cm]{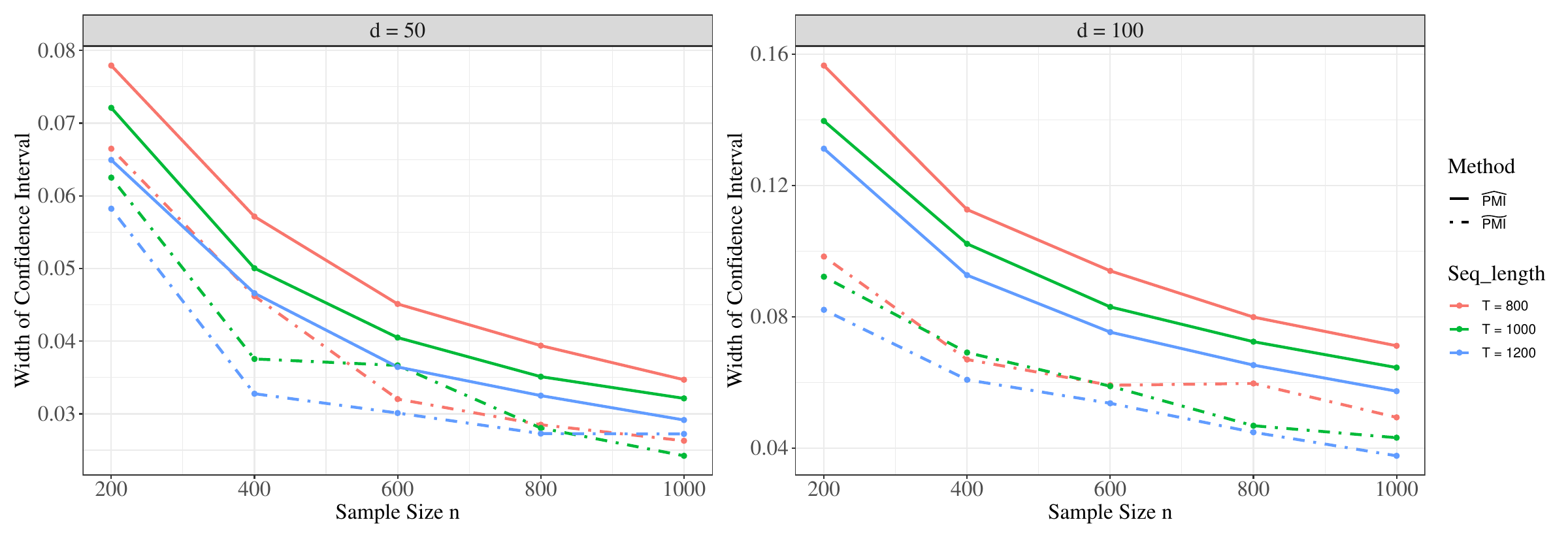} \\
\includegraphics[width=14cm]{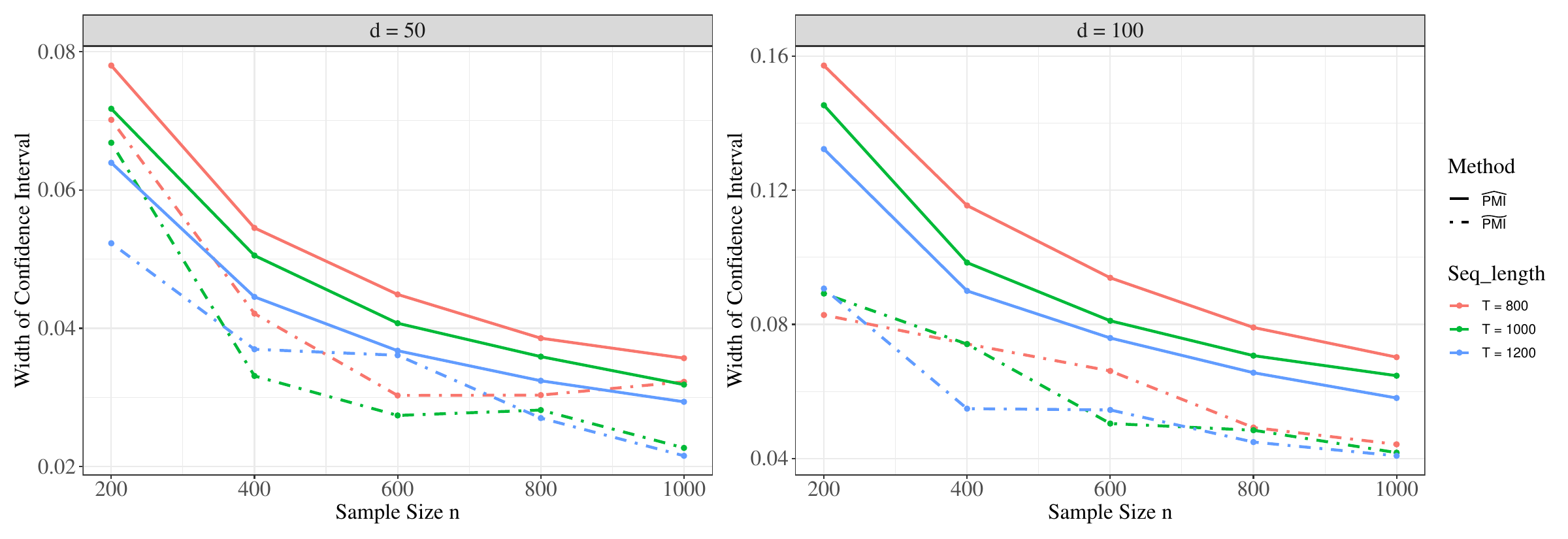} 
    \caption{Plots of the average half width of $95\%$ confidence interval of the first ten entries respectively constructed by the empirical estimator and the low-rank estimator. The first row is for $\kappa = 1$ and the second row is for $\kappa = 2$.}%
    \label{fig:ci_width}%
\end{figure}

The second and third plots in Figure \ref{fig:qqplot} depict the QQ plots of both the empirical estimator and the low-rank estimator. They suggest that both estimators are asymptotically normal. In Figure \ref{fig:ci_width}, we illustrate that the confidence intervals of $\tilde\PMIbb$ are consistently narrower than those of $\hat{\PMIbb}$. This indicates that the low-rank estimator $\tilde\PMIbb$ outperforms the empirical estimator which also validates our theoretical analysis on the order of variances.

{We further study the relation between the statistical power of our proposed methods and the signal strength \((d^{-\kappa})\). Figure \ref{fig:newtable1_and_power_analysis} shows that as the signal strength grows, the power emerges from near zero to one.}

\section{Applications to Electronic Health Records}
\label{sec:app}

We validate the efficacy of our algorithm using EHR data from the Veterans Affairs (VA) Healthcare System consisting of $12.6$ million patients who had at least one visit between $2000$ and $2019$. Patient information in the EHR data is structured as triplets (patient, date, clinical entity), which can be interpreted as sequences of EHR entities. The dataset includes four specific domains of codified information: PheCodes for diseases, Current Procedural Terminology (CPT) codes for procedures, lab test codes, and RxNorm codes for medication prescriptions. Codes occurring fewer than $5000$ times were excluded to minimize noise from infrequent codes, resulting in a final count of $1776$ PheCodes, $224$ CCS codes, $6025$ lab codes (including 382 LOINC codes and 5643 Local lab codes), and $1561$ RxNorm codes as detailed in \cite{hong2021clinical}. Consequently, the dataset contains $d=9586$ EHR entities of interest. These codes were then utilized to construct a co-occurrence matrix, adopting a window size, $q$, of $30$ days following the analysis of \cite{hong2021clinical}.

To assess our method's performance, we employ known similar and related pairs of these codes as established connections in the KG. Similar code pairs denote codes that embody highly analogous clinical entities according to existing ontologies. In contrast, related code pairs represent codes with more intricate relationships, such as `may cause' or `may treat'. For similarity assessments, we leverage the hierarchies of several ontologies to define similar pairs. These include disease-disease pairs from the PheCode Hierarchy 
and drug-drug pairs from the ATC classification system within the Unified Medical Language System \citep{mcinnes2007using}. We identified related pairs from online knowledge sources, categorizing them into four types: disease-disease, disease-drug, disease-procedure, and disease-lab. The counts of these pairings are detailed in Table \ref{tab:table_relation}, with details provided in \cite{zhou2022multiview}.

\begin{table}[htb]
    \centering
    \begin{tabular}{lcccc}
       \toprule
       Relation Type & Relation Category & Code pairs Number \\
        \hline
        \multirow{2}{*}{Similarity} & PheCode Hierarchy (Disease-Disease)  &  4094 \\
        & RxNorm-RxNorm (Drug-Drug) & 3647 \\
        \hline
        \multirow{3}{*}{Relatednedd} & PheCode-PheCode (Disease-Disease) & 2430 \\
        & PheCode-RxNorm  (Disease-Drug)  & 5523 \\
        & PheCode-CCS (Disease-Procedure) & 2545 \\
        & PheCode-Lab (Disease-Lab) & 334 \\
        \bottomrule
    \end{tabular}
    \caption{Number of curated relationship pairs categorized by seven categories.}
    \label{tab:table_relation}
\end{table}

We begin by evaluating the quality of the estimators. In each known-relation pair category, an equal number of negative pairs are randomly sampled. The predicted scores are then determined using our low-rank estimator $\tilde \PMIbb(w,w')$, compared with the empirical estimator $\PMIbbhat(w,w')$. Following this, the Area Under the Receiver Operating Characteristic Curve (AUC) is calculated for both methods, utilizing the positive and negative pairs.

Furthermore, we conduct a comparison with neural network-based embeddings, specifically those generated by various Bidirectional Encoder Representations from Transformer (BERT)-based algorithms \citep{devlin2018bert} trained on clinical corpora such as Self-aligning Pretrained BERT (SAPBERT) \citep{liu2020self}, BERT for Biomedical Text Mining (BioBERT) \citep{lee2020biobert}, and BERT pretrained with PubMed (PubMedBERT) \citep{pubmedbert}. These algorithms take code descriptions as inputs and produce embeddings as outputs.

The results, as displayed in Table \ref{tab:AUC}, reveal that $\tilde \PMIbb$ consistently outperforms the other methods in most of the categories, underlining its capability to effectively capture clinical knowledge from extensive EHR data. This demonstrates that $\tilde \PMIbb$ is a more precise estimator for representing clinical knowledge derived from massive EHR data.

\begin{table}[htb]
    \centering
    \setlength{\tabcolsep}{4pt}
    \begin{tabular}{l|c|cccccc}
        \toprule
        Relation Type & Category & $\tilde \PMIbb$ & $\PMIbbhat$ & Sap & PubMed & Bio & Bert \\
        \hline
        \multirow{3}{*}{Similarity} & PheCode Hierarchy & \textbf{0.970} & 0.856 & 0.763 & 0.616 & 0.580 & 0.592\\
        & Drug-Drug & \textbf{0.835} & 0.757 & 0.611 & 0.491 & 0.520 & 0.471\\
        \cline{2-8}
        & average & \textbf{0.906} & 0.809 & 0.691 & 0.557 & 0.552 & 0.535\\
        \hline
        \multirow{5}{*}{Relatedness} & Disease-Disease & \textbf{0.843} & 0.826 & 0.635 & 0.612 & 0.573 & 0.564\\
        & Disease-Drug & \textbf{0.823} & 0.810 & 0.604 & 0.631 & 0.610 & 0.597\\
        & Disease-Procedure & 0.743 & \textbf{0.760} & 0.635 & 0.594 & 0.565 & 0.531\\
        & Disease-Lab & \textbf{0.805} & 0.790 & 0.529 & 0.488 & 0.602 & 0.552\\
        \cline{2-8}
        & average & \textbf{0.808} & 0.801 & 0.616 & 0.614 & 0.591 & 0.573\\
        \bottomrule
        \end{tabular}
    \caption{The AUC of detecting known relation pairs with different methods. Sap denotes SapBert, PubMed denotes PubMedBert, Bio denotes BioBert. The row named `average' exhibits the results of the average AUC weighted by the number of pairs within each category.}
    \label{tab:AUC}
\end{table}

We then implement our testing procedure to identify significant edges, setting the target FDR at $0.05$ and $0.1$ within each category. To showcase the effectiveness of our proposed algorithm, we compare its power with other benchmark methods. Given that no existing methods provide statistical uncertainty quantification for the association between two entities, we rank the similarities of the pairs and select the top-ranking pairs as related pairs for other methods. The number of selected pairs matches that chosen by our testing procedure. This approach allows us to contrast the power of our algorithm with the benchmark 
methods.

From the results presented in Table~\ref{tab:power_pairs}, it is evident that our algorithm demonstrates a high power in detecting known-relation pairs. In particular, for several pair categories, such as PheCode-PheCode pairs from PheCode Hierarchy, and related disease-disease pairs, our algorithm achieves an exceptional power of over $85\%$ while all other methods have power lower than $75\%$. Furthermore, our method consistently outperforms the selection based on empirical PMI or cosine-similarity from other benchmarks for all pair categories. This underscores the capability of our algorithm to not only establish a threshold for selecting significant edges but also to leverage the variance information of the estimator, surpassing the performance of methods reliant solely on cosine similarity. Similar patterns can be found in Table~\ref{tab:app_power_pairs} in Supplementary \ref{app:additional} with target FDR set as $0.1$.

\begin{table}[htb!]
    \begin{tabular}{l|c|cccccc}
        \toprule
        Relation Type & Category & KNIT & $\PMIbbhat$ & Sap & PubMed & Bio & Bert \\
        \hline
        \multirow{3}{*}{Similarity} & PheCode Hierarchy & \textbf{0.917} & 0.816 & 0.723 & 0.514 & 0.468 & 0.478\\
        & Drug-Drug & \textbf{0.790} & 0.587 & 0.679 & 0.732 & 0.785 & 0.764\\
        \cline{2-8}
        & average & \textbf{0.857} & 0.708 & 0.702 & 0.617 & 0.617 & 0.613 \\
        \hline
        \multirow{5}{*}{Relatedness} & Disease-Disease & \textbf{0.872} & 0.833 & 0.546 & 0.505 & 0.467 & 0.442\\
        & Disease-Drug & \textbf{0.805} & 0.715 & 0.367 & 0.426 & 0.383 & 0.381\\
        & Disease-Procedure & 0.752 & \textbf{0.760} & 0.521 & 0.472 & 0.369 & 0.355\\
        & Disease-Lab & \textbf{0.677} & 0.560 & 0.394 & 0.363 & 0.333 & 0.312\\
        \cline{2-8}
        & average & \textbf{0.804} & 0.747 & 0.444 & 0.453 & 0.397 & 0.386\\
        \bottomrule
    \end{tabular}
    \caption{Power of detecting known relation pairs with our algorithm compared with other benchmarks, under target FDR being $0.05$. Sap denotes SapBert, PubMed denotes PubMedBert, Bio denotes BioBert. The row named `average' exhibits the results of the average power in detecting similar (related) pairs weighted by the number of pairs within each category.}
    \label{tab:power_pairs}
\end{table}

\begin{figure}[htbp!]
    \centering
    \includegraphics[width=0.9\textwidth]{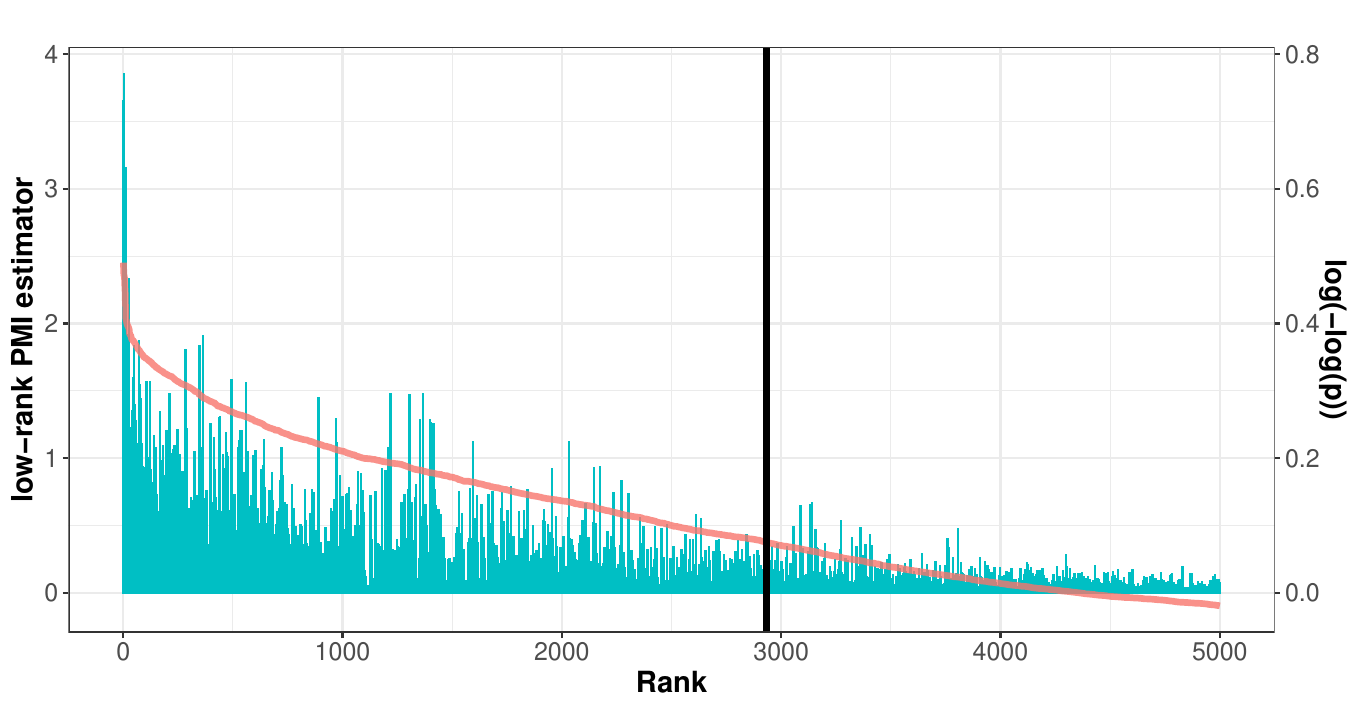}
    \caption{The estimated low-rank PMI of the top $5,000$ entities with the smallest $p$-values when quantifying their relationships with the Alzheimer's Disease (PheCode:290.11). The x-axis exhibits the rank of the $p$-values, the blue bars exhibit the estimated low-rank PMI and the red line exhibits $\log( - \log(p$-$\text{value}))$. The codes located to the left of the black line are identified as significantly related to the target feature, AD,  at a target FDR of $0.01$.}
    \label{fig:pmi_vs_rank}
\end{figure}

To showcase the efficacy of our algorithm in constructing a KG for clinical entities, we focus specifically on Alzheimer's Disease (AD) as a case study. We present in Figure~\ref{fig:pmi_vs_rank} the $p$-values and $\tilde{\PMIbb}_{w,w'}$, where $w$ is AD and $w'$'s are the other top $5,000$ entities with the smallest $p$-values in relation to AD. This visual representation highlights an important observation: entities $w'$ with high $\tilde{\PMIbb}_{w,w'}$ values do not necessarily correspond to small $p$-values. This finding underscores the limitations of existing methods that rely solely on the ranking of inner product values from KG embeddings to predict connections or edges between entities.

 To validate the accuracy of the KG, we conduct a manual review of the top twenty entities with the smallest $p$-values. This group includes eight PheCodes (diseases) and twelve RXNORMs (drugs). These entities are enumerated in Table~\ref{tab:AD_test_app} in Supplementary \ref{app:additional}. A substantial body of literature \citep{sosa2012epidemiology,bathgate2001behaviour,nussbaum2003alzheimer,reisberg2003memantine,tsuno2009donepezil,olin2002galantamine,xu2008demonstration,rabinowitz2007treating} confirms that these entities have a strong association with AD, thereby validating the precision of our KG.

To further assess the effectiveness of the proposed algorithm in identifying entities associated with the given disease AD, we incorporate a comparison with advanced language models ChatGPT \citep{OpenAI2023chatgpt} and GPT-4 \citep{OpenAI2023GPT4TR}. Specifically, we first select the top $50$ entities related to AD based on different methods, and then randomly sample an equal number of entities from the remaining set as negative samples. We then prompt ChatGPT and GPT-4 to rate the clinical relatedness between these $100$ entities and the target disease AD, outputting a score on a scale of $0$ to $1$ for each entity. The evaluation involves computing the Spearman rank correlation between the negative $p$-values generated by our algorithm and the scores from ChatGPT/GPT4. Additionally, we examine correlations between cosine similarities obtained from alternative benchmarks and the ChatGPT/GPT scores, as detailed in Table~\ref{tab:rank_corr_AD}. The results indicate that the rank correlation coefficients for KNIT with ChatGPT and GPT-4 are $0.553$ and $0.582$, respectively, outperforming other methods which all showed correlations less than $0.45$ with the language models. For reference, the rank correlation between ChatGPT and GPT-4 themselves is $0.610$.

\begin{table}[htp!]
    \centering
    \begin{tabular}{c|c|cccccc}
        \toprule
    & Statistics & KNIT & $\hat{ \mathbb{PMI}}$ & Sap & PubMed & Bio & Bert \\
        \hline
        \multirow{2}{*}{ChatGPT} & cor & \textbf{0.553} & 0.428 & 0.368 & 0.023 & -0.267 & -0.157\\
        \cline{2-8}
        & $p$-value & 0.001 & 0.001 & 0.001 & 0.397 & 0.996 & 0.945\\
        \hline
        \multirow{2}{*}{GPT-4} & cor & \textbf{0.582} & 0.303 & 0.307 & 0.085 & 0.026 & -0.152\\
        \cline{2-8}
        & $p$-value & 0.001 & 0.002 & 0.002 & 0.206 & 0.419 & 0.930\\
        \bottomrule
    \end{tabular}
    \caption{Spearman rank correlation test between the score given by different methods and that given by ChatGPT for $100$ pairs between AD and other clinical entities pairs. The correlation between the ChatGPT and GPT-4 is $0.610$.}
    \label{tab:rank_corr_AD}
\end{table}

\begin{figure}
    \centering
    \includegraphics[width=0.95\linewidth]{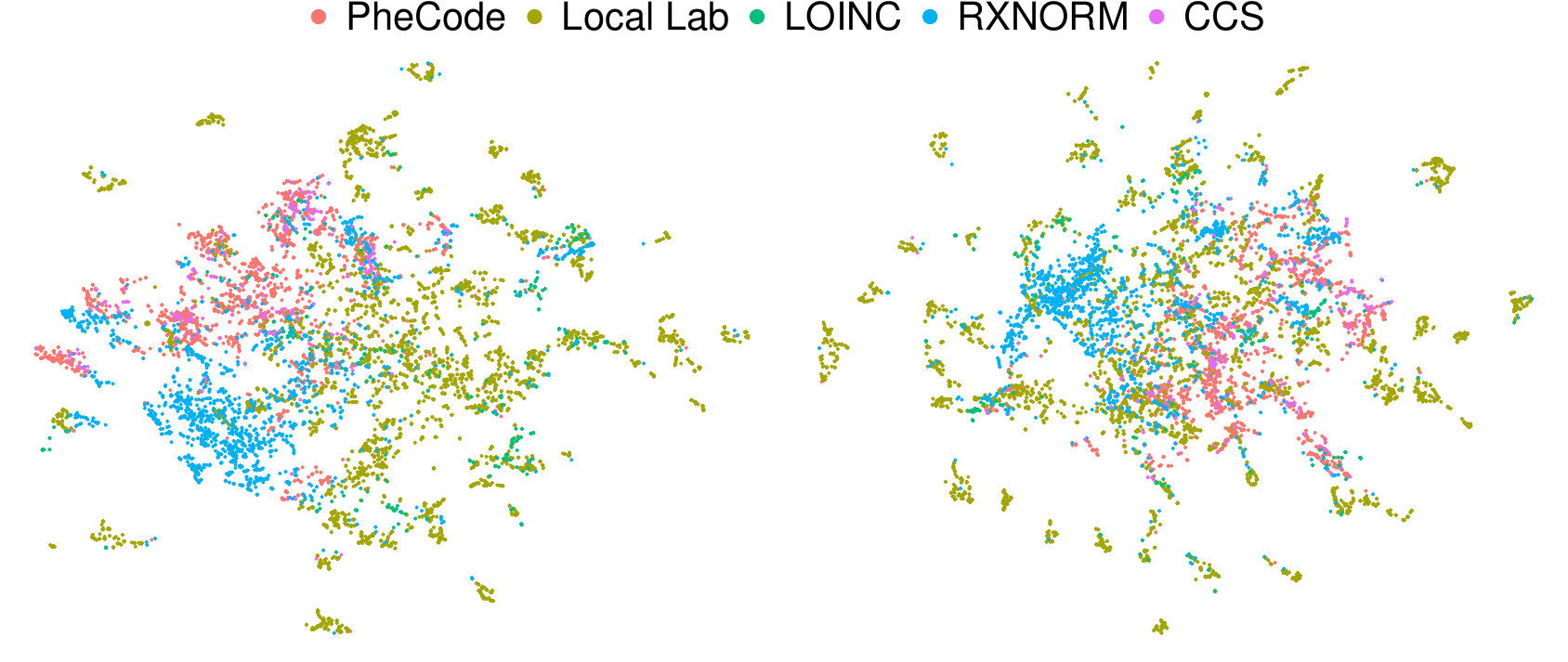}
    \caption{Visualization of the generated embedding via t-SNE. The left figure exhibits the first dimension and the second dimension; while the right figure exhibits the second dimension and the third dimension. The points are colored the group of the codified codes.}
    \label{fig:embed_visual}
\end{figure}

To better understand the generated embedding, we apply t-SNE \citep{van2008visualizing} to further reduce the embedding dimension and visualize the embeddings. From Figure~\ref{fig:embed_visual}, we can see that the generated embedding does help cluster codified codes into different types of codes. The visualization demonstrates that features with closer embedding distances exhibit higher correlations and stronger relationships, suggesting meaningful topological preservation in the latent space. These findings highlight the practical value of the learned embeddings for various downstream applications, including feature clustering analysis and predictive modeling for disease outcomes.

\section{Conclusion}
\label{sec:con}
In this paper, we prove the asymptotic normality of the low-rank estimator $\tilde \PMIbb$ and propose an algorithm for hypothesis testing on the low-rank KG $\Vb \Vb^{\top}$. When the patient-level data is approachable, the mean and variance of the entries of the knowledge matrix can be well estimated; when the patient-level data is not approachable, we estimate the covariance matrices for the low-rank estimator under the global null distribution, using them as a proxy for the actual covariance structure. The main idea of proposing the low-rank estimator is to utilize the low-rank structure of the KG and reduce the variance of the estimator to give a precise estimation of the KG.

\section*{Acknowledgments}

Tianxi Cai was supported by NIH grants R01 LM013614 and R01 NS098023.  
Junwei Lu was supported by NSF grant DMS 2434664, the William F. Milton Fund, NIH/NINDS grant R01 NS098023, and the AWS Impact Computing Project at the Harvard Data Science Initiative.  
Doudou Zhou was supported by the NUS Startup Grant A-0009985-00-00 and the MOE AcRF Tier 1 Grant A-8003569-00-00.

\bibliographystyle{chicago}

\bibliography{Bibliography-MM-MC}



\setcounter{section}{0}
\renewcommand{\thesection}{S.\arabic{section}}
\setcounter{equation}{0}
\counterwithout{equation}{section}
\renewcommand{\theequation}{S.\arabic{equation}}
\setcounter{theorem}{0}
\counterwithout{theorem}{section}
\renewcommand{\thetheorem}{S.\arabic{theorem}}
\renewcommand{\thelemma}{S.\arabic{theorem}}

\renewcommand{\thesection}{S\arabic{section}}  
\renewcommand{\thetable}{S\arabic{table}}  
\renewcommand{\thefigure}{S\arabic{figure}}
\renewcommand{\theequation}{S\arabic{equation}}

\section{Properties of the PMI Estimator}

\subsection{Additional Notation}
For two sequences $\{x_n\}$ and $\{y_n\}$, we say $x_n \sim y_n$ if $x_n = O(y_n)$ and $y_n = O(x_n)$. Denote the total variation distance between distribution $P$ and $Q$ as $\|P-Q\|_{\text{tv}}$. Given two matrices $\Ub_1,\Ub_2 \in \mathbb{R}^{d \times p}$ satisfying $\Ub_1^{\top}\Ub_1 = \Ub_2^{\top}\Ub_2 =\Ib_p$, we measure the distance of two matrices' column space by $\text{dist}(\Ub_1,\Ub_2) = \min_{\Rb \in \mathcal{O}^{p \times p}}\|\Ub_1 \Rb - \Ub_2\|$.
\subsection{Preliminary Results on PMI}\label{subsec:notation}
To state the theorems and lemmas more strictly, in this section we introduce some new notations and rewrite some notations occurred in the paper.
In our model, there are two layers of random variables: the occurrence of code $w$ at time $t$ in patient $i$'s EHR is denoted as $X_{i,w}(t)$. $X_{i,w}(t)$ follows a Bernoulli distribution with occurrence probability
\begin{equation*}
    p_{i,w}(t) = \E[X_{i,w}(t)|\bc_{i,t}] = \frac{\exp(\langle \Vb_w, \bc_{i,t}\rangle)}{\sum_{w'}\exp(\langle \Vb_{w'}, \bc_{i,t} \rangle)}.
\end{equation*}
Here $p_{i,w}(t)$ is a function of discourse vector $\bc_{i,t}$. The co-occurrence of code $w$ at time $t$ and code $w'$ at time $t+u$ in $i$-th patient's EHR is denoted as $X_{i,w,w'}(t,t+u)$, which follows a Bernoulli distribution with co-occurrence probability 
\begin{equation*}
    p_{i,w,w'}(t,t+u) = \E[X_{i,w,w'}(t,t+u)|\bc_{i,t}, \bc_{i,t+u}] = \frac{\exp(\langle \Vb_w, \bc_{i,t}\rangle)}{\sum_{w''}\exp(\langle \Vb_{w'}, \bc_{i,t}\rangle)}\cdot \frac{\exp(\langle \Vb_{w'}, \bc_{i,t+u}\rangle)}{\sum_{w''}\exp(\langle \Vb_{w'}, \bc_{i,t+u}\rangle)}.
\end{equation*}
Denote $X_{i,w} = \sum_{t=1}^{\zx{T_i}} X_{i,w}(t), S_{i,w} = \sum_{t=1}^{\zx{T_i}} p_{i,w}(t);$ and $X_{i,w,w'}^{(u)} = \sum_{t=1}^{\zx{T_i-q}}X_{i,w,w'}(t,t+u), S_{i,w,w'}^{(u)} = \sum_{t=1}^{\zx{T_i-q}} p_{i,w,w'}(t,t+u).$ Denote $p_w = \E_{\bc_{i,t}}[p_{i,w}(t)],$ which is irrelevant to subscripts $i,t$ since the distribution of $\{\bc_{i,t}\}$ is identical for each $i$ and is assumed to be stationary at the beginning. Denote $p_{w,w'}^{(u)} = \E_{\bc_{i,t},\bc_{i,t+u}}[p_{i,w,w'}(t,t+u)]$. Further we denote \zx{$N_{i,w} = \E[S_{i,w}] = T_ip_w$}, and \zx{$N_{i,w,w'}^{(u)} = \E[S_{i,w,w'}^{(u)}] = (T_i-q)p_{w,w'}^{(u)}$}. Denote $X_{i,w,w'}^{[q]} = \sum_{u=1}^q(X_{i,w,w'}^{(u)} + X_{i,w',w}^{(u)})$ and $X_{i,w}^{[q]} = \sum_{w' = 1}^d X_{i,w,w'}^{[q]}$. \zx{Then we have
\begin{equation*}
\begin{split}
        \sum_{w=1}^d X_{i,w}^{[q]} &= \sum_{u=1}^q \sum_{w,w'} (X_{i,w,w'}^{(u)} + X_{i,w',w}^{(u)}) = 2\sum_{u=1}^q \sum_{w,w'} \sum_{t=1}^{T_i-q}X_{i,w,w'}(t, t+u)\\
        &= 2\sum_{u=1}^q \sum_{t=1}^{T_i-q}\sum_w \sum_{w'}X_{i,w,w'}(t, t+u) = 2\sum_{u=1}^q \sum_{t=1}^{T_i-q}\sum_w X_{i,w}(t) = 2q(T_i-q).
\end{split}
\end{equation*}
}

Correspondingly, denote $S_{i,w,w'}^{[q]} = \sum_{u=1}^q(S_{i,w,w'}^{(u)} + S_{i,w',w}^{(u)})$ and $S_{i,w}^{[q]} = \sum_{w' = 1}^d S_{i,w,w'}^{[q]}$.
Denote \zx{$N_{i,w,w'}^{[q]} = \E[X_{i,w,w'}^{[q]}] = 2\sum_{u=1}^q (T_i-q)p_{w,w'}^{(u)},$ and $N_{i,w}^{[q]} = \E[X_{i,w}^{[q]}] = 2\sum_{u=1}^q \sum_{w'}(T_i-q)p_{w,w'}^{(u)}.$} Note that for any $t,u$, we have 
\begin{equation*}
    \sum_{w'=1}^d p_{i,w,w'}(t,t+u) = p_{i,w}(t).
\end{equation*}
Take expectation on $(\bc_{i,t},\bc_{i,t+u})$ of both sides, we have $\sum_{w'=1}^d p_{w,w'}^{(u)} = p_w.$
Then we can simplify the expression of \zx{$N_{i,w}^{[q]}$} with \zx{$N_{i,w}^{[q]} =2\sum_{u=1}^q (T_i-q)p_{w} = 2q(T_i - q)p_w.$} Finally, denote \zx{$N_i^{[q]} = \sum_{w=1}^d N_{i,w}^{[q]} = 2q(T_i-q)$.}
Thus we have
\zx{
\begin{equation*}
    \frac{N_i^{[q]}N_{i,w,w'}^{[q]}}{N_{i,w}^{[q]}N_{i,w'}^{[q]}} = \frac{2q(T_i - q)2\sum_{u=1}^q(T_i-q)p_{w,w'}^{(u)}}{(2qT_i - 2q^2)^2 p_w p_{w'}} = \frac{\sum_{u=1}^qp_{w,w'}^{(u)}}{q p_w p_{w'}},
\end{equation*}
which does not depend on \(i\) and \(T_i\).
}
Recall the definition of the population PMI matrix in (\ref{eq: pmi def}). \zx{Equivalently,} we have $p(w,w')= \sum_{u=1}^q p_{w,w'}^{(u)}/q$, $p(w) = p_w$, and
\zx{
\begin{equation}\label{def:PMI_detailed}
    \PMIbb(w,w') = \log\Big( \frac{\sum_{u=1}^q p_{w,w'}^{(u)}}{q p_w p_{w'}}\Big)= \log\frac{N_i^{[q]}N_{i,w,w'}^{[q]}}{N_{i,w}^{[q]}N_{i,w'}^{[q]}}.
\end{equation}
}
\zx{Define \(T = \sum_{i=1}^n T_i/n\). Define the error matrix $\Eb = \PMIbbhat - \PMIbb$. Define
\[
N^{[q]} = \sum_i N_i^{[q]}/n, N_{w}^{[q]} = \sum_i N_{i,w}^{[q]}/n, N_{w,w'}^{[q]} = \sum_i N_{i,w,w'}^{[q]}/n; \quad w,w' \in [d].
\]}
Define $\bq = (q_1,\cdots,q_d)^{\top} \in \mathbb{R}^{d}$ with $q_w = \exp\left(\|\Vb_w\|^2/2p\right)/Z$ and $Z = \sum_w \exp(\|\Vb_w\|^2/2p)$. Here $q_w$ is an approximation of $p_w$ that assists some proof in the Supplementary.
Note that we have defined the occurrence numbers in Section \ref{sec:dat_generation}: $\CC^{(i)}_{w,w'} = X_{i,w,w'}^{[q]}$, $\CC_{w,w'} = \sum_{i=1}^n X_{i,w,w'}^{[q]}$, $\CC_{w} = \sum_{i=1}^n X_{i,w}^{[q]}$, and $\overline{\CC} = nN^{[q]}$. For the rest part of the Supplementary, we will use notations introduced in this section to emphasize the window size $q$.
We will also omit the script $i$ when we only analyze one patient.

\subsection{Low Rank Approximation of PMI}

\begin{lemma}
Suppose $(\bc,\bc')\sim (\bc_{t},\bc_{t+u})$ for some fixed $u\in [q]$ and $t \geq 1$. We have
$$
\P \Big(\|\bc -\bc'\| \leq \frac{\sqrt{2u\log d}}{p} \Big) \geq 1 - \exp \left(-\frac{p}{3}\right).
$$
\end{lemma}
\begin{proof}
The proof is essentially the same as the concentration of chi-square distribution. Note that $(\bc,\bc') \stackrel{d}{=} (\bc, \alpha^{u/2}\bc + \sqrt{1-\alpha^u}\br)$, where $\bc$ and $\br$ are independent, and both following $N(0, \Ib_p /p)$, and in turn  
$$
\bc -\bc'  \stackrel{d}{=} \sqrt{2(1 - \alpha^{u/2}) }\br,
$$
and in turn by the concentration inequality of chi-square distribution,
\begin{align*}
    &\P\big( \|\bc -\bc'\| > \sqrt{2(1 - \alpha^{u/2}) }\sqrt{1 + t} \big) \le \exp\big(-\frac{p}{2}(t^2 - t^3/3)\big).
\end{align*}
Let $t = 1$. By Bernoulli's inequality, we have $1-\alpha^{u/2} \le u\log (d)/(2p^2)$. Then we have 
\begin{align*}
    &\PP\big( \|\bc -\bc'\| > \frac{\sqrt{2u\log d}}{p} \big) \le \exp\big(-\frac{p}{3}\big).
\end{align*}
\end{proof}

\begin{lemma}\label{lem:a_loose_upper_bound_of_pw}
   Suppose Assumptions \ref{assump1} and \ref{assump:eigen_bound} hold. Let $\bp = (p_1, \cdots, p_d)$ and $\boldsymbol{1} = (1,\cdots, 1) \in \RR^d$. Then
   \[
   \|\bp - \frac{1}{d}\boldsymbol{1}\|_{\max} \leq \frac{\log d}{2d}\|\Vb\|_{2,\infty}.
   \]
\end{lemma}
\begin{proof}
Note that $\bc \sim N(0,\Ib_p/p)$.
Denote event $\mathcal{F} = \{\|\bc\| \leq \frac{1}{8}\log d\}$, then $\P(\mathcal{F}) = 1 - \exp(-\Omega(\log^2 d))$. For each $w\in [d]$. we have
    \begin{equation*}
        \begin{split}
            \big|p_w - \frac{1}{d} \big| &= \Big|\E_{\bc}\Big[\frac{\exp(\langle \Vb_w, \bc\rangle)}{\sum_{w'}\exp(\langle \Vb_{w'}, \bc\rangle)} - \frac{1}{d}\Big]\Big| \leq \E_{\bc}\Big[\frac{\sum_{w'}|\exp(\langle \Vb_w, \bc\rangle) - \exp(\langle \Vb_{w'}, \bc\rangle)|}{d\sum_{w'}\exp(\langle \Vb_{w'}, \bc\rangle)} \Big]\\
            &\leq \E_{\bc}\Big[\frac{\sum_{w'}|\exp(\langle \Vb_w, \bc\rangle) - \exp(\langle \Vb_{w'}, \bc\rangle)|}{d\sum_{w'}\exp(\langle \Vb_{w'}, \bc\rangle)} I\{\cF\}\Big] + \P(\cF^c)\\
            &\leq \frac{d\|\Vb\|_{2,\infty}\log (d)/4}{d^2\exp(-\|\Vb\|_{2,\infty}\log (d)/8)} + \exp(-\Omega(\log^2 d)) \leq \frac{\log d}{2d}\|\Vb\|_{2,\infty}
        \end{split}
    \end{equation*}
for sufficiently large $d$. Here the second inequality uses the law of total expectation, the third inequality uses $|\langle \Vb_w, \bc\rangle I(\cF)| \leq \|\Vb\|_{2,\infty}\log(d)/8$ with probability $1$, and the last inequality uses $|\exp(x) - 1| \leq 2x$ for $|x| \leq 1/2$ and $\|\Vb\|_{2,\infty} \leq \mu_1\kappa \sqrt{p}/\sqrt{d\alpha_p} = o(1/d)$, which is from Assumption \ref{assump1} and \ref{assump:eigen_bound}.
\end{proof}

\begin{lemma}\label{lem:concen_of_Z}
Suppose Assumptions \ref{assump1} and \ref{assump:eigen_bound} hold. Define function $Z(\bc) = \sum_w \exp(\Vb_w^{\top}\bc)$. Then we have
$$
\P_{\bc\sim\mathcal{C}}\left((1-\epsilon_d)Z \leq Z(\bc) \leq (1+\epsilon_d)Z \right) = 1 - \exp(-\Omega(\log^2d))
$$
where $Z = \E_{\bc}[Z(\bc)],$ and $ \epsilon_d=\frac{\kappa^2 p^{3/2} \log^2 d}{d}.$
\end{lemma}
\begin{proof}
This Lemma's result is similar to Lemma F.2 in \cite{xxx2021}, but the proof idea is different. Lemma F.2 in \cite{xxx2021} first fixes $\bc$ and assumes $\Vb$ follows Gaussian distribution, {whereas $\Vb$ is deterministic in our case.}
Note that $\bc \sim N(0,\Ib_p/p)$.
Denote event $\mathcal{F} = \{\|\bc\| \leq 0.25\log (d)\}$, then $\P(\mathcal{F}) = 1 - \exp(-\Omega(\log^2 d))$.
Let $f(\bc) = \sum_{w=1}^d\exp\left(\Vb_w^{\top}\bc \right)I\{\mathcal{F}\}$, then with probability 1 we have that
\begin{equation*}
    \begin{split}
        \|\nabla f\|^2 =& \sum_{i=1}^p\big(\sum_{w=1}^d V_{w i}\exp\left(\Vb_w^{\top}\bc\right)\big)^2 =   \sum_{i=1}^p\big(\sum_{w=1}^d V_{w i}(\exp(\Vb_w^{\top}\bc)-1 + 1 - dp_w)\big)^2\\
        \leq & 2d\sum_{i=1}^p\sum_{w=1}^d V_{w i}^2 [|2\Vb_w^{\top}\bc|^2 + (1-dp_w)^2] \\
        \leq & 2d\sum_{i=1}^p\sum_{w=1}^d V_{w i}^2 \Big[\big(\frac{\log d}{2}\|\Vb_w\| \big)^2+ \big( \frac{\log d}{2}\|\Vb\|_{2,\infty}\big)^2 \Big]\\
        \leq & d\sum_{i=1}^p\sum_{w=1}^d V_{w i}^2 \log^2(d) \|\Vb\|_{2,\infty}^2 
        = d\log^2(d)
        \sum_{w=1}^d\|\Vb_w\|^2\|\Vb\|_{2,\infty}^2  
        \leq d^2\log^2(d)\|\Vb\|_{2,\infty}^4
    \end{split}
\end{equation*}
for large $d$. Here the second equation uses $\sum_{w=1}^d p_w \Vb_{w} = \boldsymbol{0}$;
the first inequality uses $|\exp(\Vb_w^{\top}\bc)-1| \leq 2|\Vb_w^{\top}\bc| \leq 2\|\Vb_w\|\cdot\|\bc\| \leq \|\Vb_w\|\cdot \log (d)$, which holds with probability $1$ for large $d$ under the event $\cF$ by Assumptions \ref{assump1} and \ref{assump:eigen_bound}, and the inequality $(\sum_{w=1}^d x_w)^2 \leq d\sum_{w=1}^d x_w^2$; the second inequality uses Lemma \ref{lem:a_loose_upper_bound_of_pw}. Note that
\begin{equation*}
    \E[Z(\bc)^2] \leq d\E\big[\sum_{w=1}^d\exp(2\Vb_w^{\top}\bc)\big] = d\sum_{w=1}^d\exp\Big(\frac{2\|\Vb_w\|^2}{p}\Big) \leq d^2\exp\big(\frac{2\kappa^2\mu_1^2}{\alpha_p d}\big),
\end{equation*}
where the last inequality is from Assumption \ref{assump1}.
Again by applying the Cauchy-Schwarz inequality, we have
$$
0 \leq \E[Z(\bc)] - \E[Z(\bc)I\{\mathcal{F}\}] \leq \sqrt{\P(\mathcal{F}^c) d^2\exp\big(\frac{2\kappa^2\mu_1^2}{\alpha_p d}\big)} = \exp (-\Omega(\log^2 d)).
$$
Since $\exp(\Vb_w^{\top}\bc) + \exp(-\Vb_w^{\top}\bc) \geq 2$, and $\bc$ follows a symmetric distribution, we have $\E_{\bc \sim \mathcal{C}}[\exp(\Vb_w^{\top}\bc)]\geq 1$. Then we have 
$$
Z = \E_{\bc \sim \mathcal{C}}[Z(\bc)] = \sum_{w=1}^d\E[\exp(\Vb_w^{\top}\bc)] \geq d.
$$
Then by Lemma F.1 in \cite{xxx2021}(or Corollary 2.5 in \cite{L1999}), we have
\begin{equation*}
    \begin{split}
        \P \Big(\Big|\frac{Z(\bc) - Z}{Z}\Big| \geq 2\epsilon \Big)&= \P \Big(\Big|\frac{Z(\bc) - Z}{Z}\Big| \geq 2\epsilon, \cF^c \Big) + \P \Big(\Big|\frac{Z(\bc) - Z}{Z}\Big| \geq 2\epsilon, \cF \Big)\\
        &\leq  \P(\cF^c) \!+\! \P \Big(\Big| \frac{Z(\bc)I\{\mathcal{F}\} - \E[Z(\bc)I\{\mathcal{F}\}]}{Z} \Big| \geq 2\epsilon \!-\! \Big|\frac{Z - \E[Z(\bc)I\{\mathcal{F}\}]}{Z}\Big|\Big)\\
        &\leq \P(\mathcal{F}^c) + \P \left(\left| \frac{Z(\bc)I\{\mathcal{F}\} - \E[Z(\bc)I\{\mathcal{F}\}]}{Z} \right| \geq \epsilon \right)\\
        &\leq \exp(-\Omega(\log^2 d)) + 2\exp\Big(-\frac{p\epsilon^2}{\log^2(d)\|\Vb\|_{2,\infty}^4} \Big)
    \end{split}
\end{equation*}
for $\epsilon = \exp (-O(\log^2 d))$. By Assumption \ref{assump1}, we have $\|\Vb\|_{2,\infty} \leq \kappa \mu_1 \sqrt{p/d\alpha_p} \lesssim \kappa p/\sqrt{d}$.
Let $\epsilon = \kappa^2 p^{3/2} \log^2 (d)/d.$ It yields that
$$
\P \Big(\Big|\frac{Z(\bc) - Z}{Z}\Big| \geq 2\epsilon \Big) \leq \exp(-\Omega(\log^2 d)) + 2\exp(-\Omega(\log^2 d)), = \exp(-\Omega(\log^2 d)),
$$
which completes the proof.
\end{proof}

\begin{lemma}\label{lem:B37}
Suppose Assumptions \ref{assump1} and \ref{assump:eigen_bound} hold. Then we have
$$
 \max_{w} \left|\log p_w -  \left(\frac{\|\Vb_w\|^2}{2p} - \log Z \right) \right| \leq C_1\frac{\kappa^4p^2}{d^2}
$$
for some constant $C_1 > 0$.
\end{lemma}
\begin{proof}
Denote $A_w = \exp(\Vb_w^{\top}\bc)$. Denote functions $f_i : \mathbb{R}^p \rightarrow \mathbb{R}, i=4,5$ as $$f_4(\bX) := \exp\left(\frac{\|\Vb_w + \bX\|^2}{2p}\right),$$ 
$$f_5(\bX) := f_4(\bX)/f_4(\boldsymbol{0}) = \exp\left(\frac{\|\bX\|^2+2\bX^{\top}\Vb_w}{2p} \right).$$
Then we have
\begin{equation*}
    \E \left[A_w\right] = f_4(\boldsymbol{0});
\end{equation*}
\begin{equation*}
        \E \left[A_w Z(\bc) \right] = \E\left[\sum_{w'}\exp\left((\Vb_w + \Vb_{w'})^{\top}\bc \right) \right] = \sum_{w'}f_4(\Vb_{w'});
\end{equation*}
\begin{equation*}
    \begin{split}
        \E \left[A_w Z^2(\bc) \right] = \E\left[\sum_{w',w''}\exp\left(\left(\Vb_w + \Vb_{w'} + \Vb_{w''} \right)^{\top}\bc \right) \right] = \sum_{w',w''}f_4(\Vb_{w'}+\Vb_{w''}).
    \end{split}
\end{equation*}
Denote a set $$
F=\Big\{\Big|\frac{Z( \boldsymbol{c})}{Z}-1\Big|<\epsilon_d\Big\},
$$
where $\epsilon_d = \frac{\kappa^2 p^{3/2} \log^2 d}{d}.$ By Lemma \ref{lem:concen_of_Z}, we know $\mathbb{P}(F)=1-\exp (-\Omega(\log ^{2} d)).$
For any integer $m>0$, we have the following decomposition:
\begin{equation}\label{eq:Z_decom}
    \frac{1}{Z(\bc)} = \sum_{k=1}^m\frac{(Z-Z(\bc))^{k-1}}{Z^k} + \frac{(Z-Z(\bc))^m}{Z(\bc)Z^m}.
\end{equation}
Denote 
\begin{equation*}
    g_{m,w} = \sum_{k=1}^m\frac{\E \big[ A_w  (Z-Z(\bc))^{k-1} \big]}{Z^k};
    b_{m,w} = \E \Big[\frac{A_w(Z-Z(\bc))^m}{Z(\bc)Z^m} \Big], m \geq 1.
\end{equation*}
Then we have $p_w = g_{m,w}+b_{m,w}$ for any integer $m > 0$. When $m = 2k+1, k \geq 0$, we have
\begin{equation*}
     \begin{split}
         b_{2k+1,w} &= \E \Big[\frac{A_w(Z-Z(\bc))^{2k+1}}{Z(\bc)Z^{2k+1}} I(F) \Big] + \E \Big[\frac{A_w(Z-Z(\bc))^{2k+1}}{Z(\bc)Z^{2k+1}}I(F^c) \Big]\\
         &\leq \E \Big[\frac{A_w \epsilon_d^{2k+1}}{Z(1-\epsilon_d)} I(F) \Big] + \E \Big[\frac{A_w}{Z(\bc)}I(F^c) \Big] \leq \E \Big[\frac{A_w \epsilon_d^{2k+1}}{Z(1-\epsilon_d)} \Big] + P(F^c)\\
         &= \frac{g_{1,w}\epsilon_d^{2k+1}}{1-\epsilon_d} + \exp(-\Omega(\log^2 d)),
     \end{split}
\end{equation*}
where the first inequality comes from the definition of set $F$ and $(Z-Z(\bc))^{2k+1} \leq Z^{2k+1}$; the second inequality is due to $A_w = \exp(\langle \bv_w, \bc \rangle) \leq Z(\bc)$ and $I(F) \leq 1.$ On the other side, we have
\begin{equation*}
        b_{2k+1,w} \geq \E \Big[\frac{A_w(Z-Z(\bc))^{2k+1}}{Z(\bc)Z^{2k+1}}I(F^c) \Big] \geq \E \Big[\frac{(-Z(\bc))^{2k+1}}{Z^{2k+1}}I(F^c) \Big].
\end{equation*}
By Cauchy-Schwarz inequality and $Z \geq d$, we have
\begin{equation*}
        b_{2k+1,w} \geq - \frac{\sqrt{\E[(Z(\bc))^{4k+2}]P(F^c)}}{Z^{2k+1}} \geq - \frac{\exp(-\Omega(\log^2 d))\sqrt{\E[(Z(\bc))^{4k+2}]}}{d^{2k+1}}.
\end{equation*}
By Holder's inequality, we have $(\sum_{i=1}^d x_i)^m \leq d^{m-1}\sum_{i=1}^d|x_i|^m$. Thus we have
\begin{equation*}
    \E[(Z(\bc))^{4k+2}] \leq d^{4k+1}\sum_{w=1}^d \E[\exp(\langle (4k+2)\Vb_w, \bc \rangle)] = d^{4k+1}\sum_{w=1}^d \exp \Big(\frac{(4k+2)^2\|\Vb_w\|^2}{2p} \Big).
\end{equation*}
By Assumption \ref{assump1}, we have
\begin{equation*}
   \E[(Z(\bc))^{4k+2}] \lesssim d^{4k+2}\exp((4k+2)^2\kappa^2p/d).
\end{equation*}
It yields that when $m^2 \kappa^2 p = o(d\log^2 d)$,
\begin{equation*}
     b_{2k+1,w} \geq - \exp(-\Omega(\log^2 d - (4k+2)^2\kappa^2 p/d)) = - \exp(-\Omega(\log^2 d)).
\end{equation*}
Combine the properties of $b_{m,w}$ when $m$ is odd, we have
\begin{equation*}
    g_{m,w} - \exp(-\Omega(\log^2 d)) \leq p_w \leq g_{m,w} + \frac{g_{1,w}\epsilon_d^{m}}{1-\epsilon_d} + \exp(-\Omega(\log^2 d)).
\end{equation*}
Then it yields that $|p_w - g_{m,w}| \leq 2g_{1,w}\epsilon_d^m + \exp(-\Omega(\log^2 d))$ for sufficiently large $d$. Let $m = 1$, we obtain
\begin{equation}\label{ineq:dist_pw_qw}
    |p_w - q_{w}| \leq 2g_{1,w}\frac{\kappa^2p^{3/2}\log^2 d}{d} + \exp(-\Omega(\log^2 d)) = o(g_{1,w}),
\end{equation}
where the last equation uses Assumption \ref{assump:eigen_bound} and $q_w = g_{1,w} \gtrsim 1/d$, which uses Assumptions \ref{assump1} and \ref{assump:eigen_bound}, and
\[
g_{1,w} = f_4(\boldsymbol{0})/Z \geq 1/Z \geq 1/(d\exp(\|\Vb\|_{2,\infty}^2/2p))\gtrsim 1/d
\]
A near-optimal $m$ can be taken as $m^{\star} = 2\lfloor \sqrt{\frac{d\log^2 d}{\kappa^2p}} \rfloor + 1$, and correspondingly we have for any $m \leq m^{\star}$,
\begin{equation}\label{ineq:pw_gmw}
    |p_w - g_{m,w}| \leq Cp_w\big(\epsilon_d^{m}+\exp(-\Omega(\log^2 d))\big)
\end{equation}
since $p_w = g_{1,w} + o(g_{1,w}) \gtrsim 1/d$. Denote $\varepsilon_d^m = \epsilon_d^{m}+\exp(-\Omega(\log^2 d))$. Then we have
\begin{equation}\label{ineq:logpw_gmw}
    |\log p_w - \log g_{m,w}| \leq \max \big(\log(1+C\varepsilon_d^{m}), -\log(1-C\varepsilon_d^{m})\big) \leq 2C\varepsilon_d^{m}.
\end{equation}
Next we calculate $g_{m,w}$. Note that
\begin{equation*}
    \sum_{k=1}^m\frac{(Z-Z(\bc))^{k-1}}{Z^k} = \frac{1}{Z(\bc)}\Big[ 1 - \Big(1 - \frac{Z(\bc)}{Z} \Big)^m\Big] = \sum_{k=1}^m \binom{m}{k}(-1)^{k-1}\frac{Z(\bc)^{k-1}}{Z^k}.
\end{equation*}
We have 
\begin{equation*}
    \begin{split}
        g_{m,w} &= \sum_{k=1}^m \binom{m}{k}\frac{(-1)^{k-1}}{Z^k}\E\big[A_w Z(\bc)^{k-1} \big]
        =\sum_{k=1}^m \binom{m}{k}\frac{(-1)^{k-1}}{Z^k}\sum_{w^{(1)},\cdots,w^{(k-1)}}f_4\Big(\sum_{i=1}^{k-1} \Vb_{w^{(i)}}\Big)\\
        &=\frac{f_4(\boldsymbol{0})}{Z} \Big[ \sum_{k=1}^m\binom{m}{k}\frac{(-1)^{k-1}}{Z^{k-1}}\sum_{w^{(1)},\cdots,w^{(k-1)}} f_5 \Big(\sum_{i=1}^{k-1} \Vb_{w^{(i)}}\Big)\Big].
    \end{split}
\end{equation*}
Note that $f_5(\boldsymbol{0}) = 1$, and
\begin{equation*}
    \sum_{k=1}^m\binom{m}{k}\frac{(-1)^{k-1}}{Z^{k-1}}Z^{k-1} = \sum_{k=1}^m\binom{m}{k}(-1)^{k-1} = 1 -\sum_{k=0}^m\binom{m}{k}(-1)^{k} = 1.
\end{equation*}
We can further transform $g_{m,w}$ to
\begin{equation}\label{eq:gmw_transform}
    g_{m,w}= \frac{f_4(\boldsymbol{0})}{Z} \Big[1 +  \sum_{k=2}^m\binom{m}{k}\frac{(-1)^{k}}{Z^{k-1}}\Big(Z^{k-1} - \sum_{w^{(1)},\cdots,w^{(k-1)}} f_5 \Big(\sum_{i=1}^{k-1} \Vb_{w^{(i)}}\Big)\Big)\Big].
\end{equation}
Here to make sure the transformation of $\Vb$ still preserves the low-rank structure, we take $m = 3$. 
Combine (\ref{ineq:logpw_gmw}) and (\ref{eq:gmw_transform}), we obtain
\begin{equation*}
        \Big|\log p_w - \log\frac{f_4(\boldsymbol{0})}{Z}- \log\Big(1 + 3\frac{Z - \sum_{w'}f_5(\Vb_{w'})}{Z} + \frac{\sum_{w',w''}f_5(\Vb_{w'}+\Vb_{w''})-Z^2}{Z^2} \Big) \Big| \leq C \varepsilon_d^3
\end{equation*}
for sufficiently small $\kappa$. Noting that $\Vb^{\top}\bp = 0$, we have
\begin{equation}\label{ineq:upper_bound_for_Vq}
    \|\Vb^{\top}\bq\| = \|\Vb^{\top}(\bq - \bp)\| \leq \|\Vb\|\cdot \|\bq - \bp\| \leq \|\Vb\|\cdot \sqrt{d}\|\bq - \bp\|_{\max} \leq \frac{\kappa}{\sqrt{\alpha_p}}\sqrt{d}\frac{\kappa^2 p^{3/2}\log^2d}{d^2}
\end{equation}
by \eqref{ineq:dist_pw_qw}. Applying Taylor's expansion theorem, we have
\begin{equation}\label{term1_expansion}
    \begin{split}
        \frac{Z - \sum_{w'}f_5(\Vb_{w'})}{Z} &= -\frac{\sum_{w'}\exp(\|\Vb_{w'}\|^2/2p)\big(\exp( \Vb_{w}^{\top}\Vb_{w'}/p)-1 \big)}{Z}\\
        &= -\sum_{w'}\frac{\exp(\|\Vb_{w'}\|^2/2p)}{Z}\Big[\frac{\Vb_w^{\top}\Vb_{w'}}{p} + O(\kappa^4 p^2/d^2)\Big] \\
        &= -\frac{\Vb_{w}^{\top}\overline{\bv}}{p}+ O(\kappa^4p^2/d^2) = O(\frac{\kappa^4p^2\log^2 d}{d^2}),
    \end{split}
\end{equation}
where the second equation uses $(\Vb_{w}^{\top}\Vb_{w'}/p)^2 = O(\|\Vb\|_{2,\infty}^4/p^2) = O(\kappa^4p^2/d^2)$, which is from Assumption \ref{assump1}; $\overline\bv = \Vb^{\top}\bq$ and 
\begin{equation}\label{ineq:upper_bound_for_Vwbarv}
    |\Vb_{w}^{\top}\overline{\bv}| \leq \|\Vb_w\|\cdot \|\Vb^{\top}\bq\| \leq \frac{\kappa}{\sqrt{\alpha_p}}\mu_1 \sqrt{\frac{p}{d}}\frac{\kappa^3p^{3/2}\log^2 d}{\sqrt{\alpha_p}d^{3/2}} = O(\frac{\kappa^4 p^3 \log^2d}{d^2})
\end{equation}
according to \eqref{ineq:upper_bound_for_Vq}.
And for the second part, we have
\begin{equation}\label{term2_expansion}
    \begin{split}
        &\frac{\sum_{w',w''}f_5(\Vb_{w'}+\Vb_{w''})-Z^2}{Z^2} \\
        &= \frac{\sum_{w',w''}\exp\left(\frac{\|\Vb_{w'}\|^2+\|\Vb_{w''}\|^2}{2p}\right)\left(\exp\left( \frac{\Vb_{w'}^{\top}\Vb_{w''}+\Vb_{w}^{\top}\left(\Vb_{w'}+\Vb_{w''}\right)}{p}\right)-1 \right)}{Z^2}\\
        &= \sum_{w',w''}\frac{\exp\left(\frac{\|\Vb_{w'}\|^2+\|\Vb_{w''}\|^2}{2p}\right)}{Z^2}\frac{\Vb_{w'}^{\top}\Vb_{w''}+\Vb_{w}^{\top}\left(\Vb_{w'}+\Vb_{w''}\right)}{p} + O(\kappa^4p^2/d^2)\\
        &= \frac{\overline{\bv}^{\top}\overline{\bv}+2\Vb_{w}^{\top}\overline{\bv}}{p}+ O(\kappa^4p^2/d^2) = O(\frac{\kappa^4p^2\log^2 d}{d^2}),
    \end{split}
\end{equation}
where the second equation uses $|\Vb_{w'}^{\top}\Vb_{w''}+\Vb_{w}^{\top}\left(\Vb_{w'}+\Vb_{w''}\right)|/p \leq 3\|\Vb\|_{2,\infty}^2/p \lesssim \kappa^2 p/d$.
Then by Taylor's expansion theorem, we have
\begin{equation}\label{ineq:orginal_dist_logpw_logqw}
        \left|\log p_w -  \left(\frac{\|\Vb_w\|^2}{2p} - \log Z \right) \right| \leq 4\epsilon_d^3 + O(\frac{\kappa^4p^2\log^2 d}{d^2}) \leq O(\frac{\kappa^4p^2\log^2 d}{d^2}),
\end{equation}
 where the first inequality uses \eqref{term1_expansion}, \eqref{term2_expansion}, and the second inequality is from Assumption \ref{assump:eigen_bound}. Noting that $\log q_w = \|\Vb_w\|^2/(2p) - \log Z$, equivalently, it can be expressed as for each $w \in [d]$,
\[
|\log p_w - \log q_w| = O(\frac{\kappa^4p^2\log^2 d}{d^2}).
\]
Reversely we have $|p_w - q_w| = q_w |\exp(\log p_w - \log q_w) - 1| = O(\kappa^4p^2\log^2(d)/d^3)$. It further yields that $\|\bp - \bq\|_{\max} = O(\kappa^4p^2\log^2(d)/d^3)$, which provides a better rate than \eqref{ineq:dist_pw_qw}. Thus \eqref{ineq:upper_bound_for_Vq} can be updated to
\begin{equation*}
    \|\Vb^{\top}\bq\| \leq \|\Vb\|\cdot \sqrt{d}\|\bq - \bp\|_{\max} = O(\frac{\kappa^5p^2\log^2 d}{\sqrt{\alpha_p}d^{5/2}}).
\end{equation*}
Consequently, \eqref{ineq:upper_bound_for_Vwbarv} can be updated to
\begin{equation}\label{ineq:revised_upper_bound_for_Vwvbar}
    |\Vb_{w}^{\top}\overline{\bv}| \leq \|\Vb_w\|\cdot \|\Vb^{\top}\bq\| \lesssim \frac{\kappa}{\sqrt{\alpha_p}}\mu_1 \sqrt{\frac{p}{d}}\frac{\kappa^5p^{2}\log^2 d}{\sqrt{\alpha_p}d^{5/2}} = o(\frac{\kappa^4 p^2}{d^2})
\end{equation}
for any $w\in [d]$. Here the last equation is from Assumption \ref{assump:ndTp} and \ref{assump:eigen_bound}. Finally, \eqref{ineq:orginal_dist_logpw_logqw} can be updated to
\[
\left|\log p_w -  \left(\frac{\|\Vb_w\|^2}{2p} - \log Z \right) \right| \leq 4\epsilon_d^3 + O(\kappa^4p^2/d^2) \leq C_1\frac{\kappa^4p^2}{d^2}
\]
for some universal constant $C_1>0$.
\end{proof}

\begin{lemma}\label{lem:B36}
Suppose Assumptions \ref{assump1} and \ref{assump:eigen_bound} hold. Then we have 
\begin{equation*}
    \begin{split}
        \max_{w,w'} &\Big|\log \big(p_{w, w^{\prime}}^{(u)}\big) - \Big(\frac{\|\Vb_w\|^2 + \|\Vb_{w'}\|^2 + 2\alpha^{u/2}\Vb_w^{\top}\Vb_{w'}}{2p} - 2\log Z \Big) \Big| \leq C_2 \frac{\kappa^4 p^2}{d^2};
\end{split}
\end{equation*}
\begin{equation*}
    \begin{split}
        \max_{w,w'} &\Big|\log \big(\sum_{u=1}^q p_{w, w^{\prime}}^{(u)}/q \big) - \Big(\frac{\|\Vb_w\|^2 + \|\Vb_{w'}\|^2 + 2p\alpha_p\Vb_w^{\top}\Vb_{w'}}{2p} - 2\log Z\Big) \Big| \leq C_2 \frac{\kappa^4 p^2}{d^2}.
\end{split}
\end{equation*}
for some constant $C_2 > 0$ and sufficiently large $d,p$.
\end{lemma}
\begin{proof}
Denote functions $f_i: \mathbb{R}^{p}\times\mathbb{R}^p \rightarrow \mathbb{R}, i=1,2,3$ as
\begin{equation*}
    \begin{split}
        f_1(\bX,\Yb) &= \exp\Big( \frac{\| \Vb_w + \Xb\|^2 + \| \Vb_{w'} + \Yb\|^2 + 2\alpha^{u/2}\left(\Vb_w + \bX \right)^{\top}\left(\Vb_{w'} +\Yb\right)}{2p} \Big);\\
        f_2(\bX,\Yb) &= f_1(\bX,\Yb)/f_1(\boldsymbol{0},\boldsymbol{0})\\
        &= \exp\Big( \frac{\| \Xb\|^2 + \| \Yb\|^2 + 2\Vb_{w}^{\top}\Xb + 2\Vb_{w'}^{\top}\Yb + 2\alpha^{u/2}\left(\Vb_w^{\top}\Yb+\Xb^{\top}\Vb_{w'}+\bX^{\top}\Yb \right)}{2p} \Big)\\
f_3(\bX,\Yb) &= \exp\Big( \frac{ 2\Vb_{w}^{\top}\Xb + 2\Vb_{w'}^{\top}\Yb + 2\alpha^{u/2}\left(\Vb_w^{\top}\Yb+\Xb^{\top}\Vb_{w'}+\bX^{\top}\Yb \right)}{2p} \Big).        
    \end{split}
\end{equation*}
Denote $A_w = \exp(\Vb_w^{\top}\bc), A_{w'}^{'} = \exp(\Vb_{w'}^{\top}\bc^{\prime})$. Then we have
\begin{equation*}
    \begin{split}
        \E[A_w A_{w'}^{'}] &= \mathbb{E}\Big[\exp\left(\left(\Vb_w + \alpha^{u/2}\Vb_{w'} \right)^{\top}\bc + \sqrt{1-\alpha^u}\Vb_{w'}^{\top}\br \right)\Big]\\
        &= \exp\Big(\frac{\|\Vb_w\|^2 + \|\Vb_{w'}\|^2 + 2\alpha^{u/2}\Vb_w^{\top}\Vb_{w'}}{2p} \Big) = f_1(\boldsymbol{0},\boldsymbol{0});
    \end{split}
\end{equation*}
and
\begin{equation*}
    \begin{split}
        &\E[A_w A_{w'}^{'}Z(\bc)Z(\bc^{\prime})] = \sum_{w'',w'''} f_1(\Vb_{w''},\Vb_{w'''});
    \end{split}
\end{equation*}
\begin{equation*}
    \begin{split}
        \E[A_w A_{w'}^{'}Z^2(\bc)Z^2(\bc^{\prime})]= \sum_{w^{''}_1,w^{''}_2,w^{'''}_1,w^{'''}_2} f_1(\Vb_{w_1^{''}}+\Vb_{w_2^{''}}, \Vb_{w_1^{'''}}+\Vb_{w_2^{'''}}).
    \end{split}
\end{equation*}
By Lemma \ref{lem:concen_of_Z}, we know $\mathbb{P}(F )=1-\exp \left(-\Omega\left(\log^{2} d\right)\right)$, where 
$$
F=\left\{\left|\frac{Z( \boldsymbol{c})}{Z}-1\right|<\epsilon_d,\left|\frac{Z\left( \boldsymbol{c}^{\prime}\right)}{Z}-1\right|<\epsilon_d\right\}.
$$
Here $\epsilon_d = \frac{\kappa^2 p^{3/2} \log^2 d}{d}.$ Similar to (\ref{eq:Z_decom}) in Lemma \ref{lem:B37}, we have the following decomposition:
\begin{equation*}
   \frac{1}{Z(\bc)Z(\bc^{\prime})} = \sum_{k=1}^m\frac{(Z^2-Z(\bc)Z(\bc^{\prime}))^{k-1}}{Z^{2k}} + \frac{(Z^2-Z(\bc)Z(\bc^{\prime}))^m}{Z(\bc)Z(\bc^{\prime})Z^{2m}}.
\end{equation*}
To simplify the expression, we denote 
\begin{equation*}
    g_{m,w,w'} = \sum_{k=1}^m\frac{\E \big[ A_w A_{w'}^{'} (Z^2-Z(\bc)Z(\bc^{\prime}))^{k-1} \big]}{Z^{2k}};
    b_{m,w,w'} = \E \Big[\frac{A_w A_{w'}^{'}(Z^2-Z(\bc)Z(\bc^{\prime}))^m}{Z(\bc)Z(\bc^{\prime})Z^{2m}} \Big], m \geq 1.
\end{equation*}
Then we have $p_{w,w'}^{(u)} = g_{m,w,w'}+b_{m,w,w'}$ for any integer $m > 0$. When $m = 2k+1, k \geq 0$, following the same procedure in Lemma \ref{lem:B37}, we have
\begin{equation*}
    -\exp(-\Omega(\log^2 d)) \leq b_{m,w,w'} \leq \frac{g_{1,w,w'}(3\epsilon_d)^{m}}{(1-\epsilon_d)^2} + \exp(-\Omega(\log^2 d)).
\end{equation*}
Then it yields that
\begin{equation}\label{ineq:p_ww'u_two_side_bound}
    |p_{w,w'}^{(u)} - g_{m,w,w'}| \leq 2g_{1,w,w'}(3\epsilon_d)^{m} + \exp(-\Omega(\log^2 d)).
\end{equation}
We can see that both $p_{w,w'}^{(u)}$ and $g_{m,w,w'}$ are of order $1/d^2$ with similar analysis in Lemma \ref{lem:concen_of_Z}. Denote $\varepsilon_{w,w'}^m = (3\epsilon_d)^m + \exp(-\Omega(\log^2 d))$. Then we have
\begin{equation}\label{ineq:logpww_gmww}
        |\log p_{w,w'}^{(u)} - \log g_{m,w,w'}| = \Big|\log \Big(1 + \frac{p_{w,w'}^{(u)} - g_{m,w,w'}}{g_{m,w,w'}}\Big)\Big| \leq C\varepsilon_{w,w'}^{m}. 
\end{equation}
Next we calculate $g_{m,w,w'}$. With the identity
\begin{equation*}
    \sum_{k=1}^m\frac{(Z^2-Z(\bc)Z(\bc^{\prime}))^{k-1}}{Z^{2k}} = \sum_{k=1}^m \binom{m}{k}(-1)^{k-1}\frac{(Z(\bc)Z(\bc^{\prime}))^{k-1}}{Z^{2k}},
\end{equation*}
we have
\begin{equation*}
    \begin{split}
        g_{m,w,w'} &= \sum_{k=1}^m \binom{m}{k}\frac{(-1)^{k-1}}{Z^{2k}}\E\big[A_wA_{w'} (Z(\bc)Z(\bc^{\prime}))^{k-1} \big]\\
        &=\sum_{k=1}^m \binom{m}{k}\frac{(-1)^{k-1}}{Z^{2k}}\sum_{\substack{w^{(1)}_1,\cdots,w^{(k-1)}_1\\ w^{(1)}_2,\cdots,w^{(k-1)}_2}}f_1\Big(\sum_{i=1}^{k-1} \Vb_{w^{(i)}_1},\sum_{j=1}^{k-1} \Vb_{w^{(j)}_2}\Big)\\
        &=\frac{f_1(\boldsymbol{0},\boldsymbol{0})}{Z^2} \Big[ \sum_{k=1}^m\binom{m}{k}\frac{(-1)^{k-1}}{Z^{2(k-1)}}\sum_{\substack{w^{(1)}_1,\cdots,w^{(k-1)}_1\\ w^{(1)}_2,\cdots,w^{(k-1)}_2}} f_2 \Big(\sum_{i=1}^{k-1} \Vb_{w^{(i)}_1}, \sum_{j=1}^{k-1} \Vb_{w^{(j)}_2}\Big)\Big].
    \end{split}
\end{equation*}
Note that $f_2(\boldsymbol{0},\boldsymbol{0}) = 1$, and
\begin{equation*}
    \sum_{k=1}^m\binom{m}{k}\frac{(-1)^{k-1}}{Z^{2(k-1)}}Z^{2(k-1)} = \sum_{k=1}^m\binom{m}{k}(-1)^{k-1} = 1 -\sum_{k=0}^m\binom{m}{k}(-1)^{k} = 1.
\end{equation*}
We can further equivalently write $g_{m,w,w'}$ as
\begin{equation*}
    g_{m,w,w'} = \frac{f_1(\boldsymbol{0},\boldsymbol{0})}{Z^2} \Big[1 + \sum_{k=2}^m\binom{m}{k}\frac{(-1)^{k}}{Z^{2(k-1)}}\Big(Z^{2(k-1)}- \sum_{\substack{w^{(1)}_1,\cdots,w^{(k-1)}_1\\ w^{(1)}_2,\cdots,w^{(k-1)}_2}} f_2 \Big(\sum_{i=1}^{k-1} \Vb_{w^{(i)}_1}, \sum_{j=1}^{k-1} \Vb_{w^{(j)}_2}\Big)  \Big) \Big].
\end{equation*}
To derive a low-rank approximation for $\PMIbb$, we set $m = 3$. Under this setting, the inequality \eqref{ineq:logpww_gmww} can be equivalently expressed as
\begin{equation}\label{eq:pww'_leading_term}
    \begin{split}
       &\left|\log\big(p_{w,w'}^{(u)}\big) - \log \frac{f_1(\boldsymbol{0},\boldsymbol{0})}{Z^2} -\right.\\ 
       &\log\Big(3 - 3\frac{\sum_{w'',w'''} f_2(\Vb_{w''},\Vb_{w'''})}{Z^2} + \left.\frac{\sum_{w^{''}_1,w^{''}_2,w^{'''}_1,w^{'''}_2}f_2(\Vb_{w_1^{''}}+\Vb_{w_2^{''}}, \Vb_{w_1^{'''}}+\Vb_{w_2^{'''}})}{Z^4} \Big) \right| \leq C \varepsilon_{w,w'}^3.
    \end{split}
\end{equation}
Note that $Z^2 = \sum_{w'',w'''}\exp\left(\frac{\|\Vb_{w''}\|^2 + \|\Vb_{w'''}\|^2 }{2p}\right)$. Applying Taylor's expansion, we have
\begin{equation}\label{eq:taylor_1}
    \begin{split}
        &\frac{Z^2 - \sum_{w'',w'''} f_2(\Vb_{w''},\Vb_{w'''}) }{Z^2} = -\frac{\sum_{w'',w'''} \exp\big(\frac{\|\Vb_{w''}\|^2 + \|\Vb_{w'''}\|^2}{2p}\big)\left( f_3(\Vb_{w''},\Vb_{w'''}) - 1\right) }{Z^2}\\
        &= -\sum_{w'',w'''}\frac{\exp\left(\frac{\|\Vb_{w''}\|^2 + \|\Vb_{w'''}\|^2}{2p}\right)}{Z^2}\cdot \frac{\Vb_{w}^{\top}\Vb_{w''}+ \Vb_{w'}^{\top}\Vb_{w'''}  +\alpha^{u/2}S_{w,w'}(w'',w''')}{p} + O(\frac{\|\Vb\|_{2,\infty}^4}{p^2} )\\
        &= -\frac{1+\alpha^{u/2}}{p} \left(\Vb_w + \Vb_{w'}\right)^{\top}\overline{\bv} - \frac{\alpha^{u/2}}{p}\overline{\bv}^{\top}\overline{\bv} + O\big(\frac{\kappa^4p^2}{d^2}\big) = O\big(\frac{\kappa^4p^2}{d^2}\big),
    \end{split}
\end{equation}
where the last equation uses $\|\overline{\bv}\| = o(\kappa^4p^2/(d^2\|\Vb\|_{2,\infty}))$ according to \eqref{ineq:revised_upper_bound_for_Vwvbar}, and $S_{w,w'}(w'',w''') = \Vb_w^{\top}\Vb_{w'''}+\Vb_{w'}^{\top}\Vb_{w''} + \Vb_{w''}^{\top}\Vb_{w'''}$.
Denote
\begin{equation*}
    \begin{split}
       A_w^{(4)} &= \|\Vb_{w^{''}_1}\|^2+\|\Vb_{w^{''}_2}\|^2+\|\Vb_{w^{'''}_1}\|^2+\|\Vb_{w^{'''}_2}\|^2 ;\\
       B_w^{(4)} &= \langle\Vb_{w^{''}_1}, \Vb_{w^{''}_2}\rangle + \langle\Vb_{w^{'''}_1}, \Vb_{w^{'''}_2}\rangle.
    \end{split}
\end{equation*}
Then another term in \eqref{eq:pww'_leading_term} can be written as
\begin{equation}\label{eq:taylor_2}
    \begin{split}
        &\frac{1}{Z^4}\Big(\sum_{w^{''}_1,w^{''}_2,w^{'''}_1,w^{'''}_2}f_2(\Vb_{w_1^{''}}+\Vb_{w_2^{''}}, \Vb_{w_1^{'''}}+\Vb_{w_2^{'''}})-Z^4\Big)\\
        & = \sum_{w^{''}_1,w^{''}_2,w^{'''}_1,w^{'''}_2} \frac{\exp\big(A_w^{(4)}/2p\big)}{Z^4}\cdot\Big( \exp\big(\frac{B_w^{(4)}}{p} \big)f_3(\Vb_{w_1^{''}}+\Vb_{w_2^{''}}, \Vb_{w_1^{'''}}+\Vb_{w_2^{'''}}) - 1\Big).
    \end{split}
\end{equation}
Applying Taylor's expansion to equation \eqref{eq:taylor_2}, we obtain:
\begin{equation*}
 \begin{split}
     &\frac{1}{Z^4}\Big(\sum_{w^{''}_1,w^{''}_2,w^{'''}_1,w^{'''}_2}f_2(\Vb_{w_1^{''}}+\Vb_{w_2^{''}}, \Vb_{w_1^{'''}}+\Vb_{w_2^{'''}})-Z^4\Big)\\
    &= \sum_{w^{''}_1,w^{''}_2,w^{'''}_1,w^{'''}_2} \frac{\exp\big(A_w^{(4)}/2p\big)}{Z^4}\cdot \Big(\frac{B_w^{(4)}}{p} + \log f_3(\Vb_{w_1^{''}}+\Vb_{w_2^{''}}, \Vb_{w_1^{'''}}+\Vb_{w_2^{'''}}) \Big) + O(\|\Vb\|_{2,\infty}^4/p^2)\\
        &= \frac{(2+4\alpha^{u/2})\overline{\bv}^{\top}\overline{\bv} + 2(1+\alpha^{u/2})(\Vb_w + \Vb_{w'})^{\top}\overline{\bv}}{p}+O(\kappa^4p^2/d^2)= O(\kappa^4p^2/d^2).
 \end{split}
\end{equation*}
Combining results from (\ref{eq:pww'_leading_term}),(\ref{eq:taylor_1}) and (\ref{eq:taylor_2}), we have
\begin{equation}\label{ineq:bound_pwwu}
    \begin{split}
        &\left|\log \left(p_{w, w^{\prime}}^{(u)}\right) - \left(\frac{\|\Vb_w\|^2 + \|\Vb_{w'}\|^2 + 2\alpha^{u/2}\Vb_w^{\top}\Vb_{w'}}{2p} - 2\log Z\right)        \right| = O(\varepsilon_{w,w'}^3 + \frac{\kappa^4p^2}{d^2}) \leq C_2 \frac{\kappa^4 p^2}{d^2}
\end{split}
\end{equation}
for any $(w,w')\in [d]\times [d]$. Here the last inequality uses the definition of $\varepsilon_{w,w'}$ and Assumption \ref{assump:eigen_bound}, and $C_2 > 0$ is some universal constant. Further, applying \eqref{ineq:p_ww'u_two_side_bound} for $u \in [q]$, we have
\begin{equation*}
   |\sum_{u=1}^q p_{w,w'}^{(u)}/q - \sum_{u=1}^q g_{3,w,w'}^{(u)}/q| \leq 2g_{1,w,w'}(3\epsilon_d)^{3} + \exp(-\Omega(\log^2 d)),
\end{equation*}
where $g_{3,w,w'}^{(u)} = \sum_{k=1}^3 \E[A_w A_{w'}'(Z^2 - Z(\bc)Z(\bc'))^{k-1}]/Z^{2k}$ and $\bc' \stackrel{d}{=} \alpha^{u/2}\bc + \sqrt{1 - \alpha^{u}}\br$.
Replicating the same analysis above, we have
\begin{equation*}
    \begin{split}
        &\Big|\log \big(\sum_{u=1}^q p_{w, w^{\prime}}^{(u)}/q \big) - \Big(\frac{\|\Vb_w\|^2 + \|\Vb_{w'}\|^2 + 2p\alpha_p\Vb_w^{\top}\Vb_{w'}}{2p} - 2\log Z\Big)        \Big| \leq C_2 \frac{\kappa^4 p^2}{d^2}
\end{split}
\end{equation*}
for any $(w,w')\in [d]\times [d]$.
\end{proof}
\begin{remark}
    We do not use $u \leq q$ for this proof. Thus \eqref{ineq:bound_pwwu} holds for any positive integer $u$.
\end{remark}

\subsubsection{Proof of Theorem \ref{thm0:PMI_lowrank}}
 \label{proof of 3.4}
By Lemma \ref{lem:B36} and Lemma \ref{lem:B37}, we have
\begin{equation}\label{eq:pmi_u_concen}
       \max_{w,w'} \Big|\log \frac{\sum_{u=1}^q p_{w,w'}^{(u)}/q}{p_w p_{w'}} - \alpha_p \Vb_w^{\top}\Vb_{w'}  \Big|  
        \leq C \frac{\kappa^4p^2}{d^2},
\end{equation}
where $C = \max\{C_1,C_2\}$. According to \eqref{def:PMI_detailed}, the definition of population PMI is
\begin{equation*}
        \PMIbb\left(w, w^{\prime}\right) =\log \frac{N^{[q]} N_{w, w^{\prime}}^{[q]}}{N_{w}^{[q]} N_{w^{\prime}}^{[q]}} = \log \frac{\sum_{u=1}^q p_{w,w'}^{(u)} }{q p_w p_{w'}},
\end{equation*}
Thus (\ref{eq:pmi_u_concen}) becomes
$$
\max_{w,w'}\left|\PMIbb (w,w') -  \alpha_p\Vb_w^{\top}\Vb_{w'} \right| \leq C\frac{\kappa^4p^2}{d^2},
$$
or equivalently,
\begin{equation*}
        \|\PMIbb - \alpha_p\Vb\Vb^{\top}\|_{\max} \leq C\frac{\kappa^4p^2}{d^2}
\end{equation*}
for sufficiently large $d,p$. By Weyl's inequality, we have
\begin{equation*}
    \lambda_{p}(\PMIbb) \geq \lambda_{p}( \alpha_p\Vb\Vb^{\top}) - \|\PMIbb - \alpha_p\Vb\Vb^{\top}\| \geq \kappa^2\xi^2 - C \frac{\kappa^4 p^{2} }{d} \geq \kappa^2\xi^2/2
\end{equation*}
for sufficiently large $d$. Here the last inequality comes from Assumption \ref{assump:eigen_bound}. For the largest eigenvalue and the $(p+k)$-th largest eigenvalue of $\PMIbb$, we have
\begin{equation*}
    \lambda_1(\PMIbb) \leq \lambda_1(\alpha_p\Vb\Vb^{\top})+ \|\PMIbb -  \alpha_p\Vb\Vb^{\top}\| \leq \kappa^2 + C \frac{\kappa^4 p^{2} }{d} \leq 2\kappa^2;
\end{equation*}
\begin{equation*}
    |\lambda_{p+k}(\PMIbb)| \leq |\lambda_{p+k}( \alpha_p\Vb\Vb^{\top})| + \|\PMIbb -  \alpha_p\Vb\Vb^{\top}\| \leq C \frac{\kappa^4 p^{2} }{d}, \quad 1 \leq k \leq d-p,
\end{equation*}
for sufficiently large $d$.

\subsection{Consistency of PMI Estimators}
This subsection extends results in \cite{xxx2021} to the setting of $n$ independent and identically distributed patients.

\begin{lemma}\label{lem:Sw_to_Nw}
Suppose Assumptions~\ref{assump:ndTp}-\ref{assump:eigen_bound} hold. Then we have
$$
\mathbb{P}\Big(\max_w \Big|\frac{\sum_{i=1}^{n}
S_{i,w}}{n N_{w}}-1\Big| \geq (nT)^{-1/2}dp \log^2 d \Big)=\exp(-\Omega(\log ^{2}d)).
$$
\end{lemma}
\begin{proof}
According to Lemma \ref{lem:rate_chernoff_method}, we have \zx{
\begin{equation}\label{ineq:Sw_Nw}
        \mathbb{P}\Big(\Big|\sum_{i=1}^{n} \sum_{t=1}^{T_i} p_{i,w}(t)-n T p_{w}\Big| \geq n T \epsilon \Big) \leq 2\exp(-\frac{\lambda nT\varepsilon^2}{4mM_1})
\end{equation}
for $\varepsilon \geq 0$. Here $\lambda = 1-2/e, M_1>0$ is some constant} and $ m = 4p^2\log^2d$. By Lemma \ref{lem:B37}, there exists a constant $c > 0$ such that $p_w \geq c/d, \forall w \in [d]$. Setting $\varepsilon = p_w (nT)^{-1/2} d p\log^2 d$, we obtain
$$
\mathbb{P}\Big(\Big|\frac{1}{n T p_{w}} \sum_{i=1}^{n}  \zx{\sum_{t=1}^{T_i}}  p_{i,w}(t)-1\Big| \geq (nT)^{-1/2}dp \log^2 d \Big) = \exp(-\Omega(\log^2 d))
$$
for sufficiently large $n$ and $T$. Applying the union bound, we conclude that
\begin{equation*}
    \mathbb{P}\Big(\max_w \Big|\frac{\sum_{i=1}^{n}
S_{i,w}}{n N_{w}}-1\Big| \geq (nT)^{-1/2}d p\log^2 d \Big) \leq d \exp(-\Omega(\log^2 d) ) = \exp\left(-\Omega(\log^2 d) \right).
\end{equation*}
\end{proof}

\begin{lemma}\label{lem:Sww'_to_Nww'}
Suppose Assumptions~\ref{assump:ndTp}-\ref{assump:eigen_bound} hold, for a given window size $u \in [q]$, it holds that\zx{
$$
\mathbb{P}\Big(\max_{w,w'}\Big|\frac{\sum_{i=1}^{n}S_{i,w, w^{\prime}}^{(u)}}{n N_{w, w^{\prime}}^{(u)}}-1\Big| \geq (nT)^{-1/2}d^2p \log^2 d \Big)=\exp (-\Omega(\log ^{2}d)),
$$
and
\begin{equation*}
\mathbb{P}\Big(\max_{w,w'}\Big|\frac{\sum_{i=1}^{n}S_{i,w, w^{\prime}}^{[q]}}{n N_{w, w^{\prime}}^{[q]}}-1\Big| \geq (nT)^{-1/2}d^2 p\log^2 d \Big)=\exp (-\Omega(\log ^{2}d)).
\end{equation*}}
\end{lemma}
\begin{proof}
We first separate $S_{i,w,w'}^{(u)}$ into $u+1$ subsequences and then follow the proof procedure in Lemma~\ref{lem:Sw_to_Nw}. Note that
\begin{equation*}
    S_{i,w,w'}^{(u)} = \zx{\sum_{t=1}^{T_i-q}}  p_{i,w,w'}(t,t+u) = \sum_{k=1}^{u+1}\sum_{l=0}^{\lfloor \zx{\frac{T_i-k-q}{u+1}} \rfloor}p_{i,w,w'}(k+l(u+1), k+l(u+1) + u).
\end{equation*}
Denote $\zx{T_{i,k}^{\star} = \lfloor \frac{T_i-k-q}{u+1} \rfloor + 1}$. We have for any \zx{$i,k,u$},
\begin{equation}\label{ineq:Tk_lower_bound}
        \zx{T_{i,k}^{\star} \geq \frac{T_i-k-q}{u+1} \geq \frac{T_i-2q-1}{q+1} \geq \frac{T_i}{q+2}}
\end{equation}
for $T_i \geq (q+2)(2q+1)$. Note that for any given $k \in [u+1],$ $\{(\bc_{i,k+l(u+1)},\bc_{i,k+l(u+1)+u})\}_{l=0}^{T_{i,k}^{\star}-1}$ is a stationary Markov chain. Denote $p_{i,w,w'}^{(k,u)}(l) = p_{i,w,w'}(k+l(u+1), k+l(u+1) + u).$ According to Lemma \ref{lem:rate_chernoff_method}, we have that for each $k$ and $\epsilon \geq 4m/\lambda T$,
\begin{equation}\label{ineq:Sww'_Nww'}
        \mathbb{P}\Big(\Big|\sum_{i=1}^{n}  \zx{\sum_{l=0}^{T_{i,k}^{\star}-1} p_{i,w,w'}^{(k,u)}(l)-n T_k^{\star} p_{w,w'}^{(u)}\Big| \geq n T_k^{\star} \varepsilon \Big) \leq 2\exp \big(-\frac{\lambda nT_k^{\star}\varepsilon^2}{4M_1m} \big) ,}
\end{equation}
where \zx{$T_k^{\star}=\sum_i T_{i,k}^{\star}/n, \lambda = 1-2/e, M_1>0 $ is some constant} and $ m = 4p^2\log^2d$. Let $\varepsilon = p_{w,w'}^{(u)}(nT)^{-1/2}d^2p\log^2 d$. By Lemma \ref{lem:B36}, there exists a constant $c > 0$ such that $p_{w,w'}^{(u)}\geq c/d^2, \forall (w,w') \in [d]\times[d]$.
Then for sufficiently large $n$ and $T$, we have
\begin{equation*}
    \begin{split}
        &\P\Big(\Big|\frac{1}{n(T-\zx{q})p_{w,w'}^{(u)}}\sum_{i=1}^n\sum_{t=1}^{T-\zx{q}}p_{i,w,w'}(t,t+u) - 1 \Big| \geq (nT/(q+2))^{-1/2}d^2p\log^2 d\Big) \\
        &\leq \sum_{k=1}^{u+1}\P\Big( \Big|\sum_{i=1}^{n}  \sum_{l=0}^{T_k-1} p_{i,w,w'}^{(k,u)}(l)-n T_k p_{w,w'}^{(u)}\Big| \geq \frac{1}{u+1}(nT/(q+2))^{-1/2}d^2p_{w,w'}^{(u)}p\log^2 d \Big) \\
        &\leq 2(u+1)\exp\Big(-\frac{1}{(u+1)^2(q+2)}\Omega(\log^2 d)\Big) = \exp(-\Omega(\log^2 d)),
    \end{split}
\end{equation*}
where the first inequality uses the union bound, the second inequality uses \eqref{ineq:Sww'_Nww'} and \eqref{ineq:Tk_lower_bound}, and the last equation is from Assumption \ref{assump:ndTp}. Denote $\epsilon_0 = (nT)^{-1/2}d^2p \log^2 d$. Then by the union bound, we have
\begin{equation*}
\mathbb{P}\Big(\max_{1\leq u \leq q}\Big|\frac{\sum_{i=1}^{n}S_{i,w, w^{\prime}}^{(u)}}{n N_{w, w^{\prime}}^{(u)}}-1\Big| \geq \epsilon_0 \Big) \leq q \exp(-\Omega(\log^2 d)) = \exp(-\Omega(\log^2 d) ).
\end{equation*}
Recall the definition of $S_{i,w,w'}^{[q]}$ and $N_{w,w'}^{[q]}$, we have
\begin{equation*}
\mathbb{P}\Big(\Big|\frac{\sum_{i=1}^{n}S_{i,w, w^{\prime}}^{[q]}}{n N_{w, w^{\prime}}^{[q]}}-1\Big| \geq \epsilon_0 \Big) \leq \mathbb{P}\Big(\max_{1\leq u \leq q}\Big|\frac{\sum_{i=1}^{n}S_{i,w, w^{\prime}}^{(u)}}{n N_{w, w^{\prime}}^{(u)}}-1\Big| \geq \epsilon_0 \Big) \leq \exp(-\Omega(\log^2 d)).
\end{equation*}
Again by the union bound, we have
\begin{equation*}
\mathbb{P}\Big(\max_{w,w'}\Big|\frac{\sum_{i=1}^{n}S_{i,w, w^{\prime}}^{[q]}}{n N_{w, w^{\prime}}^{[q]}}-1\Big| \geq \epsilon_0 \Big) \leq d^2\exp(-\Omega(\log^2 d)) = \exp(-\Omega(\log^2 d)).
\end{equation*}
\end{proof}

\begin{lemma}\label{lem:X_w_to_cond_exp}
Suppose Assumptions~\ref{assump:ndTp}-\ref{assump:eigen_bound} hold, it holds that
$$
\mathbb{P}\Big(\max _{w}\Big|\frac{\sum_{i=1}^{n}X_{i,w}}{\sum_{i=1}^{n} S_{i,w}}-1\Big| \geq (nT)^{-1/2}\sqrt{d} \log (d) \Big)=\exp (-\Omega\left(\log^{2} d\right)).
$$
\end{lemma}
\begin{proof}
The proof of this lemma is similar to the proof of Lemma G.6 in \cite{xxx2021}. Apply
Lemma G.5 in \cite{xxx2021} to $X_{i,w}(t) \mid\left\{\boldsymbol{c}_{i,t}\right\}$ which are independent $\operatorname{Ber}\left(p_{i,w}(t)\right)$ variables conditional on $\left\{\boldsymbol{c}_{i,t}\right\}_{t>0}$, and recall that $$S_{i,w}=\sum_{t=1}^{T} p_{i,w}(t)=\sum_{t=1}^{T} \mathbb{E}\left[X_{i,w}(t) \mid\left\{\boldsymbol{c}_{i,t}\right\}\right]$$ 
is a function of $\left\{\boldsymbol{c}_{i,t}\right\}_{t>0}$, we have that
\begin{equation*}
    \mathbb{P}\Big(\sum_{i=1}^{n}\sum_{t=1}^{T} X_{i,w}(t) \geq(1+\delta) \sum_{i=1}^{n}S_{i,w} \mid\left\{\bc_{i,t}\right\}_{i>0,t>0}\Big) \leq \exp \Big(-\frac{\delta^{2} \sum_{i=1}^{n} S_{i,w}}{2+\delta}\Big), \delta>0 ;
\end{equation*}
\begin{equation*}
    \mathbb{P}\Big(\sum_{i=1}^{n}\sum_{t=1}^{T} X_{i,w}(t) \leq(1-\delta)\sum_{i=1}^{n} S_{i,w} \mid\left\{\bc_{i,t}\right\}_{i>0,t>0}\Big) \leq \exp \Big(-\frac{\delta^{2} \sum_{i=1}^{n}S_{i,w}}{2}\Big), 0<\delta<1.
\end{equation*}
Denote 
\begin{equation*}
    \cF = \Big\{\{\bc_{i,t}\}_{i>0,t>0} : \Big|\frac{\sum_{i=1}^n S_{i,w}}{nN_w} - 1 \Big| < \epsilon  \Big\},
\end{equation*}
where $\epsilon = (nT)^{-1/2}dp\log^2 d.$ By Lemma \ref{lem:Sw_to_Nw}, $\P(\cF) = 1 - \exp(-\Omega(\log^2 d)).$ By Lemma \ref{lem:B37}, there exists a constant $c > 0$ such that $p_w \geq c/d, \forall w \in [d]$. Therefore we have $N_w \geq cT/d$. By the law of total expectation, we have
\begin{equation*}
    \begin{split}
        \P\Big(\Big|\frac{\sum_{i=1}^n X_{i,w}}{\sum_{i=1}^n S_{i,w}} - 1 \Big| \geq \delta \Big) &\leq \P\Big(\Big|\frac{\sum_{i=1}^n X_{i,w}}{\sum_{i=1}^n S_{i,w}} - 1 \Big| \geq \delta \Big| \cF \Big) + \P(\cF^c)\\
        & \leq 2\exp\Big(\frac{\delta^2 nN_w(1-\epsilon)}{2+\delta} \Big) + \exp (-\Omega\left(\log^{2} d\right)).
    \end{split}
\end{equation*}
Let $\delta = (nT)^{-1/2}\sqrt{d}\log d$. Then we have
\begin{equation*}
    \P\Big(\Big|\frac{\sum_{i=1}^n X_{i,w}}{\sum_{i=1}^n S_{i,w}} - 1 \Big| \geq \delta \Big) \leq 3\exp(-\Omega(\log^2 d)) =  \exp(-\Omega(\log^2 d)).
\end{equation*}
Then by the union bound, we have
\begin{equation*}
    \P\Big(\max_w \Big|\frac{\sum_{i=1}^n X_{i,w}}{\sum_{i=1}^n S_{i,w}} - 1 \Big| \geq \delta \Big) \leq d\exp(-\Omega(\log^2 d)) =  \exp(-\Omega(\log^2 d)).
\end{equation*}
\end{proof}

\begin{lemma}\label{lem:X_ww'_to_cond_exp}
Consider a fixed windows size $q$. Suppose Assumptions~\ref{assump:ndTp}-\ref{assump:eigen_bound} hold. Then for $u=1, \ldots, q$, it holds that
\begin{equation*}
    \mathbb{P}\Big(\max _{w, w^{\prime}}\Big|\frac{\sum_{i=1}^{n}X_{i,w, w^{\prime}}^{(u)}}{\sum_{i=1}^{n}S_{i, w, w^{\prime}}^{(u)}}-1\Big| \geq (nT)^{-1/2}d \log d \Big)=\exp (-\Omega(\log ^{2} d )),
\end{equation*}
\begin{equation*}
    \mathbb{P}\Big(\max _{w, w^{\prime}}\Big|\frac{\sum_{i=1}^{n}X_{i,w, w^{\prime}}^{[q]}}{\sum_{i=1}^{n}S_{i, w, w^{\prime}}^{[q]}}-1\Big| \geq (nT)^{-1/2}d \log d \Big)=\exp (-\Omega(\log ^{2} d)).
\end{equation*}
\end{lemma}
\begin{proof}
The proof is essentially the same as Lemma \ref{lem:X_w_to_cond_exp} after utilizing the separation trick in Lemma \ref{lem:Sww'_to_Nww'}.
\end{proof}

\begin{lemma}\label{lem:rate_chernoff_method}
Suppose Assumption \ref{assump:ndTp} holds. For any given features $w,w'$, let $S_{iT} = \sum_{t=1}^{\zx{T_i}} p_{i,w}(t)$, $S_{iT}^{(u)} = \sum_{t=1}^{\zx{T_i}} p_{i,w,w'}(t(u+1),t(u+1)+u)$. Then \zx{we have
\begin{equation*}
     \mathbb{P} (\sum_{i=1}^n (S_{iT} - \zx{T_i} p_w) \geq nT \varepsilon)  \leq  \exp(-\frac{\lambda nT\varepsilon^2}{4mM_1}),
\end{equation*}
\begin{equation*}
     \mathbb{P} (\sum_{i=1}^n (S_{iT}^{(u)} - \zx{T_i} p_{w,w'}^{(u)}) \geq nT \varepsilon)  \leq \exp(-\frac{\lambda nT\varepsilon^2}{4mM_1}),
\end{equation*}}
where $\lambda = 1-2/e $ and $ m = 4p^2\log^2d$.
\end{lemma}

\begin{proof}
Denote $Z_i := S_{iT} - \zx{T_i} p_w$, then we have the Chernoff bound
\begin{equation}\label{ineq:markov_ineq_1}
    \mathbb{P} (\sum_{i=1}^n Z_i \geq \sum_{i=1}^n T_i \varepsilon | \{\bc_{i,1}\}_{i=1}^n) \leq 
\exp(-\lambda_0 \sum_{i=1}^n T_i \varepsilon)\prod_{i=1}^n\E [\exp(\lambda_0 Z_i)|\bc_{i,1}]
\end{equation}
for any $\lambda_0 \geq 0.$
  By Lemma D.4 and D.6 in \cite{xxx2021}, we have
  \begin{align}\label{ineq: geo_ergodic}
      \|f_{\bc_{i,T_i+1}|\bc_{i,1}}(\cdot)&-\pi(\cdot)\|_{\text{tv}} \leq \big(\frac{2}{e}\big)^{\lfloor T_i/m \rfloor};\\
      \|f_{\bc_{i,T_i+1},\bc_{T_i+u+1}|\bc_{i,1}}(\cdot,\cdot)&-\pi_u(\cdot,\cdot)\|_{\text{tv}} \leq \big(\frac{2}{e}\big)^{\lfloor T_i/m \rfloor}
  \end{align}
for $T_i \geq m$ and \(\|\bc_{i,1}\| \le 2\sqrt{T_i/p^3}\) using Assumption \ref{assump:ndTp}. Here $m = 4p^2\log^2 d$, $\pi(\cdot)$ is the density of the distribution $N(0,\Ib_p/p)$, $\pi_u(\cdot,\cdot)$ is the density of the distribution of $(\bc_{i,t},\bc_{i,t+u})$, $f_{\bc_{i,T+1}|\bc_{i,1}}(\cdot)$ is the density of distribution of $\bc_{i,T+1}$ given $\bc_{i,1}$, i.e. $N(\alpha^{T/2}\bc_1,(1-\alpha^m)\Ib_p/p)$, and $f_{\bc_{i,T+1},\bc_{i,T+u+1}|\bc_{i,1}}(\cdot,\cdot)$ is the density of distribution of $(\bc_{i,T+1},\bc_{i,T+u+1})$ given $\bc_{i,1}$. Thus the Markov chain \(\{\bc_{i,t}\}_{t=1}^{T_i}\) is geometrically ergodic. Below we show the proof for $S_{iT}$, and the proof for $S_{iT}^{(u)}$ is similar to that of $S_{iT}$. Note that $p_w(t)$ is a function of $\bc_t$. Denote $h(\bc_t) = p_w(t)$, and $h_c(\bc_t) = h(\bc_t)-p_w$.\zx{ Then \(Z_i = \sum_{t=1}^{T_i}h_c(\bc_{i,t})\). According to results in \cite{dedecker2015}, we have
\[
\E[\exp(\lambda_0 Z_i)|\bc_{i,1}] \leq \exp(\lambda_0^2 M_1(1-(2/e)^{1/m})^{-1} T_i) \le \exp(m\lambda_0^2 M_1 T_i/\lambda), \quad 1 \le i \le n,
\]
where the second inequality is from Bernoulli's inequality and \(M_1>0\) is some constant independent of \(i\). Combining it with \eqref{ineq:markov_ineq_1}, we have
\[
\mathbb{P} (\sum_{i=1}^n Z_i \geq nT \varepsilon | \{\bc_{i,1}\}_{i=1}^n) \leq 
\exp(-\lambda_0 \sum_{i=1}^n T_i\varepsilon + \frac{mM_1}{\lambda} \sum_{i=1}^n T_i \lambda_0^2).
\]
Let \(\lambda_0 = \varepsilon\lambda/(2mM_1)\), we have
\[
\mathbb{P} (\sum_{i=1}^n Z_i \geq nT \varepsilon | \{\bc_{i,1}\}_{i=1}^n) \leq 
\exp(-\frac{\lambda nT\varepsilon^2}{4mM_1}).
\]
for $\|\bc_{i,1}\| \leq 2\sqrt{\zx{T_i}/p^3}, \forall i \in [n]$. Further applying the results of Theorem 0.2 in the \cite{dedecker2015}, we obtain that for any \(\varepsilon>0\),
\[
\mathbb{P} (\sum_{i=1}^n Z_i \geq nT \varepsilon) \leq 
\exp(-\frac{\lambda nT\varepsilon^2}{4mM_1}).
\]
}
\end{proof}

The following theorem provides the statistical rate of $\hat{\PMIbb}$ for $\PMIbb$.

\begin{theorem}\label{Concentration of empirical PMI to stationary PMI}
Suppose Assumptions~\ref{assump:ndTp}-\ref{assump:eigen_bound} hold. For a fixed window size $q\geq 2$, we have
$$
\|\Eb\|_{\max}  \leq 3\frac{d^2 p\log^2 d}{\sqrt{nT}}, 
\|\Eb\|  \leq 3\frac{d^3 p\log^2 d}{\sqrt{nT}}
$$
with probability at least $1-\exp(-\Omega(\log^2 d) )$ for sufficiently large $n$ and $T$. Here $\Eb = \PMIbbhat - \PMIbb$.
\end{theorem}
\begin{proof}
    Recall the definition of $\PMIbbhat$ and $\PMIbb$, we have the following decomposition:\zx{
\begin{equation}\label{eq:E_decompose}
    \begin{split}
        \mathbf{E}_{w, w^{\prime}}
        &= \PMIbbhat(w, w^{\prime}) - \PMIbb(w, w^{\prime}) \\
        &= \log\frac{\sum_i N_i^{[q]}\sum_{i=1}^n X_{i,w,w'}^{[q]}}{\sum_{i=1}^n X_{i,w}^{[q]}\sum_{i=1}^n X_{i,w'}^{[q]}} - \log\frac{\sum_i N_i^{[q]} \sum_i N_{i,w,w'}^{[q]}}{\sum_i N_{i,w}^{[q]}\sum_i N_{i,w'}^{[q]}} + \log\frac{\sum_i N_i^{[q]} \sum_i N_{i,w,w'}^{[q]}}{\sum_i N_{i,w}^{[q]}\sum_i N_{i,w'}^{[q]}} - \log\frac{\sum_{u=1}^q p_{w,w'}^{(u)}}{q p_w p_{w'}}\\
        &=\log \frac{\sum_{i=1}^n X_{i, w, w^{\prime}}^{[q]}}{\sum_i N_{i,w, w^{\prime}}^{[q]}}-\log \frac{\sum_{i=1}^n X_{i, w}^{[q]}}{\sum_i N_{i,w}^{[q]}}-\log \frac{\sum_{i=1}^n X_{i, w^{\prime}}^{[q]}}{\sum_i N_{i,w^{\prime}}^{[q]}}\\
        &=\log \frac{\sum_{i=1}^n X_{i, w, w^{\prime}}^{[q]}}{n N_{w, w^{\prime}}^{[q]}}-\log \frac{\sum_{i=1}^n X_{i, w}^{[q]}}{nN_{w}^{[q]}}-\log \frac{\sum_{i=1}^n X_{i, w^{\prime}}^{[q]}}{n N_{w^{\prime}}^{[q]}}.
    \end{split}
\end{equation}}
\zx{Denote} $\epsilon = (nT)^{-1/2}dp\log^2 d$. By Lemma~\ref{lem:Sw_to_Nw} and Lemma~\ref{lem:X_w_to_cond_exp}, we have that with probability $1 - \exp(-\Omega(\log^2 d))$,
    \begin{equation}\label{ineq:X_wN_w}
        (1 - \epsilon)(1 - \epsilon/\sqrt{md}) \leq \min_w \frac{\sum_{i=1}^n X_{i,w}}{nN_w} \leq \max_w \frac{\sum_{i=1}^n X_{i,w}}{nN_w} \leq (1 + \epsilon)(1 + \epsilon/\sqrt{md}).
    \end{equation}
    By Lemma~\ref{lem:Sww'_to_Nww'} and Lemma~\ref{lem:X_ww'_to_cond_exp}, we have that with probability $1 - \exp(-\Omega(\log^2 d))$,
    \begin{equation}\label{ineq:X_ww'_qN_ww'}
        (1 - d\epsilon)(1 - \epsilon/\sqrt{m}) \leq \min_{w,w'}\frac{\sum_{i=1}^n X_{i,w,w'}^{[q]}}{nN_{w,w'}^{[q]}} \leq \max_{w,w'}\frac{\sum_{i=1}^n X_{i,w,w'}^{[q]}}{nN_{w,w'}^{[q]}} \leq (1 + d\epsilon)(1 + \epsilon/\sqrt{m}).
    \end{equation}
According to Section B.1 in \cite{xxx2021}, we have
\begin{equation}\label{eq:Xwq_to_Xw}
    \begin{split}
        X_{i,w}^{[q]} &= \sum_{w'}X_{i,w,w'}^{[q]}=\sum_{u=1}^q \sum_{t=1}^{T-\zx{q}}\sum_{w'}(X_{i,w,w'}(t,t+u) + X_{i,w',w}(t,t+u))
        \\
        &= \sum_{u=1}^q \sum_{t=1}^{T-\zx{q}}(X_{i,w}(t)+X_{i,w}(t+u)) \\
        &= \zx{2\sum_{u=1}^q \sum_{t=1}^{T}X_{i,w}(t) - \sum_{u=1}^q \sum_{t=T-q+1}^{T}X_{i,w}(t) - \sum_{u=1}^q \sum_{t=1}^{u}X_{i,w}(t) - \sum_{u=1}^{q-1}\sum_{t=T-q+u+1}^T X_{i,w}(t) }\\
        &= 2q\sum_{t=1}^T X_{i,w}(t) - \delta_i = 2qX_{i,w} - \delta_i,
    \end{split}
\end{equation}
where $\delta_i = \sum_{u=1}^q \sum_{t=T-\zx{q}+1}^{T}X_{i,w}(t) + \sum_{u=1}^q \sum_{t=1}^{u}X_{i,w}(t) \zx{+ \sum_{u=1}^{q-1}\sum_{t=T-q+u+1}^T X_{i,w}(t)} \in [0, \zx{2}q^2]$. Since $q$ is a fixed number, and $X_{i,w}$ is of order $T$, $\delta_i$ will not affect the order of variance and expectation of $X_{i,w}^{[q]}$.
Concretely we have
\begin{equation}\label{ineq:Xw_q_Xw}
    \Big|\frac{\sum_{i=1}^n X_{i,w}^{[q]}}{nN_w^{[q]}} - \frac{\sum_{i=1}^n X_{i,w}}{nN_w}\Big| = \Big|\frac{-\bar\delta N_w + \E[\delta]\bar X_{w}  }{(2q N_w - \E[\delta]) N_w}\Big| = \Big| \frac{(\E[\delta] - \bar\delta)N_w + \E[\delta](\bar X_{w} - N_w)}{(2q N_w - \E[\delta]) N_w}\Big|
\end{equation}
where $\bar\delta = \sum_{i=1}^n\delta_i/n, \text{ and } \bar X_w = \sum_{i=1}^n X_{i,w}/n$. By Lemma \ref{lem:B37}, we know that $p_w$ is of order $1/d$, and $N_w$ is of order $T/d$ accordingly. Combined with (\ref{ineq:X_wN_w}), we have
\begin{equation}\label{ineq:Xw_q_Xw_1}
    \frac{\E[\delta]|\bar X_{w} - N_w|}{(2q N_w - \E[\delta]) N_w} \leq \frac{6\epsilon q^2}{2q N_w - q^2} \lesssim \frac{d^2p \log^2 d}{n^{1/2}T^{3/2}}
\end{equation}
holds with probability $1-\exp(-\Omega(\log^2 d) )$. Similarly we have
\begin{equation}\label{ineq:Xw_q_Xw_2}
    \frac{|\E[\delta] - \bar\delta|N_w}{(2q N_w - \E[\delta]) N_w} \lesssim  \frac{\log d }{\sqrt{n}(2q N_w - \E[\delta])} \lesssim \frac{d\log d}{n^{1/2}T}
\end{equation}
holds with probability $1-\exp(-\Omega(\log^2 d) )$, where the first inequality uses $\P(|\E[\delta]-\bar\delta| > \log(d)/\sqrt{n}) \leq \exp(-\Omega(\log^2 d))$, which is from Hoeffding's inequality. Combine (\ref{ineq:X_wN_w}), (\ref{ineq:Xw_q_Xw}), (\ref{ineq:Xw_q_Xw_1}) and (\ref{ineq:Xw_q_Xw_2}), then we have
\begin{equation*}
    (1 - \epsilon)^2 \leq \min_{w}\frac{\sum_{i=1}^n X_{i,w}^{[q]}}{nN_{w}^{[q]}} \leq \max_{w}\frac{\sum_{i=1}^n X_{i,w}^{[q]}}{nN_{w}^{[q]}} \leq (1 + \epsilon)^2, \forall w \in [d],
\end{equation*}
for sufficiently large $n$ and $T$. Combined with (\ref{ineq:X_ww'_qN_ww'}), we have that with probability $1 - \exp(-\Omega(\log^2 d))$,
\begin{equation*}
      \zx{\|\Eb\|_{\max}} \leq \max\Big\{\log \Big(\frac{(1 + d\epsilon)(1 + \epsilon)}{(1 - \epsilon)^4}\Big), -\log \Big(\frac{(1 - d\epsilon)(1 - \epsilon)}{(1 + \epsilon)^4} \Big) \Big\}.
\end{equation*}
Note that for sufficiently small $\epsilon$, we have
\begin{equation*}
 \frac{(1+d\epsilon)(1+\epsilon)}{(1-\epsilon)^4} \leq 1+2d\epsilon; \text{ and }
    \frac{(1-d\epsilon)(1-\epsilon)}{(1+\epsilon)^4} \geq 1-2d\epsilon.
\end{equation*}
Then by $\log(1+x) \leq x$ and $-\log(1-x) \leq x/(1-x), x > 0$, we have
\begin{equation*}
     \zx{\|\Eb\|_{\max}} \leq \max \{2d\epsilon, \frac{2d\epsilon}{1 - 2d\epsilon} \} \le 3d\epsilon
\end{equation*}
for sufficiently large $n$ and $T$.
It follows that
\begin{equation*}
    \|\hat{\PMIbb} - \PMIbb\| \leq d\|\hat{\PMIbb} - \PMIbb\|_{\max} \leq 3d^2 \epsilon,
\end{equation*}
which completes our proof.
\end{proof}

\begin{section}{Asymptotic Normality of PMI Estimators}

\subsection{Variance Estimator in Section \ref{sec:2.2.1}}\label{sec:proof_of_sec221}
Recall that $\Eb = \PMIbbhat - \PMIbb$ and $\Wb$ is defined as $\Eb$ plus the bias term $\PMIbb - \alpha_p \Vb\Vb^{\top}$. By \eqref{eq:E_decompose},
\begin{equation*}
        \mathbf{E}_{w, w^{\prime}}
=\log \frac{\sum_{i=1}^n X_{i, w, w^{\prime}}^{[q]}}{n N_{w, w^{\prime}}^{[q]}}-\log \frac{\sum_{i=1}^n X_{i, w}^{[q]}}{n N_w^{[q]}}-\log \frac{\sum_{i=1}^n X_{i, w^{\prime}}^{[q]}}{n N_{w^{\prime}}^{[q]}}.
\end{equation*}

We apply the first-order Taylor expansion for each term, and define the expanded error matrix $\Eb_S = [\Eb_S(w,w')] \in \mathbb R^{d\times d}$ as follows:
  $$
\mathbf{E}_S\left(w, w^{\prime}\right)=\frac{\sum_{i=1}^n X_{i, w, w^{\prime}}^{[q]}}{n N_{w, w^{\prime}}^{[q]}}-\frac{\sum_{i=1}^n X_{i, w}^{[q]}}{n N_w^{[q]}}-\frac{\sum_{i=1}^n X_{i, w^{\prime}}^{[q]}}{n N_{w^{\prime}}^{[q]}}+1.
$$
According to Lemma \ref{thm1}, we have
\begin{equation*}
    \Eb_{w, w^{\prime}} = \Eb_S(w,w') + o_p(\frac{d}{\sqrt{nT}}).
\end{equation*}
Then we can estimate the covariance structure of $\Eb$ by estimating the covariance matrix of $\Eb_S$. For the $i$-th patient, denote $\mathbf{E}^{(i)}$ as
$$
\mathbf{E}^{(i)}_{w, w^{\prime}}=\frac{X_{i, w, w^{\prime}}^{[q]}}{N_{w, w^{\prime}}^{[q]}}-\frac{X_{i, w}^{[q]}}{N_w^{[q]}}-\frac{X_{i, w^{\prime}}^{[q]}}{N_{w^{\prime}}^{[q]}}+\zx{\frac{T_i-q}{T-q}}
, \quad 1 \leq i \leq n .
$$
Then $\Eb_S = \sum_{i=1}^n \Eb^{(i)}/n$, and $\sum_i\text{Cov}(\Eb^{(i)}_{w,\cdot})/n^2$ as an asymptotically equivalent replacement of the covariance matrix $\bSigma_w = \text{Cov}(\Eb_{w,\cdot})$ and $\sum_i\text{Cov}(\Eb^{(i)}_{w,\cdot}, \Eb^{(i)}_{w',\cdot})/n^2$ as an asymptotically equivalent replacement of the covariance matrix $\bSigma_{w,w'} = \text{Cov}(\Eb_{w,\cdot},\Eb_{w',\cdot})$.
Given the patient-level co-occurrence matrices, we are able to calculate the co-occurrence matrices $\CC_i, 1\leq i\leq n$ and estimate $\Eb^{(i)}$ by
\begin{equation*}
    \hat \Eb^{(i)}_{w,w'} =\frac{n X_{i, w, w^{\prime}}^{[q]}}{\sum_{j=1}^n X_{j, w, w^{\prime}}^{[q]}}-\frac{n X_{i, w}^{[q]}}{\sum_{j=1}^n X_{j, w}^{[q]}}-\frac{n X_{i, w^{\prime}}^{[q]}}{\sum_{j=1}^n X_{j, w^{\prime}}^{[q]}}+\zx{\frac{T_i-q}{T-q}}
    , \quad 1 \leq i \leq n,
\end{equation*}
and the same applies to $\Wb^{(i)}$.

\subsection{Asymptotic Normality of Empirical PMI estimator}\label{sec:asym_normality_empirical_PMI}
Recall that
\begin{equation*}
       \Eb_{w,w'} = \log \frac{\sum_{i} X_{i,w, w^{\prime}}^{[q]} }{n N_{w, w^{\prime}}^{[q]} } - \log \frac{\sum_{i} X_{i,w}^{[q]}}{n N_{w}^{[q]}} - \log \frac{\sum_{i} X_{i,w^{\prime}}^{[q]}}{n N_{w^{\prime}}^{[q]}}.
\end{equation*}

\begin{proof} [Proof of Lemma \ref{thm1}]
Recall that earlier for simplicity, we defined $\CC_{w,w'}^{(i)} = X_{i,w,w'}^{[q]}, $ and $ \CC_{w}^{(i)} = X_{i,w}^{[q]}$.
We first show that
\begin{equation}\label{Taylor_Lim_Asy_Normal}
    \frac{\sqrt{n}}{\bar\sigma_{w,w'}}\Big( \frac{\sum_{i} X_{i,w, w^{\prime}}^{[q]}}{n N_{w, w^{\prime}}^{[q]}} - 1 \Big)  \stackrel{d}{\longrightarrow} N(0, 1),
\end{equation}
where $\bar\sigma_{w,w'}^2 = \sum_{i=1}^n (\frac{T_i-q}{T-q})^2 \text{Var}(X_{i,w,w'}^{[q]}/N_{i,w,w'}^{[q]})/n$.
Lemma \ref{lem:coccur_lower_bound_variance} and Lemma \ref{lem:coccur_upper_bound_variance} together shows that the order of $\text{Var}(X_{i,w,w'}^{[q]}/N_{i,w,w'}^{[q]})$ is $d^2/T_i$, which further implies that the order of \(\bar\sigma_{w,w'}^2\) is \(d^2/T\). Denote 
\[
Z_i = \frac{T_i-q}{\sqrt{n}\bar\sigma_{w,w'}(T-q)}\frac{X_{i,w,w'}^{[q]}-N_{i,w,w'}^{[q]}}{N_{i,w,w'}^{[q]}}, \quad i \in [n].
\]
Then we have \(\E[Z_i]=0, \text{Var}(Z_i)=\Theta(\frac{T_i}{n T}), |Z_i|=O(dT_i/\sqrt{nT})\) with probability \(1\) for \(i \in [n]\), and
\[
\frac{\sqrt{n}}{\bar\sigma_{w,w'}}\Big(\frac{\sum_i X_{i,w,w'}^{[q]}}{nN_{w,w'}^{[q]}} - 1\Big) = \sum_i  Z_i.
\]
For any \(\varepsilon>0\), combining results in Lemma \ref{lem:Sww'_to_Nww'}, \ref{lem:X_w_to_cond_exp} and \ref{lem:lemforlindeberg}, we have
\begin{equation*}
    \begin{split}
        \P\Big(\Big| \frac{X_{i,w,w'}^{[q]}}{N_{i,w,w'}^{[q]}} - 1\Big| > \varepsilon \Big) &\le \P\Big(\Big| \frac{S_{i,w,w'}^{[q]}}{N_{i,w,w'}^{[q]}} - 1\Big| > \varepsilon \Big) + 2\P\Big(\Big| X_{i,w,w'}^{[q]} - S_{i,w,w'}^{[q]}\Big| > \frac{\varepsilon N_{i,w,w'}^{[q]}}{2} \mid \Big| \frac{S_{i,w,w'}^{[q]}}{N_{i,w,w'}^{[q]}} - 1\Big| \le \varepsilon\Big)\\
        &\le \exp(-\Omega(\frac{T_i\varepsilon^2}{m})) + 2\exp(-\frac{(\varepsilon N_{i,w,w'}^{[q]})^2}{2T_i}) = \exp(-\Omega(\frac{\varepsilon^2 T_i}{d^4})).
    \end{split}
\end{equation*}
It further yields that
\[
\sum_{i=1}^n \E[Z_i^2 \mathbb{I}(|Z_i|>\varepsilon)] \le  C\sum_{i=1}^n \frac{d^2 T_i^2}{nT} \exp(-\Omega(\frac{nT\varepsilon^2}{T_i d^2})) \rightarrow 0 
\]
as \(n \rightarrow \infty\). Applying Lindeberg-Feller Central Limit Theorem, we obtain \eqref{Taylor_Lim_Asy_Normal}.
Further by Lemma \ref{lem:occur_upper_bound_variance}, we know that $\text{Var}( \sum_i X_{i,w}^{[q]}/N_w^{[q]} )/n = O(d/T) = o(\bar\sigma^2_{w,w'}).$
Applying Lemma \ref{lem:taylor_expansion_of_pmi}, we obtain
\[
    \frac{\sqrt{n}}{\bar\sigma_{w,w'}} \Big( \log \frac{\sum_{i} X_{i,w, w^{\prime}}^{[q]} }{n N_{w, w^{\prime}}^{[q]} } - \log \frac{\sum_{i} X_{i,w}^{[q]}}{n N_{w}^{[q]}} - \log \frac{\sum_{i} X_{i,w^{\prime}}^{[q]}}{n N_{w^{\prime}}^{[q]}} \Big) \stackrel{d}{\longrightarrow} N(0, 1).
    \]
Then by Slutsky's theorem,
    \begin{equation*}
        \begin{split}
            \frac{\sqrt{n}}{\sigma_{w,w'}} \Wb_{w,w'}  &= \frac{\sqrt{n}}{\bar\sigma_{w,w'}} [\Eb_{w,w'} + O(\|\PMIbb - \alpha_p \Vb\Vb^{\top}\|_{\max})]\\
            &= \frac{\sqrt{n}}{\bar\sigma_{w,w'}} \Big( \log \frac{\sum_{i} X_{i,w, w^{\prime}}^{[q]} }{n N_{w, w^{\prime}}^{[q]} } - \log \frac{\sum_{i} X_{i,w}^{[q]}}{n N_{w}^{[q]}} - \log \frac{\sum_{i} X_{i,w^{\prime}}^{[q]}}{n N_{w^{\prime}}^{[q]}} \Big)  + \zx{O(\frac{\sqrt{nT}\kappa^4 p^2}{d^3})} \\
            &=\frac{\sqrt{n}}{\bar\sigma_{w,w'}} \Big( \log \frac{\sum_{i} X_{i,w, w^{\prime}}^{[q]} }{n N_{w, w^{\prime}}^{[q]} } - \log \frac{\sum_{i} X_{i,w}^{[q]}}{n N_{w}^{[q]}} - \log \frac{\sum_{i} X_{i,w^{\prime}}^{[q]}}{nN_{w^{\prime}}^{[q]}} \Big)  + o(1)\\
    &\stackrel{d}{\longrightarrow} N(0, 1),
        \end{split}
    \end{equation*}
where the second equation uses Theorem \ref{thm0:PMI_lowrank}, and the third equation is from Assumption \ref{assump:ndTp} and \ref{assump:eigen_bound}. 
In the case where all health records are of the same length, i.e. \(T_i = T, i \in [n]\), we do not require \(n \rightarrow \infty\) but instead allowing either
\(n\) or \(T \rightarrow \infty\) to  establish the asymptotic normality. Specifically, if \(n\) goes to infinity, the proof is the same as above; if \(n\) does not go to infinity, we apply the central limit theorem for the function of Markov chain \(\{\bc_t\}_{t=1}^T\), where \(\bc_t = [\bc_{1,t}^{\top}, \cdots, \bc_{n,t}^{\top}]^{\top}\). Note that
\[
\bc_{t+1} = \sqrt{\alpha}\bc_t + \sqrt{1-\alpha}\br_{t+1},
\]
where \(\br_{t+1} = [\br_{1,t}^{\top}, \cdots, \br_{n,t}^{\top}]^{\top} \sim N(0 , \Ib_{np}/p)\).
Let \(V(\bc) = \frac{\|\bc\|^2}{n}+1\). Then we have
\[
\E[V(\bc_{t+1})|\bc_t = \bc] = \alpha \|\bc\|^2/n + (1-\alpha) + 1.
\]
It follows that
\[
\Delta V(\bc) = \E[V(\bc_{t+1})|\bc_t = \bc] - V(\bc) = -(1-\alpha)(\frac{\|\bc\|^2}{n}-1)=-(1-\alpha)(V(\bc)-2).
\]
Denote a small set \(\mathcal{C}=\{\bc: \|\bc\|^2 \le 3n \}\). If \(\bc \not\in \mathcal{C}\), we have \(V(\bc) > 4\), and
\[
\Delta V(\bc) = -(1-\alpha)(V(\bc)-2) < -\frac{1-\alpha}{2}V(\bc).
\]
If \(\bc \in \mathcal{C}\), we have
\[
\Delta V(\bc) = -(1-\alpha)(V(\bc)-2) < -\frac{1-\alpha}{2}V(\bc) + 2.
\]
Combining the inequalities above, we obtain the drift condition \(\Delta V(\bc) \le -\frac{1-\alpha}{2}V(\bc) + 2\mathbb{I}(\bc \in \mathcal{C})\), thus \(\{\bc_t\}\) is geometrically ergodic. According to the central limit theorem for the function of Markov chain \citep{jones2004}, we have
\[
\sum_{t=1}^T \sum_{i=1}^n(p_{i,w}(t)-p_w) / \sqrt{nT} \stackrel{d}{\longrightarrow} N(0, \sigma_w^2),
\]
where \(\sigma_w^2\) is the limit of the variance. Denote \(\mathcal{F}_t = \sigma(\bc_1, \bc_2, \cdots, \bc_t)\). Note that
\[
\E[X_{i,w}(t)-p_{i,w}(t)|\cF_{t-1}] = 0;
\]
\[
\text{Var}(X_{i,w}(t) | \cF_{t-1}) = \E[(X_{i,w}(t)-p_{i,w}(t))^2|\cF_{t-1}]
=\E[p_{i,w}(t)(1-p_{i,w}(t))|\cF_{t-1}] \le 1.
\]
It follows that for any \(\delta>0\),
\[
\lim_{T \rightarrow \infty}\sum_{t=1}^{T}\E[(X_{i,w}(t)-p_{i,w}(t))^2\mathbb{I}(|X_{i,w}(t)-p_{i,w}(t)|>\delta \sqrt{T}) | \cF_{t-1}] = 0.
\]
Applying the Martingale central limit theorem \citep{brown71}, we have
\[
\sum_{t=1}^T \sum_{i=1}^n(X_{i,w}(t)-p_{i,w}(t))/\sqrt{nT} \stackrel{d}{\longrightarrow} N(0, \sigma_{pw}^2).
\]
Note that for any pair \((t_1, t_2), 1 \le t_1, t_2 \le T\), 
\begin{align*}
    &\text{Cov}(X_{i,w}(t_1)-p_{i,w}(t_1), p_{i,w}(t_2) - p_w) = \E[(X_{i,w}(t_1)-p_{i,w}(t_1))( p_{i,w}(t_2) - p_w)]\\
    &= \E[\E[(X_{i,w}(t_1)-p_{i,w}(t_1))( p_{i,w}(t_2) - p_w)|\bc_{i,t_1}, \bc_{i,t_2}]]\\
    &= \E[( p_{i,w}(t_2) - p_w)\E[(X_{i,w}(t_1)-p_{i,w}(t_1))|\bc_{i,t_1}, \bc_{i,t_2}]] = 0.
\end{align*}
It follows that
\[
\text{Cov}(\sum_{t=1}^T\sum_{i=1}^n(X_{i,w}(t)-p_{i,w}(t)), \sum_{t=1}^T\sum_{i=1}^n(p_{i,w}(t)-p_{w})) = 0.
\]
Then we have
\[
\frac{1}{\sqrt{nT}}\sum_{t=1}^T\sum_{i=1}^n(X_{i,w}(t)-p_w) = \frac{1}{\sqrt{nT}}\sum_{t=1}^T\sum_{i=1}^n(X_{i,w}(t)-p_{i,w}(t) + p_{i,w}(t)-p_w) \stackrel{d}{\longrightarrow} N(0, \sigma_{pw}^2) 
\]
by Slutsky's theorem and \(\sigma_w = o(\sigma_{pw})\) from Lemma \ref{var_1_upper_bound} and \ref{lem:occur_lower_bound_variance}. Following the same procedure as above, we can prove the asymptotic normality for \(\sum_i X_{i,w,w'}^{[q]} / nN_{w,w'}^{[q]}\) and \[
\frac{\sqrt{n}}{\bar\sigma_{w,w'}} \Big( \log \frac{\sum_{i} X_{i,w, w^{\prime}}^{[q]} }{n N_{w, w^{\prime}}^{[q]} } - \log \frac{\sum_{i} X_{i,w}^{[q]}}{n N_{w}^{[q]}} - \log \frac{\sum_{i} X_{i,w^{\prime}}^{[q]}}{n N_{w^{\prime}}^{[q]}} \Big) .
\]

\end{proof}
\begin{remark}
    In Lemma \ref{lem:taylor_expansion_of_pmi}, we let 
    $$
X_n = \frac{\sum_{i} X_{i,w, w^{\prime}}^{[q]}}{\sum_i N_{i,w, w^{\prime}}^{[q]}} - 1, Y_n =  \zx{\frac{\sum_i X_{i,w}^{[q]}}{\sum_i N_{i,w}^{[q]}} - 1, \text{ and } Z_n =  \frac{\sum_i X_{i,w'}^{[q]}}{\sum_i N_{i,w'}^{[q]}} - 1}
    $$
    to prove that $\Eb_{w,w'} = X_n + o_p(\sqrt{\text{Var}(X_n)})$. According to (\ref{eq:taylor_expansion_pmi_sharp_rate}), we can improve the rate to 
    $$
\Eb_{w,w'} = X_n - Y_n - Z_n + o_p(\left(\text{Var}(X_n)\right)^{1/2+\alpha})
    $$
    for any $\alpha \in (0,1/2).$
\end{remark}

\begin{lemma}\label{lem:lemforlindeberg}
\zx{
    Suppose there are random variables \(X, N, S\) with \(X\ge 0, N>0, S>0\). For any \(\varepsilon>0\), we have
    \[
    \P\Big(\Big|\frac{X}{N}-1\Big|>\varepsilon\Big) \leq 2\P\Big(\Big|\frac{S}{N}-1\Big|>\frac{\varepsilon}{2}\Big) + \P\Big(|X-S|>\frac{\varepsilon N}{2}\mid \Big|\frac{S}{N}-1\Big| \le \varepsilon\Big)
    \]}
\end{lemma}
\begin{proof}
\zx{
    \begin{equation*}
        \begin{split}
            \P\Big(\Big|\frac{X}{N}-1\Big|>\varepsilon\Big) &= \P(|X-N| > \varepsilon N) \le \P(|X-S|+|S-N|>\varepsilon N)\\
            &=\P(|X-S|+|S-N|>\varepsilon N | |S-N|>\varepsilon N)\P(|S-N|>\varepsilon N) \\
            &\quad + \P(|X-S|+|S-N|>\varepsilon N | |S-N| \le \varepsilon N)\P(|S-N| \le \varepsilon N)\\
            &\le \P(|S-N|>\varepsilon N) + \P(|X-S| >\varepsilon N/2 | |S-N| \le \varepsilon N)\\
            &\quad + \P(|S-N|>\varepsilon N/2 | |S-N| \le \varepsilon N)\P(|S-N| \le \varepsilon N)\\
            &= \P(|\frac{S}{N}-1|>\varepsilon) + \P(|X-S| >\frac{\varepsilon N}{2} | |S-N| \le \varepsilon N) + \P(|\frac{S}{N}-1|>\frac{\varepsilon}{2})\\
            &\le \P(|X-S| >\frac{\varepsilon N}{2} | |S-N| \le \varepsilon N) + 2\P(|\frac{S}{N}-1|>\frac{\varepsilon}{2}),
        \end{split}
    \end{equation*}
where the second inequality uses the property that \(P(X+Y > \varepsilon) \leq \P(X>\varepsilon/2)+\P(Y>\varepsilon/2)\) for \(\varepsilon>0\).}
\end{proof}

\begin{lemma}\label{lem:taylor_expansion_of_pmi}
    Suppose there are three series of random variables $\{X_n\},\{Y_n\}$, and $\{Z_n\}$ with lower bound $-1$, and $\E[X_n] = \E[Y_n] = \E[Z_n] = 0$, and $\text{Var}(Y_n) = o(\text{Var}(X_n)), \text{Var}(Z_n) = o(\text{Var}(X_n)).$ Further we assume $\text{Var}(X_n) \rightarrow 0$, $\left(\text{Var}(X_n)\right)^{-1/2}X_n \stackrel{d}{\longrightarrow} N(0,1)$. Then 
    \begin{equation*}
        \left(\text{Var}(X_n- Y_n - Z_n) \right)^{-1/2} \log \frac{1+X_n}{(1+Y_n)(1+Z_n)} \stackrel{d}{\longrightarrow} N(0,1)
    \end{equation*}
\end{lemma}
\begin{proof}
    First, by $\P(|Y_n| \geq \epsilon \sqrt{\text{Var}(X_n)}) \leq \frac{\text{Var}(Y_n)}{\epsilon^2 \text{Var}(X_n)} \rightarrow 0$, we know that $Y_n = o_p(\sqrt{\text{Var}(X_n)})$. Similarly, $Z_n = o_p(\sqrt{\text{Var}(X_n)})$. And $\P(|X_n|^2  \geq \epsilon \sqrt{\text{Var}(X_n)} ) \leq \sqrt{\text{Var}(X_n)}/\epsilon \rightarrow 0$ implies that $X_n^2 = o_p(\sqrt{\text{Var}(X_n)}).$ 
    We then claim that $\log(1+X_n) = X_n + o_p((\text{Var}(X_n))^{1/2+\alpha})$ for any given $\alpha \in [0,1/2)$, which can be obtained by the following inequality:
\begin{equation*}
    \begin{split}
        &\P\big(|\log(1+X_n)-X_n | \geq \epsilon (\text{Var}(X_n))^{\frac{1}{2}+\alpha}  \big) = \P\big(\log(1+X_n)-X_n  \leq -\epsilon (\text{Var}(X_n))^{\frac{1}{2}+\alpha} \big)\\
        &\leq \P\big(\log(1+X_n)-X_n  \leq -\epsilon (\text{Var}(X_n))^{\frac{1}{2}+\alpha} \mid X_n > -\frac{1}{2}\big)\P(X_n > -\frac{1}{2}) + \P(X_n \leq -\frac{1}{2})\\
        &\leq \P(-X_n^2 \leq -\epsilon (\text{Var}(X_n))^{\frac{1}{2}+\alpha} \mid X_n > -\frac{1}{2})\P(X_n > -\frac{1}{2}) + \P(X_n \leq -\frac{1}{2})\\
        &\leq \P\big(|X_n|^2 \geq \epsilon (\text{Var}(X_n))^{\frac{1}{2}+\alpha} \big) + \P(|X_n| \geq \frac{1}{2}) \leq \frac{(\text{Var}(X_n))^{\frac{1}{2} - \alpha}}{\epsilon} + 4\text{Var}(X_n) \rightarrow 0,
    \end{split}
\end{equation*}
where the first equation uses $\log(1+x)\leq x, \forall x>-1$, and the second inequality uses $\log(1+x)\geq x - x^2,  x>-\frac{1}{2}$.
    Similarly, we have $\log(1+Y_n) = Y_n + o_p((\text{Var}(Y_n))^{1/2+\alpha})$, and $\log(1+Z_n) = Z_n + o_p(\text{Var}(Z_n))^{1/2+\alpha})$. Noting that $\text{Var}(Y_n) = o(\text{Var}(X_n)), \text{Var}(Z_n) = o(\text{Var}(X_n))$, we have
    \[
    \text{Var}(X_n - Y_n - Z_n)/\text{Var}(X_n) \rightarrow 1 \text{ as } n \rightarrow \infty,
    \]
    and
    \begin{equation}\label{eq:taylor_expansion_pmi_sharp_rate}
            \log \frac{1+X_n}{(1+Y_n)(1+Z_n)} = (X_n - Y_n - Z_n) + o_p((\text{Var}(X_n))^{\frac{1}{2}+\alpha}) = X_n + o_p(\sqrt{\text{Var}(X_n)}).
    \end{equation}
    Thus by Slutsky's theorem, we have
    \begin{equation*}
        \begin{split}
            &\left( \text{Var}(X_n - Y_n - Z_n)\right)^{-1/2}\log \frac{1+X_n}{(1+Y_n)(1+Z_n)} \\
            &= \sqrt{\frac{\text{Var}(X_n)}{\text{Var}(X_n - Y_n - Z_n)}}\left(\text{Var}(X_n)\right)^{-1/2} (X_n + o_p(\sqrt{\text{Var}(X_n)})) \\
            &= \sqrt{\frac{\text{Var}(X_n)}{\text{Var}(X_n - Y_n - Z_n)}}\left(\text{Var}(X_n)\right)^{-1/2}X_n + o_p(1) \stackrel{d}{\longrightarrow} N(0,1).
        \end{split}
    \end{equation*}
\end{proof}

In the lemmas below, we omit the subscript \(i\) if only analyzing the statistics of one patient for simplicity.
\begin{lemma}\label{var_1_upper_bound}
Suppose Assumptions~\ref{assump:ndTp}-\ref{assump:eigen_bound} hold. We have that for large $T$,
\begin{equation*}
    \text{Var}\Big( \frac{S_{w}}{N_{w}} \Big) = o(\frac{1}{T}) \text{ and }
    \text{Var}\Big( \frac{S_{w,w'}^{[q]}}{N_{w,w'}^{[q]}} \Big) = o(\frac{1}{T}).
\end{equation*}
\end{lemma}
\begin{proof}
For the first equation, recall that
\begin{equation*}
    p_{w}(t) = \mathbb{E}\left[X_{w}(t) \mid \left\{\boldsymbol{c}_{t}\right\}\right]=\frac{\exp \left(\left\langle\Vb_{w}, \boldsymbol{c}_{t}\right\rangle\right)}{\sum_{w^{\prime}} \exp \left(\left\langle\Vb_{w^{\prime}}, \boldsymbol{c}_{t}\right\rangle\right)}.
\end{equation*}
First, according to \eqref{ineq:pw_gmw} in Lemma \ref{lem:B37} and \eqref{ineq:p_ww'u_two_side_bound} in Lemma \ref{lem:B36},
we have
\begin{equation*}
    \Big|p_w - \exp\Big(\frac{\|\Vb_w\|^2}{2p} - \log Z \Big)  \Big| \leq C \varepsilon_d^3 p_w + C\frac{\kappa^4p^2}{d^2},
\end{equation*}
\begin{equation*}
    \Big|p_{w,w}^{(k)} - \exp\Big(\frac{(1+\alpha^{k/2})\|\Vb_w\|^2}{p} - 2\log Z  \Big)  \Big| \leq C \varepsilon_d^3 p_{w,w}^{(k)} + C\frac{\kappa^4p^2}{d^2}
\end{equation*}
for some constant $C>0$. Note that $p_w = \E_{\bc_t}[p_w(t)], p_{w,w}^{(k)} = \E_{(\bc_t,\bc_{t+k})}[p_w(t)p_w(t+k)]$. Thus $\text{Cov} (p_w(t), p_w(t+k)) = p_{w,w}^{(k)} - p_w^2$, which can be bounded by
\begin{equation*}
    \begin{split}
        p_{w,w}^{(k)} - p_w^2 = &\exp\Big(\frac{\|\Vb_w\|^2}{p} - 2\log Z \Big) \Big[\exp\Big(\frac{\alpha^{k/2}}{p}\|\Vb_w\|^2 \Big)  - 1\Big] +  O\big(\varepsilon_d^3 p_{w,w}^{(k)} + \varepsilon_d^3 p_w + \frac{\kappa^4p^2}{d^2}\big)\\ 
        = &   \frac{p_w^2\alpha^{k/2}}{p}\|\Vb_w\|^2 +  O\big(\frac{\kappa^4 p^2}{d^2}\big)
        \leq  C\frac{\kappa^2 \alpha^{k/2}}{d} + C \frac{\kappa^4 p^{2}}{d^{2}}
    \end{split}
\end{equation*}
for some $C > 0$. Here the second equation uses the definition of $\varepsilon_d$ and Assumption \ref{assump:eigen_bound}; and the last equation uses $\|\Vb_w\|^2 /p  \lesssim \kappa^2$, which is from Assumption \ref{assump1}.
Then we have
\begin{equation*}
    \begin{split}
        \text{Var} \big(  \frac{S_w}{N_w} \big) &= \frac{1}{N_w^2} \text{Var} \big( \sum_{t=1}^{T}p_w(t)\big) = \frac{1}{N_w^2}\Big(\sum_{t=1}^{T} \text{Var}(p_w(t)) + \sum_{k=1}^{T-1}\sum_{|t-s| = k}\text{Cov}(p_w(t), p_w(s))\Big)\\
        &\lesssim \frac{d^2}{T^2} \Big[\sum_{k=0}^T (T-k)\Big( \frac{\kappa^2 \alpha^{k/2}}{d} + \frac{ \kappa^4 p^{2}}{d^{2}}\Big)\Big] \lesssim \frac{d}{T} \sum_{k=0}^{\infty} \kappa^2 \alpha^{k/2} +  \kappa^4 p^{2} \\
        &= \frac{\kappa^2 d}{T(1-\sqrt{\alpha})} + \kappa^4 p^2 = o(\frac{1}{T}),
    \end{split}
\end{equation*}
where the last equation is from Assumptions \ref{assump:ndTp} and  \ref{assump:eigen_bound}.

For the second equation, recall that $S_{w,w'}^{[q]} = \sum_{u=1}^q S_{w,w'}^{(u)}, N_{w,w'}^{[q]} = \sum_{u=1}^q N_{w,w'}^{(u)}$, and $q$ is fixed number. By $\text{Var}(\sum_{i=1}^n X_i) \leq n\sum_{i=1}^n\text{Var}(X_i)$, we have
\begin{equation*}
    \text{Var}\Big( \frac{S_{w,w'}^{[q]}}{N_{w,w'}^{[q]}} \Big) \leq q\sum_{u=1}^q\text{Var}\Big( \frac{S_{w,w'}^{(u)}}{N_{w,w'}^{[q]}} \Big) \leq q\sum_{u=1}^q\text{Var}\Big( \frac{S_{w,w'}^{(u)}}{N_{w,w'}^{(u)}}\Big).
\end{equation*}
Thus it suffices to show that for each $u \in [q]$,
\begin{equation*}
    \text{Var}\Big( \frac{S_{w,w'}^{(u)}}{N_{w,w'}^{(u)}} \Big) = o(\frac{1}{T}).
\end{equation*}
Following the notation in Lemma \ref{lem:Sww'_to_Nww'}, we can decompose $S_{w,w'}^{(u)}$ by $S_{w,w'}^{(u)} = \sum_{k=1}^{u+1}\sum_{l=0}^{T_k-1}p_{w,w'}^{(k,u)}(l)$. Again by $\text{Var}(\sum_{i=1}^n X_i) \leq n\sum_{i=1}^n\text{Var}(X_i)$, we have
\begin{equation*}
     \begin{split}
         \text{Var}\Big( \frac{S_{w,w'}^{(u)}}{N_{w,w'}^{(u)}} \Big) &\leq (u+1)\sum_{k=1}^{u+1}\text{Var}\Big(\frac{\sum_{l=0}^{T_k-1}p_{w,w'}^{(k,u)}(l)}{N_{w,w'}^{(u)}} \Big).
     \end{split}
\end{equation*}
Note that for each feature pair $(w,w')$ and each $u \in [q]$, we have
\begin{equation*}
    \begin{split}
        p_{w,w'}(t,t+u) &= \frac{\exp(\Vb_w^{\top}\bc_t+\Vb_{w'}^{\top}\bc_{t+u})}{Z(\bc_t)Z(\bc_{t+u})} = \frac{\exp((\Vb_w+\alpha^{u/2}\Vb_{w'})^{\top}\bc_t+\sqrt{1-\alpha^{u}}\Vb_{w'}^{\top}\br_t)}{\sum_{w'',w'''}\exp((\Vb_{w''}+\alpha^{u/2}\Vb_{w'''})^{\top}\bc_t+\sqrt{1-\alpha^{u}}\Vb_{w'''}^{\top}\br_t)}\\
        &=
        \frac{\exp\big([(\Vb_w+\alpha^{u/2}\Vb_{w'})^{\top},\sqrt{1-\alpha^{u}}\Vb_{w'}^{\top}]\cdot [\bc_t^{\top},\br_t^{\top}]^{\top} \big)}{\sum_{w'',w'''}\exp\big([(\Vb_{w''}+\alpha^{u/2}\Vb_{w'''})^{\top},\sqrt{1-\alpha^{u}}\Vb_{w'''}^{\top}]\cdot [\bc_t^{\top},\br_t^{\top}]^{\top} \big)} =: \tilde p_{w,w'} (t),
    \end{split}
\end{equation*}
where the first equation is from the definition of $p_{w,w'}(t,t+u)$ in Section \ref{subsec:notation}, and the last equation reveals that, for any given $k,u$, the set $\{p_{w,w'}^{(k,u)}(l)\}_{l=0}^{T_k-1}$ can be conceptualized as a sequence of new occurrence probabilities $\{\tilde p_{w,w'}(t)\}$ with embeddings in the $\mathbb R^{2p}$ space. Following similar analysis in Lemma \ref{lem:B36}, an upper bound for the covariance $\text{Cov}(p_{w,w'}^{(k,u)}(l), p_{w,w'}^{(k,u)}(l+s))$ can be obtained, hence the term $\text{Var}\big(\sum_{l=0}^{T_k-1}p_{w,w'}^{(k,u)}(l)/N_{w,w'}^{(u)} \big)$ can be bounded using the same technique above, which completes the proof.

\end{proof}

\begin{lemma}\label{lem:occur_lower_bound_variance}
Suppose Assumptions~\ref{assump:ndTp}-\ref{assump:eigen_bound} hold. \zx{There exists some constant $c>0$, }such that for sufficient large $d,p \text{ and } T$, we have \zx{
\begin{equation*}
    \text{Var}\left(\left.\frac{X_{w}^{[q]}}{N_{w}^{[q]}}\right.\right) \geq c\frac{d}{T}.
\end{equation*}}
\end{lemma}
\begin{proof}
\zx{
Recall that conditional on $\left\{\boldsymbol{c}_{t}\right\}_{t \geq 1}$, $\left\{X_{w}(t)\right\}_{t \geq 1}$ are independent with
\begin{equation}\label{xt_condition_dist}
    X_{w}(t) \mid \bc_{t} \sim \operatorname{Bernoulli}\left(p_{w}(t)\right).
\end{equation}
Thus $X_w - S_w = \sum_{t=1}^T (X_{w}(t) - p_{w}(t))$ follows the Poisson binomial distribution. Denote $Y = \sum_{t=1}^T (X_{w}(t) - p_{w}(t))$. Then we have
\begin{equation*}
    \E[Y^2|\{\bc_t\}_{t=1}^{T}] = \sum_{t=1}^{T}\text{Var}(X_w(t)|\bc_t) = \sum_{t=1}^{T} p_{w}(t)(1-p_{w}(t)) \leq \sum_{t=1}^{T} p_{w}(t) = S_w;
\end{equation*}
It follows that
\begin{equation*}
    \E[Y^2] = \E[\E[Y^2|\{\bc_t\}_{t=1}^{T}]] \leq N_w = O(\frac{T}{d});
\end{equation*}
On the other side, note that
$$
\text{Var}(X_{w}|\{\bc_t\}_{t=1}^T ) = \sum_{t=1}^T p_w(t)(1-p_w(t)).
$$
According to \eqref{eq:Xwq_to_Xw},
\[
X_{i,w}^{[q]} = 2qX_{i,w} - \delta_i.
\]
It follows that
\begin{align*}
    &\text{Var}(X_w^{[q]}|\{\bc_t\}) = 4q^2 \text{Var}(X_{w}|\{\bc_t\}) + \text{Var}(\delta|\{\bc_t\}) - 2\text{Cov}(X_{w}, \delta|\{\bc_t\})\\
    &\ge 4q^2\sum_{t=1}^T p_w(t)(1-p_w(t)) - \sqrt{16q^4(\sum_{t=1}^T p_w(t)(1-p_w(t))) (\sum_{t=1}^{q}+\sum_{t=T-q}^{T}) p_w(t)(1-p_w(t))} \\
    &\ge c \sum_{t=q+1}^{T-q} p_w(t)(1-p_w(t))
\end{align*}
for some \(c>0\) and sufficiently large \(T\). Denote $q_{w} = \E_{\bc_t}[p_{w}^2(t)]$, then by Lemma \ref{lem:B36}, $q_{w} = O(1/d^2).$ Applying the law of total variance, we obtain
\begin{equation}\label{ineq:lower_bound_X/N}
    \text{Var}\Big(\frac{X_{w}^{[q]}}{N_{w}^{[q]}}\Big) 
    \geq \E\Big[ \text{Var}\Big(\frac{X_{w}^{[q]}}{N_{w}^{[q]}}\Big|\{\bc_t\} \Big) \Big]= 
    \frac{T-2q}{4T^2p_w^2}(p_w - q_w) = \Omega \big( \frac{d}{T}\big).
\end{equation}}

\end{proof}

\begin{lemma}\label{lem:coccur_lower_bound_variance}
Suppose Assumptions~\ref{assump:ndTp}-\ref{assump:eigen_bound} hold. There exists a constant $c'>0$, such that for sufficiently large $d,p \text{ and } T$, we have
\begin{equation*}
\text{Var}\left(\left.\frac{X_{w,w'}^{[q]}}{N_{w,w'}^{[q]}}\right.\right) \geq c'\frac{d^2}{T}.
\end{equation*}
\end{lemma}
\begin{proof}
The proof idea is the same as Lemma \ref{lem:occur_lower_bound_variance}.
Define the set of discourse vectors $\mathcal{P}_{w,w'}$ :
\begin{equation*}
    \begin{split}
        \mathcal{P}_{w,w'}=\Big\{ \{\bc_t\}_{t=1}^{T} : &\Big| \sum_{t=1}^{T}\prod_{k=1}^K p_{w_k}(t+u_k) - (T-\max_k u_k) q_{\{w_k,u_k\}}^{(K)}  \Big| \leq (T-\max_k u_k) q_{\{w_k,u_k\}}^{(K)}\epsilon,\\
    &w_k \in \{w,w'\}, 0\leq u_k \leq q, 1\leq K \leq 4\Big\},
    \end{split}
\end{equation*}
where $q_{\{w_k,u_k\}}^{(K)} = \E_{\{\bc_t\}}\big[\prod_{k=1}^K p_{w_k}(t+u_k) \big], \epsilon = \frac{\log d}{\sqrt{T/m}}$, and we define $p_w(t) = 0$ for any $w\in [d]$ and $t>T$. For $K=1$, following the proof procedure of Lemma \ref{lem:Sw_to_Nw}, we have $\P\big(\big| \sum_{t=1}^{T}p_w(t+u) - (T-u) p_w \big| \leq T p_w\epsilon \big) = 1 -\exp(-\Omega(\log^2 d))$ for any $u \in [q]$. For $K=2$, following the proof procedure of Lemma \ref{lem:Sww'_to_Nww'}, we have 
\begin{equation*}
    \P\big(\big| \sum_{t=1}^{T}p_w(t+u_1)p_{w'}(t+u_2) - (T-\max\{u_1,u_2\}) p_w \big| \leq (T-\max\{u_1,u_2\} p_w\epsilon \big) = 1 -\exp(-\Omega(\log^2 d)).
\end{equation*}
for each pair $(u_1,u_2)\in [q]\times [q]$. Similarly, for $K=3,4$, and for each series $\{w_k,u_k\}_{k=1}^K$, we have
\begin{equation*}
    \P\Big( \Big| \sum_{t=1}^{T}\prod_{k=1}^K p_{w_k}(t+u_k) - (T-\max_k u_k) q_{\{w_k,u_k\}}^{(K)}  \Big| \leq (T-\max_k u_k) q_{\{w_k,u_k\}}^{(K)}\epsilon \Big) = 1 -\exp(-\Omega(\log^2 d)).
\end{equation*}
Let $\cP = \cap_{w,w'} \cP_{w,w'}$. Applying the union bound, we have
$$
P\big( \{\bc_t\} \in \mathcal{P} \big) \geq 1- d^2\sum_{K=1}^4 (2q)^K \exp(-\Omega(\log^2 d)) =1 -\exp(-\Omega(\log^2 d)).
$$
Recall that $X_{w,w'}^{[q]} = \sum_{u=1}^q\sum_{t=1}^{T-u}[X_{w}(t)X_{w'}(t+u)+X_{w'}(t)X_{w}(t+u)]$. And by Lemma \ref{lem:B36}, we know that $N_{w,w'}^{[q]}=2\sum_{u=1}^q\sum_{t=1}^{T-u}p_{w,w'}^{(u)}$ is of order $T/d^2$.
We first claim that when $\{\bc_t\} \in \mathcal{P}$,
\begin{equation*}
    \text{Var}\big(\sum_{u=1}^q\sum_{t=1}^{T-u}X_{w}(t)X_{w'}(t+u) \mid \{\bc_t\} \big) = \sum_{u=1}^q\sum_{t=1}^{T-u}p_w(t)p_{w'}(t+u) + o\big(\frac{T}{d^2}\big).
\end{equation*}
Note that for $t=s,u=v$,
\begin{equation*}
    \text{Cov}(X_{w}(t)X_{w'}(t+u), X_{w}(s)X_{w'}(s+v)|\{\bc_t\} ) = p_w(t)p_{w'}(t+u)(1-p_w(t)p_{w'}(t+u));
\end{equation*}
for $t=s, u \neq v$,
\begin{equation*}
    \text{Cov}(X_{w}(t)X_{w'}(t+u), X_{w}(s)X_{w'}(s+v) |\{\bc_t\} ) = p_w(t)p_{w'}(t+u)p_{w'}(t+v)(1-p_w(t));
\end{equation*}
for $t \neq s, t+u = s+v$,
\begin{equation*}
    \text{Cov}(X_{w}(t)X_{w'}(t+u), X_{w}(s)X_{w'}(s+v) |\{\bc_t\} ) = p_{w'}(t+u)p_{w}(t)p_{w}(s)(1-p_{w'}(t+u));
\end{equation*}
for $s = t+u,$
\begin{equation*}
    \text{Cov}(X_{w}(t)X_{w'}(t+u), X_{w}(s)X_{w'}(s+v) |\{\bc_t\} ) = -p_{w}(t)p_{w}(t+u)p_{w'}(t+u)p_{w'}(t+u+v);
\end{equation*}
for $t = s+v$,
\begin{equation*}
    \text{Cov}(X_{w}(t)X_{w'}(t+u), X_{w}(s)X_{w'}(s+v) |\{\bc_t\} ) = -p_{w}(s)p_{w}(s+v)p_{w'}(s+v)p_{w'}(s+u+v);
\end{equation*}
and for $t \neq s, t+u \neq s+v, s \neq t+u, t \neq s+v$,
\begin{equation*}
    \text{Cov}(X_{w}(t)X_{w'}(t+u), X_{w}(s)X_{w'}(s+v) |\{\bc_t\} ) = 0.
\end{equation*}
To simplify the expression, we denote $p_{w^{(1)},w^{(2)}}(u_1,u_2) = p_{w^{(1)}}(t+u_1)p_{w^{(2)}}(t+u_2)$.
Then we have
\begin{equation*}
    \begin{split}
        &\text{Var}(\sum_{u=1}^q\sum_{t=1}^{T-u}X_{w}(t)X_{w'}(t+u) \mid \{\bc_t\} )= \sum_{u=1}^q\sum_{t=1}^{T-u}\sum_{v=1}^q\sum_{s=1}^{T-v}\text{Cov}(X_{w}(t)X_{w'}(t+u), X_{w}(s)X_{w'}(s+v) )\\
        &=\sum_{u=1}^q\sum_{t=1}^{T-u}p_{w,w'}(0,u)(1-p_{w,w'}(0,u)) + \sum_{t=1}^{T-1}\sum_{1\leq u \neq v \leq \min(q,T-t)} p_{w,w'}(0,u)p_{w'}(t+v)(1-p_w(t)) \\
        &+ \sum_{t=2}^{T}\sum_{1\leq u \neq v \leq \min(q,t-1)} p_{w',w}(0,-u)p_{w}(t-v)(1-p_{w'}(t)) \\
        & -\sum_{t=1}^T\sum_{u=1}^{\min(q,T-t)}\sum_{v=1}^{\min(q,T-t-u)}p_{w,w}(0,u)p_{w',w'}(u, u+v) \\
        &- \sum_{t=1}^T\sum_{u=1}^{\min(q,T-t)}\sum_{v=1}^{\min(q,T-t-u)} p_{w',w'}(0,u)p_{w,w}(u,u+v)= \sum_{u=1}^q\sum_{t=1}^{T-u}p_{w,w'}(0, u) + o(\frac{T}{d^2})
    \end{split}
\end{equation*}
given $\{\bc_t\} \in \cP$. Here the last equation uses the definition of the set $\mathcal{P}$. After a similar discussion, we obtain that for $w \neq w'$,
\begin{equation*}
        \text{Cov}\big(\sum_{u=1}^q\sum_{t=1}^{T-u}X_{w}(t)X_{w'}(t+u), \sum_{u=1}^q\sum_{t=1}^{T-u}X_{w'}(t)X_{w}(t+u) \mid \{\bc_t\} \big) = o(\frac{T}{d^2}).
\end{equation*}
Combining two results above, we obtain that for $w \neq w'$ and $\{\bc_t\}\in \cP$,
\begin{equation}\label{ineq:B.33}
    \begin{split}
        &\text{Var}\Big( \frac{X_{w, w^{\prime}}^{[q]}}{ N_{w, w^{\prime}}^{[q]}}\Big| \{\bc_t\} \Big) \sim \frac{d^4}{T^2} \text{Var}(X_{w, w^{\prime}}^{[q]}\mid \{\bc_t\} ) = \frac{d^4}{T^2}\big(\sum_{u=1}^q\sum_{t=1}^{T-u}(p_{w,w'}(0,u) + p_{w',w}(0,u))+ o(\frac{T}{d^2}) \big) \sim \frac{d^2}{T}.
    \end{split}
\end{equation}
 Applying the law of total variance, we obtain
$$
\begin{aligned}
 \text{Var}\Big(\frac{X_{w, w^{\prime}}^{[q]}}{ N_{w, w^{\prime}}^{[q]}} \Big)
 &\geq \E_{\{\bc_t\}}\Big[ \text{Var}\Big(\frac{X_{w, w^{\prime}}^{[q]}}{ N_{w, w^{\prime}}^{[q]}}  \Big|  \{\bc_t\}\Big)
\Big] \geq \E_{\{\bc_t\}}\Big[ \text{Var}\Big(\frac{X_{w, w^{\prime}}^{[q]}}{ N_{w, w^{\prime}}^{[q]}}  \Big|  \{\bc_t\}\Big) \Big| \cP_{w,w'}
\Big] \cdot P\big( \{\bc_t\} \in \mathcal{P}_{w,w'} \big)\\
&\gtrsim \frac{d^2}{T} \cdot P\big( \{\bc_t\} \in \mathcal{P}_{w,w'} \big) = \Omega\Big(\frac{ d^2}{T}\Big)
\end{aligned}
$$
for sufficient large $d$ and $T$.
\end{proof}

\begin{lemma}\label{lem:coccur_upper_bound_variance}
Suppose Assumptions~\ref{assump:ndTp}-\ref{assump:eigen_bound} hold. There exists a constant $C'>0$, such that for sufficient large $d,p \text{ and } T$, we have
$$
\text{Var}\Big(\frac{X_{w, w^{\prime}}^{[q]}}{ N_{w, w^{\prime}}^{[q]}}\Big) \leq C'\frac{d^2}{T}.
$$
\end{lemma}
\begin{proof} 
Applying the law of total variance, we have
\begin{equation}\label{law_of_total_var}
    \text{Var}\Big(\frac{X_{w, w^{\prime}}^{[q]}}{ N_{w, w^{\prime}}^{[q]}}\Big) = \text{Var}\Big(\E\Big[ \frac{X_{w, w^{\prime}}^{[q]}}{ N_{w, w^{\prime}}^{[q]}}\left.\right| \{\bc_t\}\Big]\Big) + \E\Big[\text{Var}\Big( \frac{X_{w, w^{\prime}}^{[q]}}{ N_{w, w^{\prime}}^{[q]}}\Big| \{\bc_t\}\Big)\Big].
\end{equation}
The first term in (\ref{law_of_total_var}) can be bounded above by Lemma~\ref{var_1_upper_bound}:
\begin{equation*}
\text{Var}\Big(\E\Big[ \frac{X_{w, w^{\prime}}^{[q]}}{ N_{w, w^{\prime}}^{[q]}}\Big|\{\bc_t\}\Big]\Big) = \text{Var}\Big( \frac{S_{w, w^{\prime}}^{[q]}}{ N_{w, w^{\prime}}^{[q]}}\Big) \leq C_1\frac{d^2}{T}.
\end{equation*}
And the second term can be bounded above by
\begin{equation*}
    \begin{aligned}
\E\Big[ \text{Var}\Big( \frac{X_{w, w^{\prime}}^{[q]}}{ N_{w, w^{\prime}}^{[q]}}\Big|\{\bc_t\}\Big) \Big] &= \E\Big[ \text{Var}\Big( \frac{X_{w, w^{\prime}}^{[q]}}{ N_{w, w^{\prime}}^{[q]}}\Big|\{\bc_t\}\Big) \Big| \cP_{w,w'}\Big]\cdot \P(\cP_{w,w'}) \\
&\quad + \E\Big[ \text{Var}\Big( \frac{X_{w, w^{\prime}}^{[q]}}{ N_{w, w^{\prime}}^{[q]}}\Big|\{\bc_t\}\Big) \Big| \cP_{w,w'}^c\Big]\cdot \P(\cP_{w,w'}^c)\\
&\leq \E\Big[ \text{Var}\Big( \frac{X_{w, w^{\prime}}^{[q]}}{ N_{w, w^{\prime}}^{[q]}}\Big|\{\bc_t\}\Big) \Big| \cP_{w,w'}\Big] + Cd^4\P(\cP_{w,w'}^c) \\
&\lesssim \frac{d^2}{T} + d^4\exp(-\Omega(\log^2 d)) = O\big(\frac{d^2}{T}\big),
\end{aligned}
\end{equation*}
where the first equation uses the law of total expectation; the first inequality uses the fact that the random variable $X_{w,w'}^{[q]}/N_{w,w'}^{[q]}$ is bounded by $[0,T/N_{w,w'}^{[q]}]$, implying that the variance of that is $O(d^4)$; and the second inequality uses \eqref{ineq:B.33}.
\end{proof}

\begin{lemma}\label{lem:occur_upper_bound_variance}
Suppose Assumptions~\ref{assump:ndTp}-\ref{assump:eigen_bound} hold. There exists a constant $C'>0$, such that for sufficient large $d,p \text{ and } T$, we have
$$
\text{Var}\Big(\frac{X_{w}^{[q]}}{N_{w}^{[q]}}\Big) \leq C'\frac{d}{T}
$$
\end{lemma}
\begin{proof}
This proof is essentially the same as Lemma~\ref{lem:coccur_upper_bound_variance}. Applying the law of total variance and Lemma \ref{var_1_upper_bound} directly yields the result. So we omit it here.
\end{proof}

\subsection{Proof of Theorem \ref{thm2}}\label{B3}
Recall that we define $\Vb = \Ub \bLambda^{1/2}$ in Section \ref{sec:thm} and $\alpha_p$ in \eqref{eq:tildeV}. Let $\bLambda^{\star} = \alpha_p \bLambda = \text{diag}(\kappa_1^2, \cdots, \kappa_p^2)$ and denote the eigen-decomposition of $\hat{\PMIbb}$ by
$$
\begin{aligned}
\hat{\PMIbb} &=\left[\begin{array}{ll}
\hat{\Ub} & \hat{\Ub}_{\perp}
\end{array}\right]\left[\begin{array}{cc}
\hat\bLambda & \mathbf{0} \\
\mathbf{0} & \hat\bLambda_{\perp}
\end{array}\right]\left[\begin{array}{l}
\hat{\Ub}^{\top} \\
\hat{\Ub}^{\top}_{\perp}
\end{array}\right]
\end{aligned}
$$
 where $\hat{\Ub} \in \mathbb{R}^{d \times p}$ has orthonormal columns, and $\hat\bLambda \in \mathbb{R}^{p \times p}$ is a diagonal matrix containing the first $p$ eigenvalues $\left\{\lambda_{i}\right\}$ of $\hat{\PMIbb}$ with $\lambda_{1} \geq \lambda_{2} \geq \cdots \geq \lambda_{p}$. Denote $\Xb := \hat{\Ub}\hat\bLambda^{\frac{1}{2}}, \Xb_{\perp} := \hat\Ub_{\perp}\hat\bLambda_{\perp}^{\frac{1}{2}},$ and $\tilde\Vb = \Ub\bLambda^{\star \frac{1}{2}}$. 
 Then $\alpha_p \Vb\Vb^{\top} = \tilde\Vb \tilde\Vb^{\top}$, and $\Xb\Xb^{\top}$ is a rank-$p$ symmetric estimator of $\PMIbb$. Recall that $\Wb = \hat{\PMIbb} - \tilde{\Vb}\tilde{\Vb}^{\top}$. Then we have following decomposition:
\begin{equation*}
    \begin{split}
        \bX &= \bX \bX^{\top}\bX(\bX^{\top}\bX)^{-1}\\
        &= \Wb\bX(\bX^{\top}\bX)^{-1} + \tilde{\Vb}\tilde{\Vb}^{\top}\bX(\bX^{\top}\bX)^{-1} - \Xb_{\perp} \Xb_{\perp}^{\top}\bX(\bX^{\top}\bX)^{-1}\\
        &= \Wb\bX(\bX^{\top}\bX)^{-1} + \tilde{\Vb}\tilde{\Vb}^{\top}\bX(\bX^{\top}\bX)^{-1},
    \end{split}
\end{equation*}
where the last equation is due to $\Xb_{\perp}^{\top}\Xb = 0$. Denote $\Hb$ to be the rotation matrix that best aligns $\hat{\Ub}$ and $\Ub$, i.e. $\Hb =\arg\min_{\Rb \in \mathcal{O}^{p\times p}}\|\hat{\Ub}\Rb - \Ub\|$,
and $\overline{\bX} = \bX \Hb$. Then we have
\begin{equation*}
    \overline\bX - \tilde\Vb = \bX \Hb - \tilde{\Vb} = \Wb \tilde{\Vb} (\tilde{\Vb}^{\top}\tilde{\Vb})^{-1} + \boldsymbol{\Phi}_{X},
\end{equation*}
where the term $\boldsymbol{\Phi}_{X}$ is
\begin{equation}\label{residual_coef_matrix}
\begin{split}
     \bPhi_{X} = &\Wb [\bX (\bX^{\top}\bX)^{-1}\Hb - \tilde{\Vb} (\tilde{\Vb}^{\top}\tilde{\Vb})^{-1}] + \tilde{\Vb} [\tilde\Vb^{\top} \bX (\bX^{\top}\bX)^{-1}\Hb - \Ib]\\
     = &\Wb [\overline{\bX} (\overline{\bX}^{\top}\overline{\bX})^{-1} - \tilde{\Vb} (\tilde{\Vb}^{\top}\tilde{\Vb})^{-1}] + \tilde{\Vb} [\tilde\Vb^{\top} \overline{\bX} (\overline{\bX}^{\top}\overline{\bX})^{-1} - \Ib].
\end{split}
\end{equation}
Recall the definition of $\Eb_S$ in Section \ref{sec:proof_of_sec221}, we have \zx{
\begin{equation*}
    \Eb_S(w,w') = \sum_{i=1}^n\Big(\frac{X_{i,w,w'}^{[q]} - N_{i,w,w'}^{[q]}}{n N_{w,w'}^{[q]}} - \frac{X_{i,w}^{[q]}-N_{i,w}^{[q]}}{nN_w^{[q]}} - \frac{X_{i,w'}^{[q]}-N_{i,w'}^{[q]}}{nN_{w'}^{[q]}} \Big), \quad 1 \leq w,w' \leq d.
\end{equation*}}
We use $\Eb_S$ to construct the leading term of $\Xb\Xb^{\top} - \tilde{\Vb}\tilde\Vb^{\top}$:
\begin{equation}\label{eq:low_rank_decom}
    \begin{split}
        \Xb \Xb^{\top} - \tilde{\Vb}\tilde{\Vb}^{\top} =& (\overline{\bX} - \tilde{\Vb})\tilde{\Vb}^{\top} + \overline{\bX} (\overline{\bX}-\tilde{\Vb})^{\top} \\
        =& (\overline{\bX} - \tilde{\Vb})\tilde{\Vb}^{\top} + \tilde{\Vb} (\overline{\bX}-\tilde{\Vb})^{\top}+(\overline{\bX}-\tilde{\Vb})(\overline{\bX}-\tilde{\Vb})^{\top}\\
        =& \Wb \U \U^{\top} + \U \U^{\top} \Wb + \bPhi_{X} \tilde\Vb^{\top} + \tilde{\Vb}\bPhi_{X}^{\top} + (\overline{\bX}-\tilde{\Vb})(\overline{\bX}-\tilde{\Vb})^{\top}\\
        =& \bTheta + \Rb
    \end{split}
\end{equation}
where $\bTheta = \Eb_S \Ub \Ub^{ \top} + \Ub \Ub^{\top} \Eb_S$ and error matrix
$$
\Rb = (\Wb-\Eb_S)\U \U^{\top} + \U \U^{\top}(\Wb - \Eb_S) +\bPhi_{X} \tilde\Vb^{\top} + \tilde{\Vb}\bPhi_{X}^{\top} + (\overline{\bX}-\tilde{\Vb})(\overline{\bX}-\tilde{\Vb})^{\top}.
$$
Recall the definition of $\Pb^{\star} = [P^{\star}_{ww'}]_{1\leq w,w'\leq d} = \Ub \Ub^{\top}$. We can write the leading term as $\bTheta =\Eb_S\Pb^{\star} + \Pb^{\star}\Eb_S$. We prove Theorem \ref{thm2} by proving the following stronger theorem.

\begin{theorem}
Suppose Assumptions~\ref{assump:ndTp}-\ref{assump:eigen_bound} hold, then for the low-rank estimator $\tilde{\PMIbb} = \Xb\Xb^{\top}$, one can write:
\begin{equation*}
    \tilde{\PMIbb} - \alpha_p\Vb\Vb^{\top} = \bTheta + \Rb
\end{equation*}
where $\bTheta$ and $\Rb$ are defined as above.
And for any entry $(w,w') \in [d] \times [d],$ we have
\begin{equation*}
    \frac{1}{v_{w,w'}} \bTheta_{w,w'} \stackrel{d}{\longrightarrow} N(0, 1)  \text{ as } n,d,T,p \rightarrow \infty,
\end{equation*}
where $v_{w,w'}^2 := \sum_{w'' = 1}^d \text{Var}(A_{i,w''}^{(w,w')} + A_{i,w''}^{(w',w)})$, and \zx{
\begin{equation*}
    A_{i,w''}^{(w,w')} := \frac{\Pb_{w w^{\prime \prime}}^{\star}\left(X_{i, w^{\prime \prime}, w^{\prime}}^{[q]}-N_{i,w^{\prime \prime}, w^{\prime}}^{[q]}\right)}{n N_{w^{\prime \prime}, w^{\prime}}^{[q]}}-\frac{\Pb_{w w^{\prime \prime}}^{\star}\left(X_{i, w^{\prime \prime}}^{[q]}-N_{i,w^{\prime \prime}}^{[q]}\right)}{n N_{w^{\prime \prime}}^{[q]}}-\frac{\Pb_{w w^{\prime \prime}}^{\star}\left(X_{i, w^{\prime}}^{[q]}-N_{i,w^{\prime}}^{[q]}\right)}{n N_{w^{\prime}}^{[q]}},
\end{equation*}}
and the residual obeys $R(w,w') = o(v_{w,w'})$ with probability exceeding $1 - \exp(-\Omega(\log^2 d))$.
\end{theorem}
\begin{proof}
Denote $B_i^{(w,w')} = \sum_{w'' = 1}^d (A_{i,w''}^{(w,w')}+A_{i,w''}^{(w',w)}).$ Then $\E[B_i^{(w,w')}] = 0$. By the proof of Lemma \ref{lem_theta_var_lower_bound}, we have $v_{w,w'}^2 = \Omega(\frac{p^4}{nT})$, $v_{w,w'}^2 = O(\frac{dp^2}{nT\xi^4})$, and 
\begin{equation*}
    \text{Var}(B_i) = v_{w,w'}^2 + o(\frac{p^4}{nT}).
\end{equation*}
Since $B_i^{(w,w')}$ is a bounded random variable, applying the central limit theorem and Slutsky's theorem, we have
    \begin{equation*}
        \frac{\sqrt{n}}{v_{w,w'}}\bTheta_{w,w'} = \sqrt{\frac{\text{Var}(B_i^{w,w'})}{v_{w,w'}^2}}\cdot \frac{1}{\sqrt{n\text{Var}(B_i^{w,w'})}}\sum_{i=1}^n B_i^{(w,w')} \stackrel{d}{\longrightarrow} N(0, 1)  \text{ as } n,d,T,p \rightarrow \infty.
    \end{equation*}
By Lemma \ref{lem_theta_var_lower_bound} and (\ref{eq:B67}), we know that $\big|\be_w^{\top}\boldsymbol{\Phi}_X \tilde\Vb^{\top} \be_{w'}\big| = o(\sqrt{\text{Var}(\bTheta_{w,w'})})$ with probability exceeding $1 - \exp(-\Omega(\log^2 d))$. By (\ref{eq:B62}), we have
 \begin{equation}\label{eq:B68}
     \|(\overline{\Xb} - \tilde{\Vb})(\overline{\Xb} - \tilde{\Vb})^{\top}\| \leq \|\overline{\Xb} - \tilde{\Vb}\|^2 \lesssim \frac{\|\Wb\|^2}{\kappa^2 \xi^6} = o(\frac{p^2}{\sqrt{nT}})
 \end{equation}
with probability exceeding $1 - \exp(-\Omega(\log^2 d))$, where the last inequality comes from Assumption \ref{assump:eigen_bound}. Lastly, by the proof of Lemma \ref{thm1}, we have
\begin{equation}\label{eq:B58}
    \Eb_{w,w'} = \Eb_S(w,w') + o_p\Big(\Big(\frac{d^2}{nT}\Big)^{\frac{1}{2}+\alpha}\Big), \forall \alpha \in (0,\frac{1}{2}).
\end{equation}
Let $\alpha = \frac{1}{4}$. Then we have
\begin{equation}\label{eq:B69}
   \begin{split}
        \|\Wb - \Eb_S\|_{2,\infty} &\leq \|\Wb - \Eb\|_{2,\infty} + \|\Eb_S - \Eb\|_{2,\infty} \leq \|\PMIbb - \tilde{\Vb}\tilde{\Vb}^{\top}\|_{2,\infty} + 
        \sqrt{d}\|\Eb_S - \Eb\|_{\max}\\
        &= O(\frac{\kappa^4 p^{2}}{d^{3/2}}) + O_p((\frac{d^2}{nT})^{\frac{3}{4}}) = O(\frac{\sqrt{d}p}{\sqrt{nT}})+ o_p(\frac{d^{3/2}}{(nT)^{3/4}}) = o_p(\frac{\sqrt{d} p^{1.1}}{\sqrt{nT}}),
   \end{split}
\end{equation}
where the third inequality uses Theorem \ref{thm0:PMI_lowrank} and (\ref{eq:B58}). and the last inequality is from Assumption \ref{assump:eigen_bound}. Combine (\ref{eq:B68}), (\ref{eq:B69}) and (\ref{eq:B67}), we have that with probability $1 - o(1)$,
\begin{equation*}
    \begin{split}
        |\Rb_{w,w'}| &\leq 2|\be_w^{\top}(\Wb-\Eb_S)\Pb^{\star}\be_{w'}|  + 2|\be_{w'}^{\top}\boldsymbol{\Phi}_X \tilde\Vb^{\top} \be_{w}| + \|(\bar{\Xb} - \tilde{\Vb})(\bar{\Xb} - \tilde{\Vb})^{\top}\| \\
        &\leq 2\|\Wb-\Eb_S\|_{2,\infty}\cdot \|\Ub\|_{2,\infty} + o(\frac{p^2}{\sqrt{nT}}) = o(\frac{p^{2}}{\sqrt{nT}})  = o(v_{w,w'}),
    \end{split}
\end{equation*}
where the first equation uses Assumption \ref{assump1}. Thus we know that in the decomposition (\ref{eq:low_rank_decom}), $\bTheta$ is the leading term, which completes the proof.
\end{proof}

\begin{lemma}\label{lem_theta_var_lower_bound}
    Suppose Assumptions \ref{assump:ndTp}-\ref{assump:eigen_bound} hold. Then for any code pair $(w,w')$, we have the following lower and upper bound:
    \begin{equation*}
        \text{Var}(\bTheta_{w,w'}) \gtrsim \frac{p^4}{nT}; \quad \text{Var}(\bTheta_{w,w'}) \lesssim \frac{d p^2}{nT\xi^4}.
    \end{equation*}
\end{lemma}
\begin{proof}
    From the definition of $\bTheta$, we have the expansion \zx{
    \begin{equation*}
        \begin{split}
            \bTheta_{w,w'} &= \be_w^{\top}\big( \Eb_S \Ub \Ub^{\top} + \Ub \Ub^{\top} \Eb_S \big)\be_{w'} = \sum_{w''=1}^d (\Pb^{\star}_{ww''}\Eb_{S}(w'',w') + \Pb^{\star}_{w'w''}\Eb_S(w'',w))\\
            &= \frac{1}{n}\sum_{i=1}^n\sum_{w''}\Big(\frac{\Pb^{\star}_{ww''}(X_{i,w'',w'}^{[q]} - N_{i,w'',w'}^{[q]})}{N_{w'',w'}^{[q]}} - \frac{\Pb^{\star}_{ww''}(X_{i,w''}^{[q]}-N_{i,w''}^{[q]})}{N_{w''}^{[q]}} - \frac{\Pb^{\star}_{ww''}(X_{i,w'}^{[q]}-N_{i,w'}^{[q]})}{N_{w'}^{[q]}} \Big)\\
            &+\frac{1}{n}\sum_{i=1}^n\sum_{w''}\Big(\frac{\Pb^{\star}_{w'w''}(X_{i,w'',w}^{[q]} - N_{i,w'',w}^{q})}{N_{w'',w}^{q}} - \frac{\Pb^{\star}_{w'w''}(X_{i,w''}^{[q]}-N_{i,w''}^{[q]})}{N_{w''}^{[q]}} - \frac{\Pb^{\star}_{w'w''}(X_{i,w}^{[q]}-N_{i,w}^{[q]})}{N_{w}^{[q]}} \Big)
        \end{split}
    \end{equation*}}
We first consider the circumstance when $w = w'$. To simplify the notation, we further denote $a_{w''} = \Pb^{\star}_{ww''}/(nN_{w'',w}^{[q]}), b_{w''} = \Pb^{\star}_{ww''}/(nN_{w''}^{[q]})$. By Lemma \ref{lem:B37} and \ref{lem:B36}, we have $a_{w''}/b_{w''} = N_{w''}^{[q]}/N_{w'',w}^{[q]} \sim d$.
Given $\{\bc_t\}$ in $\mathcal{P}$, which is a set of discourse vectors defined in Lemma \ref{lem:coccur_lower_bound_variance}, we have
\begin{equation}
    \begin{split}
        \text{Var}(\bTheta_{w,w'}) &= \text{Var}(2\sum_{i=1}^n\sum_{w''}(a_{w''}X_{i,w'',w}^{[q]} - b_{w''}X_{i,w''}^{[q]}-b_w X_{i,w}^{[q]}))\\
        &=4\sum_{i=1}^n \text{Var}(\sum_{w''}(a_{w''}X_{i,w'',w}^{[q]} - b_{w''}X_{i,w''}^{[q]}-b_w X_{i,w}^{[q]}))
        \\ &=4\sum_{i=1}^n\Big[
         \sum_{w''}\text{Var}(a_{w''}X_{i,w'',w}^{[q]} - b_{w''}X_{i,w''}^{[q]}-b_w X_{i,w}^{[q]}) \\
       &\quad + \sum_{w'' \neq w'''}\text{Cov}(a_{w''}X_{i,w'',w}^{[q]} - b_{w''}X_{i,w''}^{[q]}-b_w X_{i,w}^{[q]}, a_{w'''}X_{i,w''',w}^{[q]} - b_{w'''}X_{i,w'''}^{[q]}-b_w X_{i,w}^{[q]})  \Big].
    \end{split}
\end{equation}
\begin{equation}\label{eq:var_decom}
    \begin{split}
    \frac{1}{4n}\text{Var}(\bTheta_{w,w}) &= \text{Var}\big( \sum_{w''}(a_{w''}X_{w'',w}^{[q]} - b_{w''}X_{w''}^{[q]} - b_{w}X_{w}^{[q]} ) \big)\\
    &= \sum_{w''}\text{Var}\big(a_{w''}X_{w'',w}^{[q]} - b_{w''}X_{w''}^{[q]} - b_{w}X_{w}^{[q]} \big) \\
    &+ \sum_{w'' \neq w'''}\text{Cov}\big(a_{w''}X_{w'',w}^{[q]} - b_{w''}X_{w''}^{[q]} - b_{w}X_{w}^{[q]}, a_{w'''}X_{w''',w}^{[q]} - b_{w'''}X_{w'''}^{[q]} - b_{w}X_{w}^{[q]} \big).
    \end{split}
\end{equation}
By the proof of Lemma \ref{thm1}, we have \zx{$\text{Var}(X_{i,w'',w}^{[q]}/N_{i,w'',w}^{[q]}) \sim d^2/T_i$ and $\text{Var}(X_{i,w}^{[q]}/N_{i,w}^{[q]}) = o (d^2/T_i)$. It follows that
\begin{equation}\label{eq:order_of_var}
    \begin{split}
        \sum_{i=1}^n\sum_{w''}\text{Var}\left(a_{w''}X_{i,w'',w}^{[q]} - b_{w''}X_{i,w''}^{[q]} - b_{w}X_{i,w}^{[q]} \right) &= \frac{1}{n^2}\sum_{i=1}^n\sum_{w''}\Pb_{ww''}^{\star 2}\text{Var}\left(\frac{X_{i,w'',w}^{[q]}}{N_{i,w'',w}^{[q]}} - \frac{X_{i,w''}^{[q]}}{N_{i,w''}^{[q]}} - \frac{X_{i,w}^{[q]}}{N_{i,w}^{[q]}} \right)\\
        &\sim \frac{d^2}{n^2 T^2}\sum_{i=1}^n T_i\sum_{w''}\Pb^{\star 2}_{ww''} \sim \frac{d^2}{nT}\sum_{w''} \Pb^{\star 2}_{ww''}.
    \end{split}
\end{equation}}
For the covariance terms, we first estimate the lower and upper bound of the covariance of two co-occurrences. For simplicity, we omit the subscript \(i\) when only analyzing one patient. Given $\{\bc_t\} \in \mathcal{P}$, we have
\begin{equation}\label{eq:decom_of_covXww'}
    \begin{split}
        \text{Cov}(X_{w'',w}^{[q]}, X_{w''',w}^{[q]}\mid \{\bc_t\}) &= \text{Cov}\big(\sum_{u=1}^q\sum_{t=1}^{T-u}X_{w''}(t)X_{w}(t+u), \sum_{v=1}^q\sum_{s=1}^{T-v}X_{w'''}(s)X_{w}(s+v) \mid \{\bc_t\}\big)\\
        &+ \text{Cov}\big(\sum_{u=1}^q\sum_{t=1}^{T-u}X_{w}(t)X_{w''}(t+u), \sum_{v=1}^q\sum_{s=1}^{T-v}X_{w'''}(s)X_{w}(s+v) \mid \{\bc_t\}\big)\\
        &+ \text{Cov}\big(\sum_{u=1}^q\sum_{t=1}^{T-u}X_{w''}(t)X_{w}(t+u), \sum_{v=1}^q\sum_{s=1}^{T-v}X_{w}(s)X_{w'''}(s+v) \mid \{\bc_t\}\big)\\
        &+ \text{Cov}\big(\sum_{u=1}^q\sum_{t=1}^{T-u}X_{w}(t)X_{w''}(t+u), \sum_{v=1}^q\sum_{s=1}^{T-v}X_{w}(s)X_{w'''}(s+v) \mid \{\bc_t\}\big)
    \end{split}
\end{equation}
For the first term, we have 
\begin{equation*}
    \begin{split}
        &\text{Cov}\big(\sum_{u=1}^q\sum_{t=1}^{T-u}X_{w''}(t)X_{w}(t+u), \sum_{v=1}^q\sum_{s=1}^{T-v}X_{w'''}(s)X_{w}(s+v) \mid \{\bc_t\} \big) \\
        &=\sum_{u=1}^q\sum_{v=1}^q\sum_{t=1}^{T-u}\sum_{s=1}^{T-u}\text{Cov}(X_{w''}(t)X_w(t+u),X_{w'''}(s)X_w(s+v)|\{\bc_t\})\\
        &\geq -\sum_{u=1}^q\sum_{v=1}^q\sum_{t=1}^T (p_{w''}(t)p_{w}(t+u)p_{w'''}(t)p_w(t+v) + p_{w''}(t)p_{w}(t+u)p_{w'''}(t+u)p_w(t+u+v) \\
        &+ p_{w''}(t)p_{w}(t+u)p_{w'''}(t-v)p_w(t) ) \sim -\frac{T}{d^4};
    \end{split}
\end{equation*}
Here we let the probability $p_{w}(t) = 0$ for any $1\leq w \leq d, t>T$. The order of this lower bound is obtained according to the definition of the set $\mathcal{P}_{w,w'}$. (See the definition in Lemma \ref{lem:coccur_lower_bound_variance}). On the other side, we have that for $w'' \neq w'''$,
\begin{equation*}
\begin{split}
    &\text{Cov}\big(\sum_{u=1}^q\sum_{t=1}^{T-u}X_{w''}(t)X_{w}(t+u), \sum_{v=1}^q\sum_{s=1}^{T-v}X_{w'''}(s)X_{w}(s+v) \mid \{\bc_t\} \big) \\
    &=\sum_{u=1}^q\sum_{v=1}^q\sum_{t=1}^{T-u}\sum_{s=1}^{T-u}\text{Cov}(X_{w''}(t)X_w(t+u),X_{w'''}(s)X_w(s+v)|\{\bc_t\})\\
    &\lesssim \sum_{u=1}^q\sum_{v=1}^q\sum_{t=1}^{T-u} p_{w''}(t)p_w(t+u)p_{w'''}(t+u-v)    \lesssim \frac{T}{d^3}.
\end{split}
\end{equation*}
Applying the same analysis procedure for the other three terms in \eqref{eq:decom_of_covXww'}, we have
\begin{equation}\label{ineq:lower_bound_1}
    \text{Cov}\big(X_{w'',w}^{[q]}, X_{w''',w}^{[q]}\mid \{\bc_t\}\big) \gtrsim -\frac{T}{d^4},
\end{equation}
and
\begin{equation}\label{ineq:S55}
   \text{Cov}\big(X_{w'',w}^{[q]}, X_{w''',w}^{[q]}\mid \{\bc_t\}\big) \lesssim  \frac{T}{d^3} , \quad w'' \neq w'''.
\end{equation}
Following the same analysis, we have $\text{Cov}(X_{w,w'}^{[q]}, X_{w'',w'''}^{[q]}\mid \{\bc_t\}) \gtrsim -\frac{T}{d^4}$ for any  $w,w',w'',w'''$. Then we estimate the lower and upper bound of the covariance of co-occurrence and occurrence. Given $\{\bc_t\},\{\bc_t\} \in \mathcal{P}_{w,w'}$, we have that for $w'' \neq w'''$,
\begin{equation*}
        \text{Cov}\big(X_{w'',w}^{[q]}, X_{w'''}^{[q]}\mid \{\bc_t\}\big) = \text{Cov}\Big(\sum_{u=1}^q\sum_{t=1}^{T-u}X_{w''}(t)X_{w}(t+u), 2q\sum_{s=1}^{T}X_{w'''}(s) - \delta_{w'''} \mid \{\bc_t\}\Big).
\end{equation*}
where the equation is from \eqref{eq:Xwq_to_Xw}. If $w''' \neq w$, then $\text{Cov}(X_{w'',w}^{[q]}, X_{w'''}^{[q]}\mid \{\bc_t\}) < 0$. If $w''' = w$, we have
\begin{equation}\label{ineq:lower_bound_2}
       \text{Cov}\Big(\sum_{u=1}^q\sum_{t=1}^{T-u}X_{w''}(t)X_{w}(t+u), X_{w'''}^{[q]} \mid \{\bc_t\}\Big) \lesssim \sum_{u=1}^q\sum_{t=1}^{T-u}p_{w''}(t)p_{w}(t+u) \sim \frac{T}{d^2},
\end{equation}
where the inequality uses the property that the conditional covariance $\text{Cov}(X_{w'',w}^{[q]},\delta_{w'''})$ is at most a constant. Similarly we have
\begin{equation}\label{ineq:S57}
    \text{Cov}\Big(\sum_{u=1}^q\sum_{t=1}^{T-u}X_{w''}(t)X_{w}(t+u), X_{w'''}^{[q]} \mid \{\bc_t\}\Big) \gtrsim -\sum_{u=1}^q\sum_{t=1}^{T-u}p_{w''}(t)p_{w}(t+u)p_{w'''}(t) \sim - \frac{T}{d^3}.
\end{equation}
Lastly, we estimate the bounds of the covariance between occurrences. Given $\{\bc_t\},\{\bc_t\} \in \mathcal{P}_{w,w'}$, we have that for $w'' \neq w'''$,
\begin{equation}\label{ineq:lower_bound_3}
        \begin{split}
            \text{Cov}(X_{w''}^{[q]}, X_{w'''}^{[q]}\mid \{\bc_t\}) &= 4q^2\text{Cov}\Big(\sum_{t=1}^{T}X_{w''}(t) - \frac{\delta_{w''}}{2q}, \sum_{s=1}^{T}X_{w'''}(s) - \frac{\delta_{w'''}}{2q}\mid \{\bc_t\}\Big)\\
            &\sim  \text{Cov}\Big(\sum_{t=1}^{T}X_{w''}(t), \sum_{s=1}^{T}X_{w'''}(s)\mid \{\bc_t\}\Big) = 
            - \sum_{t=1}^T p_{w''}(t)p_{w'''}(t) \sim -\frac{T}{d^2}.
        \end{split}
\end{equation}
Back to (\ref{eq:var_decom}), combining the results in (\ref{eq:order_of_var}),(\ref{ineq:lower_bound_1}), (\ref{ineq:lower_bound_2}), and (\ref{ineq:lower_bound_3}), we have
\zx{
\begin{equation}\label{eq:var_theta}
    \begin{split}
        \text{Var}(\bTheta_{w,w}) &\gtrsim \frac{d^2}{nT}\sum_{w''}\Pb^{\star 2}_{ww''} - \sum_{i=1}^n\Big(\sum_{w'' \neq w'''} a_{w''}a_{w'''}\frac{T_i}{d^4} + \sum_{w''}a_{w''}b_{w}\frac{T_i}{d^2} + \sum_{w'' \neq w'''} b_{w''}b_{w'''}\frac{T_i}{d^2}\Big)\\
        &\sim \frac{d^2}{nT}\sum_{w''}\Pb^{\star 2}_{ww''} - \frac{1}{nT}\Big(\sum_{w'' \neq w'''}\Pb^{\star}_{ww''}\Pb^{\star}_{ww'''} + \sum_{w''}\Pb^{\star}_{ww''}\Pb^{\star}_{ww} \Big).
    \end{split}
\end{equation}
Combining \eqref{eq:order_of_var}, \eqref{ineq:S55}, \eqref{ineq:S57} and \eqref{ineq:lower_bound_3}, we have
\begin{equation}\label{ineq:S60}
    \begin{split}
        \text{Var}(\bTheta_{w,w}) &\lesssim \frac{d^2}{nT}\sum_{w''}\Pb^{\star 2}_{ww''} + \sum_{i=1}^n \Big(\sum_{w'' \neq w'''} a_{w''}a_{w'''}\frac{T_i}{d^3} + \sum_{w''}a_{w''}b_{w}\frac{T_i}{d^3}\Big)\\
        &\sim \frac{d^2}{nT}\sum_{w''}\Pb^{\star 2}_{ww''} + \frac{d}{nT}\Big(\sum_{w'' \neq w'''}\Pb^{\star}_{ww''}\Pb^{\star}_{ww'''} + \frac{1}{d}\sum_{w''}\Pb^{\star}_{ww''}\Pb^{\star}_{ww} \Big).
    \end{split}
\end{equation}}
By Assumption \ref{assump1}, we have
\begin{equation*}
    \Pb^{\star 2}_{ww''} = (\be_w^{\top} \Ub \Ub^{\top} \be_{w''})^2 \leq \|\Ub\|_{2,\infty}^4 \lesssim  \frac{p^2}{\xi^4 d^2};
\end{equation*}
\begin{equation}\label{eq:var_theta_term1}
    \sum_{w''}\Pb^{\star 2}_{ww''} = \|\Ub_{w,\cdot} \Ub^{\top}\|^2 = \|\Ub_{w,\cdot}\|^2\gtrsim \frac{p^4}{d^2};
\end{equation}
\begin{equation*}
    \sum_{w''}\Pb^{\star 2}_{ww''} = \|\Ub_{w,\cdot}\|^2 \lesssim \frac{p}{d}.
\end{equation*}
By Assumption \ref{assump:ndTp} and \ref{assump:eigen_bound}, we have
\begin{equation}\label{eq:var_theta_term2}
    \sum_{w'' \neq w'''}\Pb^{\star}_{ww''}\Pb^{\star}_{ww'''} + \sum_{w''}\Pb^{\star}_{ww''}\Pb^{\star}_{ww} \leq d^2 \max_{w,w'}(\Pb^{\star}_{w,w'})^2= O(\frac{p^2}{\xi^4}) = o(p^4).
\end{equation}
Combining (\ref{eq:var_theta}), (\ref{eq:var_theta_term2}), and (\ref{eq:var_theta_term1}), we have $ \text{Var}(\bTheta_{w,w}) \gtrsim \frac{p^4}{nT}$. Combining \eqref{ineq:S60}, (\ref{eq:var_theta_term2}), and (\ref{eq:var_theta_term1}), we have $ \text{Var}(\bTheta_{w,w}) \lesssim \frac{d p^2}{nT\xi^4}$. It further yields that
\[
\text{Var}(\bTheta_{w,w'}) \lesssim \text{Var}(\bTheta_{w,w}) + \text{Var}(\bTheta_{w',w'}) \lesssim \frac{d p^2}{nT\xi^4}.
\]
Then we consider the circumstance when $w \neq w'$. Following similar procedure to (\ref{ineq:lower_bound_1}), we have
\zx{
\begin{equation}\label{ineq:lower_bound_4}
    \text{Cov}\left(X_{i,w'',w}^{[q]}, X_{i,w''',w'}^{[q]}\mid \{\bc_t\}\right) \gtrsim -\frac{T_i}{d^4}
\end{equation}
for $w \neq w'$.} Combining the results in (\ref{eq:order_of_var}),(\ref{ineq:lower_bound_2}), (\ref{ineq:lower_bound_3}), and (\ref{ineq:lower_bound_4}), we have
\begin{equation}\label{ineq:B.57}
    \begin{split}
        \text{Var}(\bTheta_{w,w'}) &\gtrsim \frac{d^2}{nT}\sum_{w''}(\Pb^{\star 2}_{ww''} + \Pb^{\star 2}_{w'w''})- \frac{1}{nT}\sum_{w'' \neq w'''} (\Pb^{\star}_{ww''} + \Pb^{\star}_{w'w''})(\Pb^{\star}_{ww'''} + \Pb^{\star}_{w'w'''})  \\
        &- \frac{1}{nT}\sum_{w''}(\Pb^{\star}_{ww''}+\Pb^{\star}_{w'w''})\Pb^{\star}_{w'w}.
    \end{split}
\end{equation}
Similar to (\ref{eq:var_theta_term2}), we have
\begin{equation*}
    \sum_{w'' \neq w'''} (\Pb^{\star}_{ww''} + \Pb^{\star}_{w'w''})(\Pb^{\star}_{ww'''} + \Pb^{\star}_{w'w'''}) +\sum_{w''}(\Pb^{\star}_{ww''}+\Pb^{\star}_{w'w''})\Pb^{\star}_{w'w} = o(d).
\end{equation*}
By Assumption \ref{assump1}, we have
\begin{equation*}
    \sum_{w''}(\Pb^{\star 2}_{ww''} + \Pb^{\star 2}_{w'w''}) \gtrsim \frac{p^4}{d^2}.
\end{equation*}
Then it follows that $\text{Var}(\bTheta_{w,w'}) \gtrsim p^4/nT$ by \eqref{ineq:B.57}.

\end{proof}

\subsection{Upper Bound of Residual Term}
We first introduce the Davis-Kahan sin $\bTheta$ theorem mentioned in \cite{chen2020sp}:
\begin{lemma}[Corollary 2.2.2. in \cite{chen2020sp}]\label{lem:wedin_sin}
Suppose Assumptions \ref{assump:ndTp}-\ref{assump:eigen_bound} hold. If $\|\Wb\|<(1-1 / \sqrt{2}) \kappa^2\xi^2$, then we have
$$
\operatorname{dist}(\hat{\Ub}, \Ub) \leq \frac{2\|\Wb\|}{\kappa^2\xi^2}.
$$
\end{lemma}
Recall that we denote $\Hb$ to be the orthogonal matrix that best aligns $\hat{\Ub}$ to $\Ub$. According to Lemma \ref{lem:wedin_sin} and Assumption \ref{assump1},
\begin{equation}\label{eq:dist_U2}
    \operatorname{dist}(\hat{\Ub}, \Ub) \leq \frac{2\|\Wb\|}{\kappa^2 \xi^2}.
\end{equation}
According to Theorem \ref{Concentration of empirical PMI to stationary PMI}, with probability at least $1 - \exp(-\Omega(\log^2 d))$,
$$
\|\Eb\| \leq 3\frac{d^3 p \log^2 d}{\sqrt{n T}}.
$$
By Theorem \ref{thm0:PMI_lowrank}, we have
\begin{equation*}
    \|\PMIbb - \tilde{\Vb}\tilde{\Vb}^{\top} \| \leq d\|\PMIbb - \tilde{\Vb}\tilde{\Vb}^{\top} \|_{\max} \leq C \frac{\kappa^4 p^{2} }{d},
\end{equation*}
and for the eigenvalues of $\PMIbb$, we have
\begin{equation}\label{ineq:PMI_eigenv}
    \begin{split}
        &\lambda_{1}(\PMIbb) \leq 2 \kappa^2,
    \lambda_{p}(\PMIbb) \geq  \frac{\kappa^2 \xi^2}{2},\\
    &|\lambda_{p+k}(\PMIbb)| \leq C \frac{\kappa^4 p^{2} }{d} < \lambda_p(\PMIbb), 1 \leq k \leq d-p,
    \end{split}
\end{equation}
for some constant $C>0.$ We can see from (\ref{ineq:PMI_eigenv}) that $\sigma_i(\PMIbb) = \lambda_i(\PMIbb), 1\leq i\leq p$ under Assumption \ref{assump:eigen_bound}. Then by Weyl's inequality,
\begin{equation}\label{ineq:pmi_hat_ev1}
    \lambda_1(\hat{\PMIbb}) \leq \lambda_1(\PMIbb) + \|\Eb\| \leq 3 \kappa^2;
\end{equation}
where the last inequality is due to $\kappa^2 = \Omega(\frac{d^3p^3\log^5 d}{\sqrt{nT}\xi^2})$, which comes from Assumption \ref{assump:eigen_bound}. Similarly we have
\begin{equation}\label{ineq:pmi_hat_evp}
    \lambda_{p}(\hat{\PMIbb}) \geq \lambda_{p}(\PMIbb) - \|\Eb\| \geq \frac{\kappa^2\xi^2}{3};
\end{equation}
\begin{equation}\label{ineq:B.66}
    |\lambda_{p+k}(\hat{\PMIbb})| \leq |\lambda_{p+k}(\PMIbb)| + \|\Eb\| \lesssim \frac{d^3 p^2 \log^4 d}{\sqrt{n T}}.
\end{equation}
Thus we can also see that $\sigma_i(\hat{\PMIbb}) = \lambda_i(\hat{\PMIbb}), 1\leq i\leq p$. We also have the bound for the operator norm of $\Wb$:
\begin{equation}\label{ineq:E_tilde_norm}
    \|\Wb\| \leq \|\Eb\| + \|\PMIbb - \tilde{\Vb}\tilde{\Vb}^{\top} \| \leq \frac{4d^3 p \log^2 d}{\sqrt{n T}}
\end{equation}
with probability at least $1 - \exp(-\Omega(\log^2 d))$. Here the last inequality is from Assumption \ref{assump:eigen_bound}.
Denote $\overline{\bLambda}^{1/2} = \Hb^{\top} \hat\bLambda^{1/2} \Hb, \text{ and }\overline{\Ub} = \hat{\Ub}\Hb$. Then we have
\begin{equation}\label{ineq:B.65}
    \begin{split}
        \|\overline{\bLambda}^{1/2} - \bLambda^{\star 1/2}\| &\leq \frac{\|\Hb^{\top}\hat\bLambda\Hb - \bLambda^{\star}\|}{\sigma_p(\hat\bLambda^{\frac{1}{2}})+\sigma_p(\bLambda^{\star \frac{1}{2}})} \lesssim \frac{1}{\kappa\xi}\|\Hb^{\top}\hat{\Ub}^{\top}\bX\bX^{\top}\hat{\Ub}\Hb - \Ub^{\top}\tilde{\Vb}\tilde{\Vb}^{\top}\Ub\| \\
        &\lesssim \frac{1}{\kappa\xi}\Big(\|\overline{{\Ub}}^{\top}(\bX\bX^{\top} - \hat{\PMIbb}) \overline{{\Ub}}\| +\|(\overline{{\Ub}} - \Ub)^{\top}\tilde{\Vb}\tilde{\Vb}^{\top} \overline{{\Ub}}\| \\
        &\quad + \| \Ub^{\top}\tilde{\Vb}\tilde{\Vb}^{\top}(\overline{{\Ub}} - \Ub)\| + \|\overline{{\Ub}}^{\top}\Wb\overline{{\Ub}} \|\Big) \\
        &\lesssim \frac{1}{\kappa\xi}\big(\sigma_{p+1}(\hat{\PMIbb}) + 2\frac{\|\Wb\|}{\xi^2} + \|\Wb\| \big) \lesssim \frac{\|\Wb\|}{\kappa\xi^3},
    \end{split}
\end{equation}
where the first inequality uses Lemma 2.2 in \cite{SCH92}; the second inequality comes from (\ref{ineq:PMI_eigenv}) and (\ref{ineq:pmi_hat_evp}); the third inequality uses triangle inequality; the fourth inequality uses $\|\Xb\Xb^{\top} - \hat{\PMIbb}\| = \sigma_{p+1}(\hat{\PMIbb})$ and \eqref{eq:dist_U2}; and the last inequality uses the fact that  $\lambda_i (\hat{\PMIbb}) = \sigma_i(\hat\PMIbb), 1\leq i \leq p$ and \eqref{ineq:B.66}.
By Lemma \ref{lem:wedin_sin} and Lemma B.24 in \cite{CLL2020}, we have
\begin{equation}\label{eq:B62}
\begin{split}
    \|\overline{\bX} - \tilde{\Vb}\| &\leq \| (\hat{\Ub}\Hb - \Ub)\overline{\bLambda}^{\frac{1}{2}} \| + \| \Ub (\overline{\bLambda}^{\frac{1}{2}} - \bLambda^{\star \frac{1}{2}}) \|\\
    & \lesssim \frac{\|\Wb\|}{\kappa^2\xi^2} \sqrt{\lambda_1(\hat{\PMIbb})} + \frac{\|\Wb\|}{\kappa\xi^3} \lesssim \frac{\|\Wb\|}{\kappa\xi^3},
\end{split}
\end{equation}
where the second inequality comes from \eqref{eq:dist_U2} and \eqref{ineq:B.65}; and the third inequality comes from (\ref{ineq:pmi_hat_ev1}).
By Theorem 3.3 in \cite{ste1977}, we have
\begin{equation}\label{ineq:B82}
       \| \overline{\bX} (\overline{\bX}^{\top}\overline{\bX})^{-1} - \tilde{\Vb} (\tilde\Vb^{\top}\tilde{\Vb})^{-1} || \lesssim \frac{1}{\kappa^2\xi^2} || \overline{\bX} - \tilde{\Vb} \| \lesssim \frac{1}{\kappa^2\xi^2} \frac{\|\Wb\|}{\kappa\xi^3},
\end{equation}
where the last inequality uses \eqref{eq:B62}. Combining (\ref{ineq:B82}) and Assumption \ref{assump1}, we have
\begin{equation}\label{ineq:B83}
        \| \tilde\Vb^{\top} \overline{\bX} (\overline{\bX}^{\top}\overline{\bX})^{-1} - \Ib \| \leq \|\tilde{\Vb} \| \cdot \|\overline{\bX} (\overline{\bX}^{\top}\overline{\bX})^{-1} - \tilde{\Vb} (\tilde{\Vb}^{\top}\tilde{\Vb})^{-1} \|  \lesssim \frac{\kappa}{\kappa\xi}\| \overline{\bX} - \tilde{\Vb} \| \lesssim \frac{\|\Wb\|}{\kappa\xi^4}.
\end{equation}
Below we will bound the norm of $\boldsymbol{\Phi}_X$ defined in (\ref{residual_coef_matrix}). Denote
$$
    \boldsymbol{\Phi}_1 = \Wb [\overline{\bX}(\overline{\bX}^{\top}\overline{\bX})^{-1} - \tilde{\Vb} (\tilde\Vb^{\top}\tilde{\Vb} )^{-1}]; \quad \boldsymbol{\Phi}_2 = \tilde{\Vb}[\tilde\Vb^{\top}\overline{\bX}(\overline{\bX}^{\top}\overline{\bX})^{-1} - \Ib].
$$ 
Then we have that for each $j \in [d]$,
\begin{equation}\label{ineq:Phi_1}
        \|\be_j^{\top} \boldsymbol{\Phi}_1\| \lesssim \max\{\frac{1}{\kappa^2\xi^2}, \frac{1}{\lambda_p} \} \|\be_j^{\top}\Wb \| \cdot \|\overline{\bX} - \tilde{\Vb}\|  \lesssim  \frac{\|\Wb\|_{2,\infty}\cdot\|\Wb\|}{\kappa^3\xi^5},
\end{equation}
where the first inequality uses Lemma \ref{MP_inverse} and the second inequality uses \eqref{eq:B62} and \eqref{ineq:pmi_hat_evp}. Further we have
\begin{equation}\label{ineq:Phi_2}
        \| \be_j^{\top} \boldsymbol{\Phi}_2\| \lesssim \|\tilde\Vb\|_{2,\infty}\cdot \frac{\|\Wb\|}{\kappa\xi^4} \lesssim \frac{\sqrt{p}\|\Wb\|}{\sqrt{d}\xi^4},
\end{equation}
where the first inequality uses \eqref{ineq:B83}. Combining \eqref{ineq:Phi_1} and \eqref{ineq:Phi_2}, it yields that with probability at least $1 - \exp(-\Omega(\log^2 d))$, 
\begin{equation}\label{eq:B67}
    \begin{split}
        |\be_i^{\top}\boldsymbol{\Phi}_X \tilde\Vb^{\top} \be_j| &= |\be_i^{\top}(\boldsymbol{\Phi}_1 + \boldsymbol{\Phi}_2 )\tilde{\Vb}_{j \cdot}| \leq  (\| \be_j^{\top} \boldsymbol{\Phi}_1\| + \| \be_j^{\top} \boldsymbol{\Phi}_2\| )\cdot \|\tilde\Vb \|_{2,\infty}   \\
        &\lesssim \frac{\kappa\sqrt{p}}{\sqrt{d}}\big(\frac{\|\Wb\|_{2,\infty}\cdot\|\Wb\|}{\kappa^3\xi^5} +  \frac{\sqrt{p}\|\Wb\|}{\sqrt{d}\xi^4}\big) \\
        &\lesssim \frac{d^5 p^{5/2} \log^4 d}{\kappa^2 \xi^5 nT} + \frac{\kappa d^2p^2\log^2 d}{\sqrt{nT}\xi^4} = o(\frac{p^2}{\sqrt{nT}}),
    \end{split}
\end{equation}
 where the last inequality uses \eqref{ineq:E_tilde_norm} and the last equation is from Assumption \ref{assump:eigen_bound}.

\begin{lemma}\label{MP_inverse}
(Perturbation of pseudo-inverses). Let $\boldsymbol{A}^{\dagger}$ (resp. $\left.\boldsymbol{B}^{\dagger}\right)$ be the pseudo-inverse (i.e. Moore-Penrose inverse) of $\boldsymbol{A}$ (resp. $\boldsymbol{B})$. Then we have
$$
\left\|\boldsymbol{B}^{\dagger}-\boldsymbol{A}^{\dagger}\right\| \leq 3 \max \left\{\left\|\boldsymbol{A}^{\dagger}\right\|^{2},\left\|\boldsymbol{B}^{\dagger}\right\|^{2}\right\}\|\boldsymbol{B}-\boldsymbol{A}\|
$$
\end{lemma} 
\begin{proof}
    See Theorem 3.3 of \cite{ste1977}.
\end{proof}

\begin{proposition}\label{prop: est rank}
    Under the condition that ${d^3 p \log^2 d}/{\sqrt{n T}} \ll \kappa^2 \xi^2$, with probability at least $1 - \exp(-\Omega(\log^2 d))$, we have that $r = p$.
\end{proposition}
\begin{proof}
    By \eqref{ineq:E_tilde_norm}, we know that the following event holds with probability at least  $1 - \exp(-\Omega(\log^2 d))$:
    $$
    \cE =  \left\{ \|\Wb\| = \|\widehat\PMIbb - \tilde{\Vb}\tilde{\Vb}^{\top} \| \leq \frac{4d^3 p \log^2 d}{\sqrt{n T}}\right\}.
    $$
    Also since $\lambda_p \big(\tilde{\Vb}\tilde{\Vb}^{\top}\big) = \kappa^2 \xi^2$, by Weyl's inequality \citep{franklin2012matrix}, under the condition that ${d^3 p \log^2 d}/{\sqrt{n T}} \ll \kappa^2 \xi^2$, when the event $\cE$ holds we have that 
    \begin{align*}
        \lambda_k \big(\widehat\PMIbb\big) & \ge \lambda_k \big(\tilde{\Vb}\tilde{\Vb}^{\top}\big) - \|\Wb\| \ge \kappa^2 \xi^2 - \frac{4d^3 p \log^2 d}{\sqrt{n T}} \ge \frac{8d^3 p \log^2 d}{\sqrt{n T}} , \quad k \le p,\\
        \lambda_{k'} \big(\widehat\PMIbb\big) & \le \lambda_{k'} \big(\tilde{\Vb}\tilde{\Vb}^{\top}\big) + \|\Wb\| \le \frac{4d^3 p \log^2 d}{\sqrt{n T}} < \frac{8d^3 p \log^2 d}{\sqrt{n T}} , \quad k' > p.
    \end{align*}
    Therefore, the claim follows.
\end{proof}

\subsection{Covariance under Null Distribution}\label{sec:b.5}
Recall that in Section \ref{sec:proof_of_sec221}, we show that $\Wb$ is defined as $\Eb$ plus the bias term $\PMIbb - \alpha_p \Vb\Vb^{\top}$ and $\|\PMIbb - \alpha_p \Vb\Vb^{\top}\|_{\max} = O(\kappa^4 p^2/d^2) = o(\sqrt{\text{Var}(\Eb)})$ holds, which allows $\Wb$ and $\Eb$ to be estimated by the same estimator. 
The null distribution in Section \ref{sec:2.2.2} assumes that patients are independent with each other, and the occurrence of each feature are independent Bernoulli random variables:
\begin{equation*}
    H_0:w_{i,t}  \stackrel{i.i.d}{\sim} {\rm Multinomial}(1, \{p_{w}\}_{w=1}^d) \text{ for }   { 1 \leq t \leq T_i }, 1 \leq i \leq n.
\end{equation*}
Following similar analysis procedure in Lemma \ref{thm1}, we have
\begin{equation*}
        \text{Var}(\Eb_{w,w'})= \text{Var}(\Eb_S(w,w'))(1+o(1)).
\end{equation*}
Then the variance of each entry of the residual matrix $\Eb$ can be estimated by
\zx{
\begin{equation*}
\begin{aligned}
\text{Var}\big(\Eb_{w, w^{\prime}}\big)& \approx \text{Var}(\Eb_S(w,w'))
=\text{Var}\Big(\frac{\sum_{i=1}^n X_{i, w, w^{\prime}}^{[q]}}{n N_{w, w^{\prime}}^{[q]}} - \frac{\sum_{i=1}^n X_{i, w}^{[q]}}{n N_w^{[q]}} - \frac{\sum_{i=1}^n X_{i, w^{\prime}}^{[q]}}{n N_{w^{\prime}}^{[q]}}\Big)\\
&= \frac{1}{n^2}\sum_{i=1}^n\text{Var}\Big(\frac{ X_{i, w, w^{\prime}}^{[q]}}{N_{w, w^{\prime}}^{[q]}} - \frac{X_{i, w}^{[q]}}{N_w^{[q]}} - \frac{X_{i, w^{\prime}}^{[q]}}{N_{w^{\prime}}^{[q]}}\Big)\\
&= \frac{1}{n^2}\sum_{i=1}^n (\frac{T_i-q}{T-q})^2 \text{Var}\Big(\frac{ X_{i, w, w^{\prime}}^{[q]}}{N_{i,w, w^{\prime}}^{[q]}} - \frac{X_{i, w}^{[q]}}{N_{i,w}^{[q]}} - \frac{X_{i, w^{\prime}}^{[q]}}{N_{i,w^{\prime}}^{[q]}}\Big)
\end{aligned}
\end{equation*}
To simplify the expression, we define $T_0 = T q - q^2$. Denote
\begin{equation*}
    \bSigma_{w w} = \text{Cov}(\frac{\sqrt{n T}}{d}\Eb_{w, \cdot});\quad  \bSigma_{w w'} = \frac{n T}{d^2}\text{Cov}(\Eb_{w, \cdot}, \Eb_{w', \cdot}).
\end{equation*}
Under $H_0$, the entries of $\bSigma_{ww}$ and $\bSigma_{w w'}$ can be estimated by:
\begin{equation*}
    \begin{split}
        &\hat\bSigma_{w w}(w,w) = \frac{n T}{d^2}\hat{\text{Var}}(\Eb(w,w)) = \frac{ (1-\hat p_w)^2}{d^2 q \hat p_w^2} \\
        &\hat\bSigma_{w w}(w,w') = \hat\bSigma_{w w'}(w,w) = \frac{n T}{d^2} \hat{\text{Cov}}(\Eb(w,w),\Eb_{w,w'}) = 
        \frac{ \hat p_w-1}{d^2  q\hat p_w}\\
        &\hat\bSigma_{w w}(w',w') = \hat\bSigma_{w w'}(w',w) = \frac{n T}{d^2} \hat{\text{Var}}(\Eb_{w,w'}) = \frac{ (1-\hat p_w-\hat p_{w'}+2\hat p_w \hat p_{w'})}{d^2 2 q \hat p_w \hat p_{w'}} \\
        &\hat\bSigma_{w w}(w',w'') = \hat\bSigma_{w w''}(w',w) = \hat\bSigma_{w' w''}(w,w) =\frac{n T}{d^2} \hat{\text{Cov}}(\Eb_{w,w'},\Eb(w,w'')) = 
    \frac{ 2\hat p_w- 1}{d^2 2 q \hat p_w}\\
     &\hat\bSigma_{w w'}(w,w') = \hat\bSigma_{w w'}(w,w'') =\hat\bSigma_{w w'}(w'',w''') =\frac{n T}{d^2}\hat{\text{Cov}}(\Eb(w,w),\Eb(w',w')) =  \frac{1}{d^2 q}.
    \end{split}
\end{equation*}
Here $w,w',w'',w'''$ are four different features, and $\hat p_w = \sum_{i=1}^n X_{i,w}^{[q]}/nT_0, 1 \leq w\leq d$. }We can further represent $\hat\bSigma_{ww}$ and $\hat\bSigma_{w w'}$ in the matrix form \eqref{eq:var_null_dist}.
According to the results in Section \ref{B3}, $\Pb^{\star}\Eb +\Eb\Pb^{\star}$ is the leading term of $\tilde\PMIbb - \PMIbb$, and
\begin{equation*}
    \begin{split}
        \text{Cov}&\left((\Pb^{\star}\Eb_S +\Eb_S\Pb^{\star})\be_i\right) = \text{Cov}(\Pb^{\star}\Eb_S(\cdot,i) + \sum_{j=1}^d \Eb_S(\cdot,j)\Pb^{\star}_{ji})\\
        =& \Pb^{\star}\text{Cov}(\Eb_S(\cdot, i))\Pb^{\star} + \sum_{j=1}^d \Pb^{\star 2}_{ji}\text{Cov}(\Eb_S(\cdot,j)) + \sum_{j \neq l} \Pb^{\star}_{ji}\Pb^{\star}_{li}\text{Cov}(\Eb_S(\cdot,j),\Eb_S(\cdot, l)) \\
        &+ \sum_{j=1}^d \big(\Pb_{ji}^{\star}\text{Cov}(\Eb_S(\cdot,j),\Eb_S(\cdot, i))\Pb^{\star} + \Pb^{\star}_{ji}\Pb^{\star}\text{Cov}(\Eb_S(\cdot,i),\Eb_S(\cdot, j)) \big).
    \end{split}
\end{equation*}
Hence given $\hat\Pb$, $\bSigma_{ww}$ and $\bSigma_{w w'}$, we can estimate the covariance matrix for each row of the low-rank estimator by
\begin{equation}\label{3.17}
\begin{split}
    \hat{\text{Cov}}(\tilde{\PMIbb}_{i \cdot}) 
    = &\hat{\Pb} \hat{\text{Cov}} (\Eb_{\cdot i}) \hat{\Pb} + \sum_{j=1}^{d}\hat{\mathbf{P}}_{j i}^2 \hat{\text{Cov}} (\Eb_{\cdot j}) + \sum_{1 \leq j \neq l \leq d}\hat{\mathbf{P}}_{j i} \hat{\mathbf{P}}_{l i} \hat{\text{Cov}} (\Eb_{\cdot j}, \Eb_{\cdot l}) +\\ &\sum_{j=1}^{d} (\hat{\mathbf{P}}_{j i} \hat{\text{Cov}} (\Eb_{\cdot j}, \Eb_{\cdot i}) \hat{\Pb}  + \hat{\mathbf{P}}_{j i}\hat{\Pb}\hat{\text{Cov}} (\Eb_{\cdot i}, \Eb_{\cdot j}) ),
\end{split}
\end{equation}
where $\tilde{\PMIbb}$ and $\hat{\Pb}$ are defined in Section \ref{sec:est}.
To reduce computation complexity, we decompose the formula (\ref{3.17}) into four main parts:
\begin{equation*}
    \begin{split}
        (1) \text{  } \hat\Pb \text{Cov}(\Eb_{i \cdot})\hat\Pb = & \frac{1}{2nT_0 p_i} \left[ (\hat\Pb \boldsymbol{1})(\boldsymbol{1}^{\top} \hat\Pb)(2 p_i - 1) - (\hat\Pb\boldsymbol{1})(\be_i^{\top} \hat\Pb) - (\hat\Pb\be_i)(\boldsymbol{1}^{\top}\hat\Pb)\right. \\
        &+ \left.\hat\Pb \Db  \hat\Pb(1-p_i) + (\hat\Pb \be_i)(\be_i^{\top}\hat\Pb)\frac{1}{p_i}  \right] \\
        (2) \text{  } \sum_{j=1}^d \hat\Pb_{ji}^2\text{Cov}(\Eb_{j \cdot}) = & \frac{1}{2nT_0}\Big[\boldsymbol{1}\boldsymbol{1}^{\top}\Big( 2\hat\Pb_{ii} - \sum_{j=1}^d a_j \Big)  - \boldsymbol{1}\boldsymbol{a}^{\top} - \boldsymbol{a} \boldsymbol{1}^{\top} +  \Db \Big(\sum_{j=1}^d a_j - \hat\Pb_{ii}\Big) + \text{diag}(\boldsymbol{b}) \Big],
    \end{split}
\end{equation*}
where $\Db = \text{diag}(1/p_1, \cdots, 1/p_d)$; 
$\boldsymbol{a} = (a_1, \cdots, a_d)^{\top}, a_j = \frac{\hat\Pb_{ji}^2}{p_j}$ and $\boldsymbol{b} = (b_1, \cdots, b_d), b_j = \frac{\hat\Pb_{ji}^2}{p_j^2}$. Denote $s_i = \sum_{k=1}^d\hat\Pb_{ki}$, then we have
\begin{equation*}
    \begin{split}
        (3)  &\sum_{1 \leq j \neq l \leq d} \hat\Pb_{ji}\hat\Pb_{li}\text{Cov}(\Eb_{j \cdot}, \Eb_{l \cdot}) = \frac{1}{2nT_0}\left[\boldsymbol{1}\boldsymbol{1}^{\top}(2s_i^2 - 2\hat\Pb_{ii})
    \right.- \boldsymbol{1}\sum_{j=1}^d\frac{\hat\Pb_{ji}(s_i - \hat\Pb_{ji})}{p_j}\be_j^{\top} \\
    &- \sum_{l=1}^d \frac{\hat\Pb_{li}(s_i - \hat\Pb_{li})}{p_l} \be_l \boldsymbol{1}^{\top} - (s_i^2 - \hat\Pb_{ii})\Db  +  \left. \Db \hat\Pb_{\cdot i} \hat\Pb_{i \cdot}  \Db - \text{diag}(\boldsymbol{b}) \right]
    \end{split}
\end{equation*}
\begin{equation*}
    \begin{split}
    &(4) \sum_{j=1}^d (\hat\Pb_{ji}\text{Cov}(\Eb_{j \cdot}, \Eb_{i \cdot})\hat\Pb + \hat\Pb_{ji}\hat\Pb\text{Cov}(\Eb_{i \cdot}, \Eb_{j \cdot})) = \frac{1}{nT_0}\left[(\boldsymbol{1}\boldsymbol{1}^{\top}\hat\Pb + \hat\Pb\boldsymbol{1}\boldsymbol{1}^{\top})s_i \right. - \boldsymbol{1}(\sum_{j=1}^d\frac{\hat\Pb_{ji}}{2 p_j}\be_j^{\top})\hat\Pb \\
    &-\! \hat\Pb(\sum_{j=1}^d\frac{\hat\Pb_{ji}}{2p_j}\be_j)\boldsymbol{1}^{\top} \!-\! \frac{s_i}{2 p_i}\left(\be_i\boldsymbol{1}^{\top}\hat\Pb + \hat\Pb\boldsymbol{1}\be_i^{\top} \right) - \frac{s_i}{2}\left(\Db\hat\Pb + \hat\Pb \Db \right)+ \frac{1}{2}\sum_{j=1}^d\frac{\hat\Pb_{ji}}{p_i p_j}\left(\be_i \be_j^{\top}\hat\Pb + \hat\Pb \be_j\be_i^{\top} \right) \Big].
    \end{split}
\end{equation*}

\end{section}

\section{Additional Numerical Results on EHR}
\label{app:additional}
\begin{table}[htb]
    \begin{tabular}{l|c|cccccc}
        \toprule
        Relation Type & Category & PMI test & $\PMIbbhat$ & Sap & PubMed & Bio & Bert \\
        \hline
        \multirow{4}{*}{Similarity} & PheCode Hierarchy & \textbf{0.924} & 0.148 & 0.727 & 0.522 & 0.476 & 0.488\\
        & Drug-Drug & \textbf{0.795} & 0.286 & 0.691 & 0.739 & 0.798 & 0.777\\
        \cline{2-8}
        & average & \textbf{0.863} & 0.213 & 0.710 & 0.624 & 0.628 & 0.624 \\
        \hline
        \multirow{5}{*}{Relatedness} & Disease-Disease & \textbf{0.877} & 0.148 & 0.553 & 0.514 & 0.480 & 0.449\\
        & Disease-Drug & \textbf{0.808} & 0.166 & 0.377 & 0.435 & 0.397 & 0.392\\
        & Disease-Procedure & \textbf{0.760} & 0.126 & 0.534 & 0.486 & 0.384 & 0.368\\
        & Disease-Lab & \textbf{0.686} & 0.114 & 0.406 & 0.376 & 0.336 & 0.323\\
        \cline{2-8}
        & average & \textbf{0.808} & 0.151 & 0.454 & 0.463 & 0.411 & 0.397\\
        \bottomrule
    \end{tabular}
    \caption{Power of detecting known relation pairs with our algorithm compared with cosine similarity, under target FDR being $0.1$. The last block shows the power of detecting similar pairs (related pairs) weighted by the number of pairs within each category.}
    \label{tab:app_power_pairs}
\end{table}

We apply the proposed algorithm to the real-world EHR data and compare it performance of detecting known relationship pairs with other benchmarks. Compared to the Table~\ref{tab:power_pairs} with target FDR being $0.05$, the power of our method under the target FDR being $0.1$ gains increase as expected. The comparison of two tables exhibit the flexibility of the proposed algorithm as users are able to adjust the threshold by prior information of the data set.

Here, we also list the top twenty features that are tested to be related with Alzhimer's Disease code (PheCode:290.11), including the feature id, feature description obtained online along with the estimated low-rank PMI and the $p$-value computed by the porposed algorithm. All the features are highly correlated with the target disease with a substantial body of research.

\begin{table}[htb]
    \centering
    \begin{tabular}{l|ll|rr}
        \toprule
        group & code & description &  $\tilde \PMIbb$ & $\log(p)$\\
        \hline
        \multirow{8}{*}{PheCode} & PheCode:290.1 & dementias & 3.652 & -211780\\
        & PheCode:290.16 & vascular dementia & 2.932 & -137002\\
        & PheCode:290 & delirium dementia & 3.127 & -130891\\
        & PheCode:290.3 & other persistent mental disorders & 2.208 & -78442 \\
        & PheCode:290.13 & senile dementia & 2.740 & -70039\\
        & PheCode:290.12 & dementia with cerebral degenerations & 3.155 & -45859\\
        & PheCode:599.4 & urinary incontinence & 1.704 & -24929\\
        & PheCode:295.3 & psychosis & 1.278 & -20861 \\
        \hline
        \multirow{12}{*}{RxNorm} & RXNORM:6719 & memantine & 2.672 & -167316 \\
        & RXNORM:135447 & donepezil & 1.864 & -119347 \\
        & RXNORM:4637 & galantamine & 1.962 & -98585 \\
        & RXNORM:183379 & rivastigmine & 2.483 & -41429 \\
        & RXNORM:51272 & quetiapine & 1.113 & -26221 \\
        & RXNORM:35636 & risperidone & 1.067 & -25219 \\
        & RXNORM:709312 & diaper & 1.913 & -24128 \\
        & RXNORM:11118 & valproate & 1.108 & -22570 \\
        & RXNORM:5093 & haloperidol & 0.869 & -22564 \\
        & RXNORM:11248 & cyanocobalamin & 0.932 & -21036 \\
        & RXNORM:25025 & finasteride & 0.878 & -20888 \\
        & RXNORM:36437 & sertraline & 0.789 & -19816\\
        \bottomrule
        \end{tabular}
    \caption{Top twenty features tested to be related with Alzheimer's Disease code (PheCode:290.11) with the lowest $p$-value computed by the proposed algorithm.}
    \label{tab:AD_test_app}
\end{table}

\end{document}